\documentclass[10pt,a4paper]{article}
\usepackage[english]{babel}
\usepackage[latin1]{inputenc}
\usepackage{amsfonts,amsbsy,bm,euscript,mathrsfs}
\usepackage{amssymb,stmaryrd,faktor}
\usepackage[tbtags]{amsmath}
\usepackage{bbm}
\usepackage{graphicx}
\usepackage[title,titletoc]{appendix}
\usepackage[bookmarks=true,colorlinks=true,linkcolor=blue,citecolor=blue,urlcolor=blue,bookmarksnumbered]{hyperref}

\usepackage{dsfont}
\usepackage{collref}
\usepackage{lmodern}
\usepackage{mathrsfs}
\usepackage{mathtools}
\usepackage{bbm}
\usepackage{braket}
\usepackage{slashed}
\usepackage{float}
\usepackage{graphicx}
\usepackage{booktabs}
\usepackage{subfig}
\usepackage{tikz}
\usepackage[symbol]{footmisc}
\usepackage{hyperref}
\usepackage{cleveref}
\usetikzlibrary{plotmarks,calc,decorations,decorations.pathmorphing}

\numberwithin{equation}{section}

\makeatletter
\renewcommand\section{\@startsection {section}{1}{\z@}
{-3.5ex \@plus -1ex \@minus -.2ex}
{2.3ex \@plus.2ex}
{\normalfont\Large\bfseries}}
\renewcommand\subsection{\@startsection{subsection}{2}{\z@}
{-3.25ex\@plus -1ex \@minus -.2ex}
{1.5ex \@plus.2ex}
{\normalfont\large\bfseries}}
\makeatother

\newcommand{\arXivlink}[1]{\href{http://arXiv.org/abs/#1}{arXiv:#1}}

\newcommand{\alg}[1]{\mathfrak{#1}}

\newcommand{\sech}{\mathrm{sech} \,}

\usepackage{blkarray}

\begin{document}

\thispagestyle{empty}
\begin{flushright}\footnotesize\ttfamily
DMUS-MP-23/16
\end{flushright}
\vspace{2em}

\begin{center}

{\Large\bf \vspace{0.2cm}
{\color{black} \large Boundary scattering in massless $AdS_3$}} 
\vspace{1.5cm}

\textrm{Daniele Bielli$^{\dag , \ddag}$, Vaibhav Gautam$^{\ddag , \S}$, Vasileios Moustakis$^{\ddag}$,
\\
Andrea Prinsloo$^{\ddag}$ and Alessandro Torrielli$^{\ddag}$ \footnote[1]{\textit{E-mail:} \texttt{d.bielli4@gmail.com, \{v.gautam@, v.moustakis@, a.prinsloo@, a.torrielli@\}surrey.ac.uk}}}
\vspace{0.8cm}
\\
\vspace{0.3cm}
$^\dag$ \textit{High Energy Physics Research Unit, Faculty of Science,
Chulalongkorn University,
\\
Bangkok 10330, Thailand}
\\
\vspace{0.3cm}
$^\ddag$ \textit{School of Mathematics and Physics, University of Surrey, Guildford, GU2 7XH, UK}
\\
\vspace{0.3cm}
$^\S$ \textit{School of Mathematical Sciences, Queen Mary University of London,
\\
Mile End Road, London, E1 4NS, United Kingdom}




\end{center}

\vspace{2em}

\begin{abstract}\noindent 
We study the boundary integrability problem of the massless sector of $AdS_3 \times S^3 \times T^4 $ string theory. Exploiting the difference-form of the massless scattering theory, we find a very simple and exhaustive list of reflection matrices for all the possible boundary coideal subalgebras - singlet and vector representations, right and left boundary - and check basic properties of our solutions, primarily the boundary Yang-Baxter equation, for all possible combinations of scattering particles.
\end{abstract}


\newpage

\overfullrule=0pt
\parskip=2pt
\parindent=12pt
\headheight=0.0in \headsep=0.0in \topmargin=0.0in \oddsidemargin=0in

\vspace{-3cm}
\thispagestyle{empty}
\vspace{-1cm}

\tableofcontents

\setcounter{footnote}{0}

\section{Introduction} \label{Sec::Intro}

\subsection{Integrability in $AdS_3$}
String theory in $AdS_3 \times S^3 \times T^4$ \cite{Bogdan}, see also the reviews \cite{rev3,Borsato:2016hud}, continues to be a very fertile playground in which to test unconventional ideas in integrability \cite{Beisertreview}. A large body of work is by now available \cite{OhlssonSax:2011ms,seealso3,Borsato:2012ss,Borsato:2013qpa,Borsato:2014hja,Borsato:2013hoa}. The important novel feature of this system, as opposed to the higher-dimensional counterparts $AdS_{5,4}$, is given by the presence of massless modes \cite{Sax:2012jv} which activate the machinery of massless $S$-matrices \cite{Zamol2,Fendley:1993jh,DiegoBogdanAle}. Various aspects of the massless sector have been explored, see for instance \cite{Lloyd:2013wza} and \cite{Ben}. More recently, the Quantum Spectral Curve (QSC) method has been formulated for massive $AdS_3 \times S^3 \times T^4$ in \cite{QSC}, see also \cite{Cavaglia:2022xld}. A massless version is however still missing. A recent body of work \cite{AleSSergey}, see also \cite{Seibold:2022mgg}, has revisited the problem $AdS_3 \times S^3 \times T^4$ and obtained a new dressing phase and the formulation of the mirror TBA. Very recent exciting new developments can now be found in \cite{recent}.

In this paper, we will study the boundary scattering theory \cite{GhoshalZamo} of the massless modes in $AdS_3 \times S^3 \times T^4$. The massive version of the boundary integrability programme was developed in \cite{Prinsloo:2015apa}, which provides guidance for our massless analysis. We refer to \cite{Prinsloo:2015apa} for a list of references on the boundary scattering literature which is relevant to $AdS_3$ and to $AdS_5$ string theory.  

When dealing with the case of $AdS_3$,  \cite{gamma1,gamma2} showed the existence of a change of variables which displays the complete difference-form of the  massless $S$-matrix, a form which is retained in the non-trivial BMN limit \cite{DiegoBogdanAle}. A massive version of this variable was introduced in \cite{AleSSergey} where it is used to rewrite portions of the massive $S$-matrix in difference form. 

The paper is organized as follows: in the rest of this introductory section \ref{Sec::Intro}, we setup the notations and conventions, and provide the bulk $R$-matrices for the symmetric co-product considered in this paper. In section \ref{sec::Ref Mat}, we solve for the reflection matrices of a right wall, in the case in which the boundary carries a trivial representation with respect to the preserved boundary subalgebra, i.e. the \textit{singlet} boundary, and also in the case where it carries a non-trivial representation, i.e. the \textit{vector} boundary. Subalgebras which are preserved by the boundary need to satisfy the \textit{coideal subalgebra} condition, which imposes non-trivial constraints and forces us to choose a different basis on the boundary. In section \ref{sec::left wall}, we repeat the calculations for the case of a left wall. In section \ref{massive-case}, we consider again a right wall and compute the reflection matrices for the massive excitations using the symmetric coproduct, so as to make comparisons with the massless case by taking appropriate limits. In section \ref{Gen Vec Bound}, we reverse our point of view, constructing an ansatz that generalises the vector boundary reflection matrices of the previous sections and exploiting it to build new vector boundary representations.

\subsection{$AdS_3$ massless conventions}

In the bulk the massless modes are characterised by the following dispersion relation:
\begin{eqnarray}
&&\mbox{right movers} \qquad E_i = 2h \sin \frac{p_i}{2}>0, \qquad p_i \in (0,\pi), \qquad x_i^\pm = e^{\pm \frac{i p_i}{2}},\nonumber\\
&&\mbox{left movers} \qquad E_i = -2h \sin \frac{p_i}{2}>0, \qquad p_i \in (-\pi,0), \qquad x_i^\pm = -e^{\pm \frac{i p_i}{2}},\label{addi}
\end{eqnarray}
with $i=1,2$ denoting the scattering particles.
In these formulae $h>0$ is the coupling constant of the theory.
The right and left nature of a massless particle is referred to as {\it chirality}, and one sets
\begin{eqnarray}
\mbox{right movers} \leftrightarrow \mbox {chirality +}, \qquad \mbox{left movers} \leftrightarrow \mbox {chirality -}.
\end{eqnarray}
The scattering matrices are of difference form if one uses the pseudo-relativistic variable, which was introduced in \cite{gamma1,gamma2} and further extended in \cite{Majumder:2021zkr,AleSSergey}: 
\begin{eqnarray}
&&\gamma_i = \log \tan \frac{p_i}{4}<0, \quad\,\,\,\, p_i \in (0,\pi),\nonumber\\
&&\gamma_i = \log \cot \frac{-p_i}{4}>0, \quad p_i \in (-\pi,0).\label{acco}
\end{eqnarray}
The basic representation for a single particle is always based on a supersymmetric (boson,fermion) doublet ordered as $\{|\phi\rangle,|\psi\rangle\}$, carrying various representations of the superalgebra $\mathfrak{psu}(1|1)^2$ which we subsequently define in the paper. 

The bulk scattering can occur in a number of ways, according to the particles' chirality. We list here the $R$-matrices in all the different cases, as presented in \cite{AleSSergey}. The basis is always ordered as $\{|\phi\rangle, |\psi\rangle\}$, with $|\phi\rangle$ a boson and $|\psi\rangle$ a fermion: for the scattering one orders according to $\{|\phi\rangle \otimes |\phi\rangle,|\phi\rangle \otimes |\psi\rangle,|\psi\rangle \otimes |\phi\rangle,|\psi\rangle \otimes |\psi\rangle\}$:
\begin{eqnarray}
&&R_{++} = \Sigma_{++}(\gamma)\begin{pmatrix}1&0&0&0\\0&-\tanh\frac{\gamma}{2}&\mbox{sech} \frac{\gamma}{2}&0\\0&\mbox{sech}\frac{\gamma}{2}&\tanh\frac{\gamma}{2}&0\\0&0&0&-1\end{pmatrix},\qquad \gamma_1,\gamma_2<0\nonumber\\
&&R_{--} =\Sigma_{--}(\gamma)\begin{pmatrix}1&0&0&0\\0&\tanh\frac{\gamma}{2}&\mbox{sech} \frac{\gamma}{2}&0\\0&\mbox{sech}\frac{\gamma}{2}&-\tanh\frac{\gamma}{2}&0\\0&0&0&-1\end{pmatrix},\qquad \gamma_1,\gamma_2>0,\nonumber\\
&&R_{+-} =\Sigma_{+-}(\gamma)\begin{pmatrix}1&0&0&0\\0&-\tanh\frac{\gamma}{2}&-i\mbox{sech} \frac{\gamma}{2}&0\\0&-i\mbox{sech}\frac{\gamma}{2}&-\tanh\frac{\gamma}{2}&0\\0&0&0&1\end{pmatrix},\qquad \gamma_1<0,\gamma_2>0,\nonumber\\
&&R_{-+} =\Sigma_{-+}(\gamma)\begin{pmatrix}1&0&0&0\\0&\tanh\frac{\gamma}{2}&i\mbox{sech} \frac{\gamma}{2}&0\\0&i\mbox{sech}\frac{\gamma}{2}&\tanh\frac{\gamma}{2}&0\\0&0&0&1\end{pmatrix},\qquad \qquad \gamma_1>0,\gamma_2<0,\label{exactsame}
\end{eqnarray}
where 
\begin{eqnarray}
\gamma := \gamma_1 - \gamma_2,
\end{eqnarray}
and $\gamma_1$ and $\gamma_2$ go according to (\ref{acco}) in the various cases. The $S$-matrices are easily obtained from the $R$-matrices by the formula
\begin{eqnarray}
S=\Pi \circ R,
\end{eqnarray}
where $\Pi$ is the graded permutation on states. Because of this relationship, we will often colloquially speak of attributes of the $S$-matrix even when we are expressing a property of the $R$-matrix - with the understanding that the formulae will instead be precise and always technically distinguish between the two.

\subsection{Symmetries \label{symma}}

The symmetry algebra underlying the scattering theory is a Yangian-type quantum group based on the centrally-extended $\mathfrak{psu}(1|1)^2$ Lie superalgebra \footnote{The complete algebra is $\mathfrak{psu}(1|1)^4$, but in this paper it will be sufficient to focus on one copy of $\mathfrak{psu}(1|1)^2$ and ignore in the first instance the additional so-called $su(2)_\circ$ index.}
\begin{eqnarray}
\{\mathfrak{Q}_L,\mathfrak{G}_L\}=\mathfrak{H}_L,\quad \{\mathfrak{Q}_R,\mathfrak{G}_R\}=\mathfrak{H}_R,\quad\{\mathfrak{Q}_L,\mathfrak{Q}_R\}=\mathfrak{P},\quad\{\mathfrak{G}_L,\mathfrak{G}_R\}=\mathfrak{K}.\label{comma}
\end{eqnarray}
The generators $\mathfrak{H}_L,\mathfrak{H}_R,\mathfrak{P},\mathfrak{K}$ are central.
We use conventions inspired by \cite{Stromwall:2016dyw} for the fundamental representation. Let us focus for a moment on the right movers, which have $p_i \in (0,\pi)$ (with $i$ being either $1$ or $2$, depending on the scattering particle), so that we get
\begin{eqnarray} \label{Lrep+}
&&\pi^{L+}_{p_i}(\mathfrak{Q}_L) = - \pi^{L+}_{p_i}(\mathfrak{G}_R) = \sqrt{\frac{h}{2}}\sqrt{\sin \frac{p_i}{2}}\begin{pmatrix}0&0\\1&0\end{pmatrix} \nonumber \\ 
&&\pi^{L+}_{p_i}(\mathfrak{G}_L) = - \pi^{L+}_{p_i}(\mathfrak{Q}_R) = \sqrt{\frac{h}{2}}\sqrt{\sin \frac{p_i}{2}}\begin{pmatrix}0&1\\0&0\end{pmatrix}, \qquad p_i\in (0,\pi).\label{one8}
\end{eqnarray}
Here, $L$ in the superscript denotes the $L-$representation of $\mathfrak{psu}(1|1)^2$ algebra (more on this in Sec.~\ref{sec:LLtilreps}) while $+$ denotes the right movers.

In this representation all the central charges $\mathfrak{H}_L,\mathfrak{H}_R$ are equal to the same matrix $\tfrac{h}{2}\sin \frac{p_i}{2} \mathbbmss{1}$, while $\mathfrak{P},\mathfrak{K}$ are equal to $-\tfrac{h}{2}\sin \frac{p_i}{2} \mathbbmss{1}$.
The dispersion relation is given by
\begin{eqnarray}\label{energy}
E(p_i) = 2 (h_L + h_R),
\end{eqnarray}
where $h_L$ is the eigenvalue of $\mathfrak{H}_L$ and $h_R$ is the eigenvalue of $\mathfrak{H}_R$.
We also adopt what is dubbed the {\it old} coproduct structure in \cite{Stromwall:2016dyw}, see \cite{OhlssonSax:2011ms,Borsato:2013qpa}:
\begin{eqnarray}
&&\Delta(\mathfrak{Q}_L)=\mathfrak{Q}_L\otimes e^{-i\frac{\mathfrak{p}}{4}} + e^{i\frac{\mathfrak{p}}{4}}\otimes \mathfrak{Q}_L, \qquad \Delta(\mathfrak{G}_L)=\mathfrak{G}_L\otimes e^{i\frac{\mathfrak{p}}{4}} + e^{-i\frac{\mathfrak{p}}{4}}\otimes \mathfrak{G}_L,\nonumber\\
&&\Delta(\mathfrak{Q}_R)=\mathfrak{Q}_R\otimes e^{-i\frac{\mathfrak{p}}{4}} + e^{i\frac{\mathfrak{p}}{4}}\otimes \mathfrak{Q}_R, \qquad \Delta(\mathfrak{G}_R)=\mathfrak{G}_R\otimes e^{i\frac{\mathfrak{p}}{4}} + e^{-i\frac{\mathfrak{p}}{4}}\otimes \mathfrak{G}_R.\label{simo}
\end{eqnarray}
The momentum generator $\mathfrak{p}$ takes the value $p_1$ when it is in the first space, $p_2$ when in the second space of the tensor product.
The coproducts of the central charges can be obtained from the commutation relations (\ref{comma}) using the homomorphism property. In Appendix \ref{App:C} we see how this symmetric co-product is related to the one used in \cite{Prinsloo:2015apa}. The symmetry condition reads
\begin{eqnarray}
\Delta^{op}(x) R_{++} = R_{++} \Delta(x)
\end{eqnarray}
for any generator $x$ in the Lie superalgebra. We then say that the $R$-matrix {\it intertwines} (is an intertwiner for) the given coproduct. 

The structure which we have just outlined can be considered as the {\it level-0} of the Yangian. To these charges we can associate {\it level-1} corresponding charges (in the so-called {\it Drinfeld's first realisation}):
\begin{eqnarray}
\widehat{\mathfrak{Q}}_L = u_i \, \mathfrak{Q}_L, \qquad \widehat{\mathfrak{G}}_L = u_i \, \mathfrak{G}_L, \qquad\widehat{\mathfrak{Q}}_R = u_i \, \mathfrak{Q}_R, \qquad\widehat{\mathfrak{G}}_R = u_i \, \mathfrak{G}_R,\qquad u_i = i \cos \frac{p_i}{2}\label{eval}
\end{eqnarray}
(again with $i=1,2$ depending on the scattering particle)
with coproducts
\begin{eqnarray}
&&\Delta(\widehat{\mathfrak{Q}}_L)=\widehat{\mathfrak{Q}}_L\otimes e^{-i\frac{\mathfrak{p}}{4}} + e^{i\frac{\mathfrak{p}}{4}}\otimes \widehat{\mathfrak{Q}}_L + \mathfrak{Q}_L \otimes e^{-i\frac{\mathfrak{p}}{4}}\mathfrak{H}_L - e^{i\frac{\mathfrak{p}}{4}}\mathfrak{H}_L \otimes \mathfrak{Q}_L, \nonumber\\
&&\Delta(\widehat{\mathfrak{G}}_L)=\widehat{\mathfrak{G}}_L\otimes e^{i\frac{\mathfrak{p}}{4}} + e^{-i\frac{\mathfrak{p}}{4}}\otimes \widehat{\mathfrak{G}}_L - \mathfrak{G}_L \otimes e^{i\frac{\mathfrak{p}}{4}}\mathfrak{H}_L + e^{-i\frac{\mathfrak{p}}{4}}\mathfrak{H}_L \otimes \mathfrak{G}_L, \nonumber\\
&&\Delta(\widehat{\mathfrak{Q}}_R)=\widehat{\mathfrak{Q}}_R\otimes e^{-i\frac{\mathfrak{p}}{4}} + e^{i\frac{\mathfrak{p}}{4}}\otimes \widehat{\mathfrak{Q}}_R + \mathfrak{Q}_R \otimes e^{-i\frac{\mathfrak{p}}{4}}\mathfrak{H}_R - e^{i\frac{\mathfrak{p}}{4}}\mathfrak{H}_R \otimes \mathfrak{Q}_R, \nonumber\\
&&\Delta(\widehat{\mathfrak{G}}_R)=\widehat{\mathfrak{G}}_R\otimes e^{i\frac{\mathfrak{p}}{4}} + e^{-i\frac{\mathfrak{p}}{4}}\otimes \widehat{\mathfrak{G}}_R - \mathfrak{G}_R \otimes e^{i\frac{\mathfrak{p}}{4}}\mathfrak{H}_R + e^{-i\frac{\mathfrak{p}}{4}}\mathfrak{H}_R \otimes \mathfrak{G}_R.\label{check}
\end{eqnarray}
One can once again check using (\ref{check}) that 
\begin{eqnarray}
\Delta^{op}(\widehat{x}) R_{++} = R_{++} \Delta(\widehat{x})
\end{eqnarray}
for any level-1 $\hat{x}$ in the Lie superalgebra. The coproduct of the level-1 central charges $\widehat{\mathfrak{H}}_L,\widehat{\mathfrak{H}}_R,\widehat{\mathfrak{P}},\widehat{\mathfrak{K}}$ can be obtained from the homomorphism property by recalling that
\begin{eqnarray}
[\widehat{x}^a,x^b\} = i f^{ab}_c \widehat{x}^c
\end{eqnarray}
for any generator $x^a$ of the Lie superalgebra. The representation (\ref{eval}) is called {\it evaluation representation}, with {\it evaluation parameter} $u_i$. 

When considering the left moving chirality, we need to change the representation: the change is actually minimal, since everything remains the same except that one needs to replace 
\begin{eqnarray}\label{Lrep-}
&&\pi^{L-}_{p_i}(\mathfrak{Q}_L) = + \pi^{L-}_{p_i}(\mathfrak{G}_R) = \sqrt{\frac{h}{2}}\sqrt{\sin \frac{-p_i}{2}}\begin{pmatrix}0&0\\1&0\end{pmatrix}, \nonumber\\
&&\pi^{L-}_{p_i}(\mathfrak{G}_L) = + \pi^{L-}_{p_i}(\mathfrak{Q}_R) = \sqrt{\frac{h}{2}}\sqrt{\sin \frac{-p_i}{2}}\begin{pmatrix}0&1\\0&0\end{pmatrix}, \quad p_i \in (-\pi,0)\label{lefto}
\end{eqnarray}
with $-$ in the superscript denoting the left movers (the central charges again being obtained using the commutation relations).
The coproducts remain the same, and one can make mixed representations where different chiralities are in the first, resp. second space of the tensor product, and one needs to adjust the representations everywhere accordingly. When considering the level-1 Yangian charges, we have systematically checked that the coproducts can again be taken as previously - that is (\ref{check}) - and the representation to be used for left movers will now be (\ref{lefto}). In addition, the evaluation representation for left movers has to be set to 
\begin{eqnarray}
\widehat{\mathfrak{Q}}_L = -u_i \, \mathfrak{Q}_L, \qquad \widehat{\mathfrak{G}}_L = -u_i \, \mathfrak{G}_L, \qquad\widehat{\mathfrak{Q}}_R = -u_i \, \mathfrak{Q}_R, \qquad\widehat{\mathfrak{G}}_R = -u_i \, \mathfrak{G}_R,\qquad u_i = i \cos \frac{-p_i}{2}.\label{eval1}
\end{eqnarray}

\subsection{Boost symmetry}

In the paper \cite{Stromwall:2016dyw} a boost symmetry was shown to complement the level-0 Lie superalgebra into a super-Poincar\'e algebra. For a right moving representation with $p_i\in (0,\pi)$ the one-particle representation of the boost generator is given by
\begin{eqnarray}
\mathfrak{J} = i  \sin \frac{p_i}{2} \frac{\partial}{\partial p_i}, \qquad i=1,2,
\end{eqnarray} 
with coproduct
\begin{eqnarray}
\Delta(\mathfrak{J}) = \mathfrak{J} \otimes e^{i \frac{\mathfrak{p}}{2}}+e^{-i\frac{\mathfrak{p}}{2}}\otimes \mathfrak{J} + \frac{1}{2} \mathfrak{Q}_L e^{-i \frac{\mathfrak{p}}{4}} \otimes \mathfrak{G}_L e^{i \frac{\mathfrak{p}}{4}}+\frac{1}{2} \mathfrak{G}_L e^{-i \frac{\mathfrak{p}}{4}} \otimes \mathfrak{Q}_L e^{i \frac{\mathfrak{p}}{4}}.
\end{eqnarray}
The complete study of the boost superalgebra and of how to achieve a covariant formulation of the commutation relations is presented in \cite{JM}. The coproduct is a homomorphism only if one adopts a different (and equally valid) coproduct as opposed to the one which we have adopted in section \ref{symma}. The coproduct which is suitable to the boost is what is called the {\it new} coproduct in \cite{Stromwall:2016dyw}, which only works thanks to the very special properties of the massless case. 

The boost coproduct is not strictly speaking an intertwiner of the $S$-matrix, but it was shown in \cite{Fontanella:2016opq} that it annihilates it\footnote{Because of the tail in the boost coproduct, annihilating the $S$ matrix is not exactly equivalent to the ordinary condition one has in relativistic theories, namely that the boost commutes with the $S$-matrix. We thank Niklas Beisert and Ben Hoare for discussions on this point.}:
\begin{eqnarray}
\Delta(\mathfrak{J}) \check{R}_{++} = 0 = \Delta^{op}(\mathfrak{J}) \check{R}_{++},
\end{eqnarray}
having defined
\begin{eqnarray}
\check{R}_{++} := \frac{R_{++}}{\Sigma_{++}}
\end{eqnarray}
(the dressing phase can be incorporated by adding a constant term to the differential operator, as further elaborated in \cite{Borsato:2017lpf,Borsato:2017icj} and in \cite{DiegoBogdanAle}, see also \cite{qseealso}).
This translates into a differential constraint which the matrix $R_{++}$ can be seen to solve, and which can be recast as a differential-geometric condition:
\begin{eqnarray}
D_i \check{R}_{++} = \Bigg[\frac{\partial}{\partial p_i} + \Gamma_i\Bigg]\check{R}_{++} = 0, \qquad i=1,2,\label{sys}
\end{eqnarray}  
where
\begin{eqnarray}
\Gamma_1 = - \frac{1}{4}\sqrt{\frac{\sin \frac{p_2}{2}}{\sin \frac{p_1}{2}}}\frac{E_{12} \otimes E_{21} + E_{21}\otimes E_{12}}{\sin \frac{p_1+p_2}{4}}, \qquad \Gamma_2 = \frac{1}{4}\sqrt{\frac{\sin \frac{p_1}{2}}{\sin \frac{p_2}{2}}}\frac{E_{12} \otimes E_{21} + E_{21}\otimes E_{12}}{\sin \frac{p_1+p_2}{4}},
\end{eqnarray}
$E_{ij}$ being the matrix with all zeroes except $1$ in row $i$, column $j$. By making use of the difference form one can further reduce the system (\ref{sys}) to a matrix-valued ordinary differential equation:
\begin{eqnarray}
\Bigg[\frac{d}{d\gamma} - \frac{1}{2}(E_{12} \otimes E_{21} + E_{21}\otimes E_{12})\,\mbox{sech} \frac{\gamma}{2} \Bigg]\check{R}_{++}=0.
\end{eqnarray}
The solution to this equation can be expressed as
\begin{eqnarray}
\check{R}_{++} = \Pi \circ e^{-(E_{12} \otimes E_{21} + E_{21}\otimes E_{12}) \, \mbox{gd}\big(\frac{\gamma}{2}\big)},
\end{eqnarray}
where the {\it Gudermannian} function is defined as
\begin{eqnarray}
\mbox{gd}(x) = \int_0^x dy \, \mbox{sech} y = 2 \, \mbox{arctanh} \tanh \frac{x}{2}.
\end{eqnarray}
Of course one has also extracted the condition \cite{gamma1} 
\begin{eqnarray}
\Big(\frac{\partial}{\partial \gamma_1} + \frac{\partial}{\partial \gamma_2}\Big)\check{R}_{++}=0,\label{ordo}
\end{eqnarray}
which states the difference-form of the $S$-matrix.

The introduction of the super-Poincar\'e picture of massless $AdS_3$ scattering is amenable to a delightful physical analogy \cite{Stromwall:2016dyw}. A {\it phonon} excitation of momentum $p_i$ in a one-dimensional harmonic chain of lattice spacing $h^{-1}$ (in suitable conventions) has a dispersion relation
\begin{eqnarray}
E(p_i) = 2 h \, \Big\vert\sin \frac{p_i}{2} \Big\vert,
\end{eqnarray}
where $p_i \in (-\pi,\pi)$ defines the {\it Brillouin zone}. This is identical to the dispersion relation (\ref{addi}) of our massless particles. The new coproduct found in \cite{Stromwall:2016dyw} is very peculiar, in that it implies that the generators $\mathfrak{H}_L$ and $\mathfrak{H}_R$ have a non-trivial coproduct. However, since they are central, they are still forced by the intertwining equation with the $R$-matrix to be cocommutative. Indeed, by a special property of the trigonometric functions, one can check that the new coproduct satisfies (if we focus on right movers)
\begin{eqnarray}
\Delta(\mathfrak{H}_L) = \sin \frac{p_1 + p_2}{2} \, \mathbbmss{1} \otimes \mathbbmss{1}= \Delta(\mathfrak{H}_R), 
\end{eqnarray}
which is clearly cocommutative. As demonstrated in \cite{Ballesteros:1999ew}, this is the way a three-phonon process works on the harmonic chain: two phonons come in (assuming for simplicity both momenta in $(0,\frac{\pi}{2})$), one with dispersion $2 h \sin\frac{p_1}{2}$ and the other with dispersion $2 h \sin\frac{p_2}{2}$, and a third emerges with dispersion $2 h \sin \frac{p_1 + p_2}{2}$. The difference in our case is that no single particle emerges from the scattering, but always two, as dictated by integrability. The analogy is therefore only mathematical. We have also neglected the velocity parameter of the phonon scattering, which is inserted to ensure energy conservation. For recent algebraic developments see also \cite{Niklas}.

In the case of the harmonic chain, the presence of a Brillouin zone is at the origin of the so-called {\it umklapp} scattering: if we instead assume that $p_1$ and $p_2$ are both close to and slightly less than $\pi$, then the emerging third phonon can be assigned a momentum $(p_1 + p_2) \, mod \, 2\pi$ which is much smaller in absolute value than each individual $p_1$ and $p_2$, and negative:
\begin{eqnarray}
p_{1,2} = \pi - \epsilon, \qquad p_1 + p_2 = 2 \pi - 2 \epsilon \sim - 2 \epsilon. 
\end{eqnarray}

\subsection{Relativistic limit}

If one takes the following limit:
\begin{eqnarray}
h \to +\infty, \qquad p_i \to 0, \qquad h p_i = c q_i \, \, \mbox{finite},\qquad c>0,
\end{eqnarray}
one obtains a relativistic scattering theory. The limit should be understood as $0^+$ (such that $q_i>0$) for right movers, $0^-$ for left movers (such that $q_i<0$). All the $S$-matrices trivialise except those of massless right with right, and left with left, movers, and the theory is projected to a critical point in this limit - the scattering theory describes a 2D CFT \cite{DiegoBogdanAle}. The particles have dispersion relation
\begin{eqnarray}
E_i \to \pm c q_i,
\end{eqnarray} 
where the sign is $+$ for right movers and $-$ for left movers. It is customary to set
\begin{eqnarray}
q_i = \pm e^{\pm\theta_i}
\end{eqnarray}
(again $+$ for right and $-$ for left movers).
The pseudo-relativistic variable $\gamma_i$ diverges, but thanks to the difference form the constant divergence cancels out in the $S$-matrix and one is left with a genuinely relativistic rapidity $\theta_i$:
\begin{eqnarray}
&&\gamma_i \to \pm \log \frac{c}{4h} + \theta_i, \qquad h \to + \infty
\end{eqnarray}
once again $+$ being for right and $-$ for left movers. The divergent constant offsets cancel out in right-right and left-left kinematics, and the two $S$-matrices (equivalently, the two $R$-matrices $R_{++}$ and $R_{--}$) therefore retain the exact same functional form as in (\ref{exactsame}), just with $\theta = \theta_1 - \theta_2$ replacing $\gamma$. On the other hand, the mixed right-left and left-right kinematics $S$-matrices trivialise: the divergent offsets add up in such a way that the sech function of a large argument goes to $0$, and the $\tanh$ function goes to $\mp 1$ (for right-left, resp. left-right chiralities). The limits can then be seen to produce
\begin{eqnarray}
R_{+-}, R_{-+} \to \mathbbmss{1} \otimes \mathbbmss{1}.
\end{eqnarray} 
 This means that the theory has been pushed towards a critical point - right and left modes are decoupeld, and the purely right-right and left-left scattering describes a conformal field theory \cite{DiegoBogdanAle}.

In the relativistic limit the level-0 symmetry algebra (after a rescaling of the generators by appropriate powers of $h$ to ensure they remain finite) can be seen to become the ordinary ${\cal{N}}=2$ relativistic massless supersymmetry: for right movers we have
\begin{eqnarray}\label{relr}
\mathfrak{Q}_L = - \mathfrak{G}_R = e^{\frac{\theta_i}{2}}\begin{pmatrix}0&0\\1&0\end{pmatrix}, \qquad \mathfrak{G}_L = - \mathfrak{Q}_R = e^{\frac{\theta_i}{2}}\begin{pmatrix}0&1\\0&0\end{pmatrix}.
\end{eqnarray}
The level-0 coproducts are all local in the same limit, since the braiding factors (which generate a non-locality) $e^{\pm i \frac{p_i}{4}} \to 1$ and the action of two-particle symmetries becomes the simple Leibniz rule $\Delta(x) = x \otimes \mathbbmss{1} + \mathbbmss{1} \otimes x$.
A very interesting observation, which was made in \cite{Torrielli:2017nab} in a different but related context, is that the level-1 Yangian charges in Drinfeld's first realisation are given by  
\begin{eqnarray}
\widehat{\mathfrak{Q}}_L = u_i \, \mathfrak{Q}_L, \qquad \widehat{\mathfrak{G}}_L = u_i \, \mathfrak{G}_L, \qquad\widehat{\mathfrak{Q}}_R = u_i \, \mathfrak{Q}_R, \qquad\widehat{\mathfrak{G}}_R = u_i \, \mathfrak{G}_R,\qquad u_i = 0.\label{evalrel}
\end{eqnarray}
The vanishing of the charge does not mean that the coproduct vanishes, but only that the coproduct is {\it entirely tail}: one can explicitly verify that the limiting $R_{++}$ intertwines the coproducts (\ref{check}) in the representation (\ref{relr}), and with all the hatted generators set to $0$ (only the tails survive). This is rather a signature of a conformal field theory, as further elaborated in \cite{Torrielli:2017nab}.

For left movers one has instead 
\begin{eqnarray}
\mathfrak{Q}_L = +\mathfrak{G}_R = e^{\frac{-\theta_i}{2}}\begin{pmatrix}0&0\\1&0\end{pmatrix}, \qquad \mathfrak{G}_L = + \mathfrak{Q}_R = e^{\frac{-\theta_i}{2}}\begin{pmatrix}0&1\\0&0\end{pmatrix}, \label{rell}
\end{eqnarray}
and again a vanishing evaluation parameter at level 1, with the tails being exactly as in the non-relativistic case but taken in the representation (\ref{rell}). This can be directly checked by brute-force computation in the case of $R_{--}$. In the case of $R_{+-}$ and $R_{-+}$ they are trivial in the relativistic limit; hence it becomes quite pointless to discuss their symmetries.

It is amusing to notice that, by virtue of the fact that $R_{++}$ and $R_{--}$ maintain the same functional form in the non-relativistic as in the relativistic case, it follows that the {\it unbraided} ${\cal{N}}=2$ supersymmetry intertwines the non-relativistic version of those two $S$-matrices, this time with $\gamma$ replacing $\theta$ everywhere. This can also be proven by showing that the non-relativistic braided symmetry can be written in terms of the unbraided (in right-right and left-left kinematics), which has been explicitly demonstrated using the free-fermion language in \cite{DeLeeuw:2020ahx}.  

Based on the above consideration, it will not be surprising that, when expressed in the $\gamma$ variable and for right moving particles, all the expressions involving the boost which we wrote in the previous subsection are exactly preserved by the BMN limit, just with $\theta$ replacing $\gamma$ everywhere.

\subsection{$L$ and $\tilde{L}$ representations}\label{sec:LLtilreps}

Even if we have not specified it explicitly, all we have said so far was in the so-called ``$L$" representation. This is only half of the story, since one also has to consider an ``$\tilde{L}$" representation (using the naming of representations in \cite{Borsato:2014hja}). From the point of view of the scattering theory which we are concerned with, these $L$ and $\tilde{L}$ modes are just an additional distinction between possible representations - a sort of flavour degree of freedom. Each of the $L$ and $\tilde{L}$ modes separately have either right or left moving chirality. Each of the $L$ and $\tilde{L}$ representations is a representation of centrally-extended $\mathfrak{psu}(1|1)^2$, and therefore displays a matrix form for all four supercharges $\mathfrak{Q}_L$, $\mathfrak{G}_L$, $\mathfrak{Q}_R$ and $\mathfrak{G}_R$. We can say that the matrices which we have shown in (\ref{one8}) and (\ref{lefto}) were $\pi^{L\pm}(\mathfrak{Q}_L)$, $\pi^{L\pm}(\mathfrak{G}_L)$, $\pi^{L\pm}(\mathfrak{Q}_R)$ and $\pi^{L\pm}(\mathfrak{G}_R)$ for right and left movers respctively, while now we will display $\pi^{\tilde{L}\pm}$ of all the four supercharges - $\pi$ denoting the act of taking a representation of the abstract algebra generators. 

The $\tilde{L}$ representation for right movers is given by
\begin{eqnarray}\label{Ltilrep+}
&&\pi^{\Tilde{L}+}_{p_i}(\mathfrak{Q}_L) = -\pi^{\Tilde{L}+}_{p_i}(\mathfrak{G}_R) =  \sqrt{\frac{h}{2}}\sqrt{\sin \frac{p_i}{2}}\begin{pmatrix}0&1\\0&0 \end{pmatrix}, \nonumber\\
&&\pi^{\Tilde{L}+}_{p_i}(\mathfrak{G}_L) = -\pi^{\Tilde{L}+}_{p_i}(\mathfrak{Q}_R) =  \sqrt{\frac{h}{2}}\sqrt{\sin \frac{p_i}{2}}\begin{pmatrix}0&0\\1&0 \end{pmatrix}. \qquad p_i \in (0,\pi)
\end{eqnarray}
To distinguish the states from the $L$ representation, we can call them $\{|\bar{\phi}\rangle,|\bar{\psi}\rangle\}$, and the ordering of the basis in which we write our formulae is the same as before, just understanding a corresponding bar whenever there is an $\tilde{L}$ state.
  
The $\tilde{L}$ representation for left movers is given by
\begin{eqnarray}\label{Ltilrep-}
&&\pi^{\Tilde{L}-}_{p_i}(\mathfrak{Q}_L) = \pi^{\Tilde{L}-}_{p_i}(\mathfrak{G}_R) = \sqrt{\frac{h}{2}}\sqrt{\sin \frac{-p_i}{2}}\begin{pmatrix}0&1\\0&0 \end{pmatrix}, \nonumber\\
&&\pi^{\Tilde{L}-}_{p_i}(\mathfrak{G}_L) = \pi^{\Tilde{L}-}_{p_i}(\mathfrak{Q}_R) = \sqrt{\frac{h}{2}}\sqrt{\sin \frac{-p_i}{2}}\begin{pmatrix}0&0\\1&0 
\end{pmatrix}.\qquad p_i \in (-\pi,0)
\end{eqnarray}
Likewise, the $R$-matrices (\ref{exactsame}) are in the $LL$ representation, that is to say that they intertwine the coproducts in the representation $\pi^L \otimes \pi^L$(with additional $\pm$ indices for right and left movers which we omit here):
\begin{eqnarray}
\pi^L \otimes \pi^L \big(\Delta(\mathfrak{Q}_L)\big), \quad \pi^L \otimes \pi^L \big(\Delta(\mathfrak{G}_L)\big), \quad \pi^L \otimes \pi^L \big(\Delta(\mathfrak{Q}_R)\big), \quad \pi^L \otimes \pi^L \big(\Delta(\mathfrak{G}_R)\big),
\end{eqnarray}
and they could be called $R^{LL}$. The mix-representation $R$-matrices $R^{L\tilde{L}}$ intertwine instead  the (very same abstract) coproducts taken in the representation $\pi^L \otimes \pi^{\tilde{L}}$: 
\begin{eqnarray}
\pi^L \otimes \pi^{\tilde{L}} \big(\Delta(\mathfrak{Q}_L)\big), \quad \pi^L \otimes \pi^{\tilde{L}} \big(\Delta(\mathfrak{G}_L)\big), \quad \pi^L \otimes \pi^{\tilde{L}} \big(\Delta(\mathfrak{Q}_R)\big), \quad \pi^L \otimes \pi^{\tilde{L}} \big(\Delta(\mathfrak{G}_R)\big).
\end{eqnarray}
Such $R$-matrices read (in the various combinations of chiralities):
\begin{eqnarray}
&&R^{L\tilde{L}}_{++} = \tilde{\Sigma}_{++}(\gamma) \begin{pmatrix}-\tanh \frac{\gamma}{2}&0&0&\mbox{sech}\frac{\gamma}{2}\\0&1&0&0\\0&0&-1&0\\\mbox{sech}\frac{\gamma}{2}&0&0&\tanh\frac{\gamma}{2}\end{pmatrix}, \qquad\,\,\, \gamma_1,\gamma_2<0,
\nonumber\\
&&R^{L\tilde{L}}_{--} = \tilde{\Sigma}_{--}(\gamma) \begin{pmatrix}-\tanh \frac{\gamma}{2}&0&0&-\mbox{sech}\frac{\gamma}{2}\\0&-1&0&0\\0&0&1&0\\-\mbox{sech}\frac{\gamma}{2}&0&0&\tanh\frac{\gamma}{2}\end{pmatrix},\qquad \gamma_1,\gamma_2>0,
\nonumber\\
&&R^{L\tilde{L}}_{+-} = \tilde{\Sigma}_{+-}(\gamma) \begin{pmatrix}-\tanh \frac{\gamma}{2}&0&0&-i\mbox{sech}\frac{\gamma}{2}\\0&1&0&0\\0&0&1&0\\-i\mbox{sech}\frac{\gamma}{2}&0&0&-\tanh\frac{\gamma}{2}\end{pmatrix},\qquad\,\,\,  \gamma_1<0,\gamma_2>0,
\nonumber\\
&&R^{L\tilde{L}}_{-+} = \tilde{\Sigma}_{-+}(\gamma) \begin{pmatrix}\tanh \frac{\gamma}{2}&0&0&i\mbox{sech}\frac{\gamma}{2}\\0&1&0&0\\0&0&1&0\\i\mbox{sech}\frac{\gamma}{2}&0&0&\tanh\frac{\gamma}{2}\end{pmatrix},\qquad \qquad \,\, \gamma_1>0,\gamma_2<0,
\end{eqnarray}
and again $\gamma = \gamma_1 - \gamma_2$.
   
We have verified that the $R$-matrices in the $\pi^{\tilde{L}} \otimes \pi^{\tilde{L}}$ representation - which we call $R^{\tilde{L}\tilde{L}}$ - and the ones in the $\pi^{\tilde{L}} \otimes \pi^L$ representation - which we call $R^{\tilde{L}L}$ - satisfy the following relations (ignoring the dressing factors for the moment):
\begin{equation}\label{compa}
\begin{aligned}
R^{\tilde{L}\tilde{L}}_{++}(\gamma) &= R_{++}(\gamma), \qquad \qquad
&&R^{\tilde{L}\tilde{L}}_{--}(\gamma) = R_{--}(\gamma), 
\\
R^{\tilde{L}\tilde{L}}_{+-}(\gamma) &= R_{-+}(-\gamma), \qquad\qquad
&& R^{\tilde{L}\tilde{L}}_{-+}(\gamma) = R_{+-}(-\gamma), 
\\
R^{\tilde{L}L}_{++}(\gamma) &= R^{L\tilde{L}}_{++}(\gamma), \qquad \qquad
&& R^{\tilde{L}L}_{--}(\gamma) = R^{L\tilde{L}}_{--}(\gamma),
\\
R^{\tilde{L}L}_{+-}(\gamma) &= R^{L\tilde{L}}_{-+}(-\gamma),
\qquad \qquad
&&R^{\tilde{L}L}_{-+}(\gamma) = R^{L\tilde{L}}_{+-}(-\gamma).
\end{aligned}
\end{equation}

We refer to  \cite{Borsato:2014hja} for comparison - in particular formulae (5.9) - (5.12) in that paper. After converting $S$- to $R$-matrices we have the exact same matrix entries as in \cite{Borsato:2014hja} except for an overall minus in $R^{L\tilde{L}}_{--}(\gamma)$. 


We have checked all the 8 mixed Yang-Baxter equations with any assortment of $L$ and $\tilde{L}$ representations in right-right moving kinematics. For instance, just to show two of them, we have 
\begin{equation}\label{scatte}
\begin{aligned}
R^{\tilde{L}\tilde{L}}_{12}(\gamma_1-\gamma_2)R^{\tilde{L}L}_{13}(\gamma_1-\gamma_3)R^{\tilde{L}L}_{23}(\gamma_2-\gamma_3)&=R^{\tilde{L}L}_{23}(\gamma_2-\gamma_3)R^{\tilde{L}L}_{13}(\gamma_1-\gamma_3)R^{\tilde{L}\tilde{L}}_{12}(\gamma_1-\gamma_2), 
\\
R^{\tilde{L}L}_{12}(\gamma_1-\gamma_2)R^{\tilde{L}L}_{13}(\gamma_1-\gamma_3)R^{LL}_{23}(\gamma_2-\gamma_3)&=R^{LL}_{23}(\gamma_2-\gamma_3)R^{\tilde{L}L}_{13}(\gamma_1-\gamma_3)R^{\tilde{L}L}_{12}(\gamma_1-\gamma_2), \quad etc.
\end{aligned}
\end{equation}
We trust that the remaining Yang-Baxter equations will be satisfied since we are using the same conventions as \cite{Borsato:2014hja}. We have also checked that all the unitarity conditions can be satisfied with an appropriate choice of dressing factors, something which again must follow from \cite{Borsato:2014hja}.

Notice that again the BMN limit, where $\gamma_i$ are sent to $\pm \infty$ according to their domain of definition (\ref{scatte}),  trivialises the mixed-chirality scattering (\ref{scatte}), while maintaining a non-trivial scattering between the same chiralities. The left and right movers decouple, which is a the hallmark of a conformal field theory.

\section{Right wall reflection matrices} \label{sec::Ref Mat}

\begin{figure}
\centerline{\includegraphics[width=10cm]{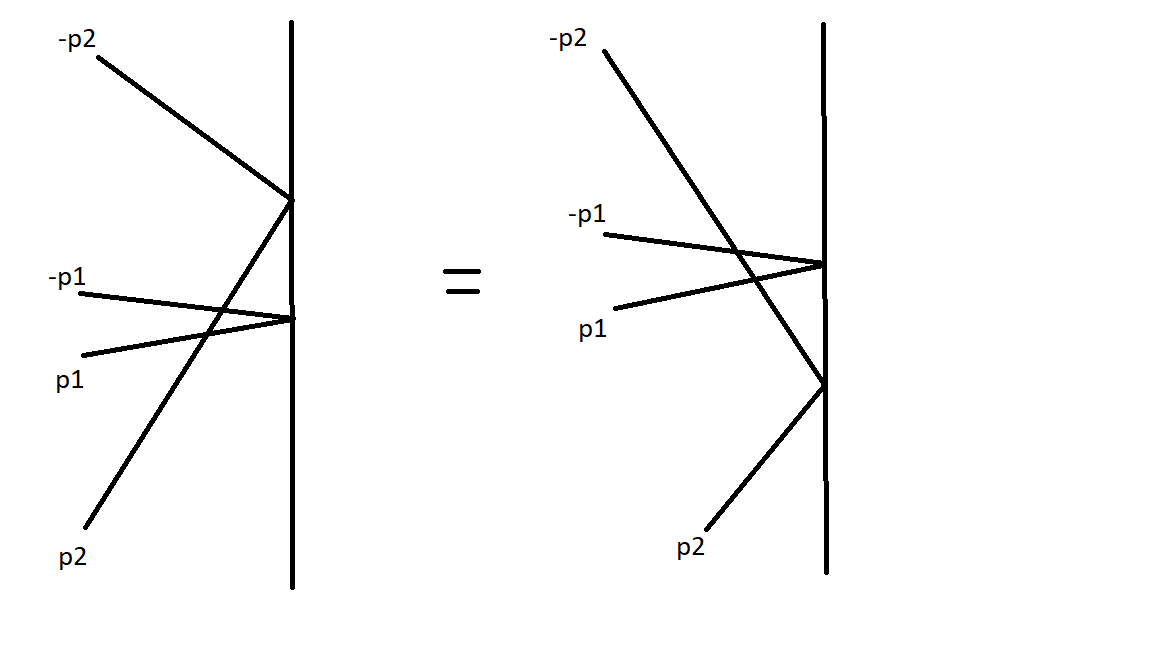}}
\caption{\label{1} The Boundary Yang-Baxter Equation (BYBE) for a right wall (hence all the incoming particles are right movers, and all the outgoing particles after reflection are left movers).}
\end{figure}

Let us write the Boundary Yang-Baxter Equation (BYBE), considering a {\it right wall}, namely one where all the dynamics happen to its left, as in figure \ref{1}. 

In the case of massless scattering, we need to keep in mind that a particle coming towards the wall has chirality $+$ (right mover), while after reflection it has chirality $-$ (left mover). This influences which bulk $S$-matrices  appear in the various parts of the BYBE. First, we can see that changing from a right to a left mover inverts the sign of $\gamma$. In fact, one can interpret (\ref{acco}) as defining the variable $\gamma$ piece-wise, and so, if $p_i \in (0,\pi)$, we have 
\begin{eqnarray}
\gamma(-p_i) = \log \cot \frac{p_i}{4} = - \log \tan \frac{p_i}{4} = - \gamma(p_i). \label{uso}
\end{eqnarray}
At this point, if we look at figure \ref{1}
and read it from top to bottom on both sides, we can write the BYBE for two particles in the $L$-representation:
\begin{eqnarray}
&&K_{2}^{L}(\gamma_2) [R^{LL}_{+-}]^{op}\Big(\gamma_2 - (-\gamma_1)\Big)K^{L}_1(\gamma_1)R^{LL}_{++}\Big(\gamma_1-\gamma_2\Big) = \nonumber\\
&&\qquad \qquad [R^{LL}_{--}]^{op}\Big((-\gamma_2)-(-\gamma_1)\Big)K^{L}_1(\gamma_1)R^{LL}_{+-}\Big(\gamma_1 - (-\gamma_2)\Big)K^{L}_2(\gamma_2).
\label{BYBE}
\end{eqnarray}
Notice that, for clarity, we made explicit the representation index in both $K$ and $R$ matrices and from now on we shall continue doing so. Since we make use of (\ref{uso}), then in (\ref{BYBE}) we should think everywhere
\begin{eqnarray}
\gamma_1 = \log \tan \frac{p_1}{4}, \qquad  \gamma_2 = \log \tan \frac{p_2}{4}, \qquad p_1, p_2 \in (0,\pi) , \qquad \gamma_1,\gamma_2 <0. 
\end{eqnarray}
We also recall that $K_1^{L} := K^{L} \otimes \mathbbmss{1}$, while $K_2^{L} := \mathbbmss{1} \otimes K^{L}$. Furthermore, if one defines
\begin{eqnarray}
R(\gamma) = r_{abcd}(\gamma) E_{ab}\otimes E_{cd}, 
\end{eqnarray}
then
\begin{eqnarray}
R^{op}(\gamma) = r_{abcd}(\gamma) \, (-1)^{\big(deg(a)+deg(b)\big)\big(deg(c)+deg(d)\big)}E_{cd}\otimes E_{ab}, 
\end{eqnarray}
where $|1\rangle = |\phi\rangle$ has $deg(1)=0$ and $|2\rangle = |\psi\rangle$ has $deg(2)=1$.

\subsection{Singlet boundary}
Let us first focus on the case in which the boundary is in a singlet representation.

Forgetting, for the moment, about dressing factors, we have found by brute force computation a family of solutions to (\ref{BYBE}) (in the ordered basis $\{|\phi\rangle,|\psi\rangle\}$) 
\begin{eqnarray}\label{refa}
K^{L}(\gamma_i) = \begin{pmatrix}1&0\\0&f^{L}(\gamma_i)\end{pmatrix}, \qquad f^{L}(\gamma_i) = -i \tanh \Big( \mbox{arccoth} (e^{\gamma_i}) + i c^{L}\Big), \qquad \gamma_i < 0 \,\, ,
\end{eqnarray}
with $c^{L}$ any complex constant. Two particularly interesting choices of $c^{L}$ are $c^{L}_{\pm} = \pi \big( m \pm \frac{1}{4} \big)$, with $m\in \mathbbmss{Z}$, which produce, upon using $\gamma_i(p_i) = \log \tan \frac{p_i}{4}$, the result
\begin{eqnarray}\label{KL-matrix-c+-}
K^{L}\Big(\gamma_i(p_i)\Big)\Big\vert_{c^{L}_{+}} = \begin{pmatrix}1&0\\0&-e^{\frac{i p_i}{2}}\end{pmatrix} \qquad \text{and} \qquad K^{L}\Big(\gamma_i(p_i)\Big)\Big\vert_{c^{L}_{-}} = \begin{pmatrix}1&0\\0&e^{-\frac{i p_i}{2}}\end{pmatrix}.
\end{eqnarray}
If we do the same now when the two bulk particles are in the $\tilde{L}$ representation, we find the $\tilde{L}$ reflection matrix - which we shall denote by $K^{\tilde{L}}$. Explicitly, we solve the boundary Yang-Baxter equation with two $\tilde{L}$ bulk particles - which is obtained from (\ref{BYBE}) by replacing $L$ with $\tilde{L}$ everywhere and using the $\tilde{L}\tilde{L}$ bulk $R$-matrices accordingly, {\it i.e.}
\begin{eqnarray}
&&K^{\tilde{L}}_2(\gamma_2) \big[R_{+-}^{\tilde{L}\tilde{L}}\big]^{op}\Big(\gamma_2 - (-\gamma_1)\Big)K^{\tilde{L}}_1(\gamma_1)R^{\tilde{L}\tilde{L}}_{++}\Big(\gamma_1-\gamma_2\Big) = \nonumber\\
&&\qquad \qquad \big[R_{--}^{\tilde{L}\tilde{L}}\big]^{op}\Big((-\gamma_2)-(-\gamma_1)\Big)K^{\tilde{L}}_1(\gamma_1)R^{\tilde{L}\tilde{L}}_{+-}\Big(\gamma_1 - (-\gamma_2)\Big)K^{\tilde{L}}_2(\gamma_2).
\label{BYBERR}
\end{eqnarray}
We find 
\begin{eqnarray}
K^{\tilde{L}}(\gamma_i) = \begin{pmatrix}1&0\\0&f^{\tilde{L}}(\gamma_i)\end{pmatrix}, \qquad f^{\tilde{L}}(\gamma_i) = i \tanh \Big( \mbox{arccoth} (e^{\gamma_i}) - i c^{\tilde{L}}\Big),\label{refb}
\end{eqnarray}
with $c^{\tilde{L}}$ any complex constant. Also in this case there are two particularly interesting choices of $c^{\tilde{L}}$, namely $c^{\tilde{L}}_{\pm} = \pi \big( n \pm \frac{1}{4} \big)$, with $n\in \mathbbmss{Z}$, leading to 
\begin{eqnarray}\label{KLtilde-matrix-c+-}
K^{\tilde{L}}\Big(\gamma_i(p_i)\Big)\Big\vert_{c^{\tilde{L}}_{+}} = \begin{pmatrix}1&0\\0&-e^{-\frac{i p_i}{2}}\end{pmatrix} \qquad \text{and} \qquad K^{\tilde{L}}\Big(\gamma_i(p_i)\Big)\Big\vert_{c^{\tilde{L}}_{-}} = \begin{pmatrix}1&0\\0&e^{\frac{i p_i}{2}}\end{pmatrix}.
\end{eqnarray}

Now we need to see whether this consistently solves the boundary Yang-Baxter equation for one bulk $L$ particle and one bulk $\tilde{L}$ particle. In such equations both $L$ and $\tilde{L}$ reflection matrices appear. For instance, if particle $1$ is $L$ and particle $2$ is $\tilde{L}$, we write 
\begin{eqnarray}
&&K_2^{\tilde{L}}(\gamma_2) \big[R_{+-}^{\tilde{L}L}\big]^{op}\Big(\gamma_2 - (-\gamma_1)\Big)K^L_1(\gamma_1)R^{L\tilde{L}}_{++}\Big(\gamma_1-\gamma_2\Big) =\nonumber\\
&&\qquad \qquad \big[R_{--}^{\tilde{L}L}\big]^{op}\Big((-\gamma_2)-(-\gamma_1)\Big)K^L_1(\gamma_1)R^{L\tilde{L}}_{+-}\Big(\gamma_1 - (-\gamma_2)\Big)K^{\tilde{L}}_2(\gamma_2),
\label{BYBER}
\end{eqnarray}
while if particle $1$ is $\tilde{L}$ and particle $2$ is $L$, we write 
\begin{eqnarray}
&&K^{L}_2(\gamma_2) \big[R_{+-}^{L\tilde{L}}\big]^{op}\Big(\gamma_2 - (-\gamma_1)\Big)K_1^{\tilde{L}}(\gamma_1)R^{\tilde{L}L}_{++}\Big(\gamma_1-\gamma_2\Big) =\nonumber\\
&&\qquad \qquad \big[R_{--}^{L\tilde{L}}\big]^{op}\Big((-\gamma_2)-(-\gamma_1)\Big)K_1^{\tilde{L}}(\gamma_1)R^{\tilde{L}L}_{+-}\Big(\gamma_1 - (-\gamma_2)\Big)K^{L}_2(\gamma_2).
\label{BYBER2}
\end{eqnarray}
Interestingly, one finds that to solve (\ref{BYBER}) and (\ref{BYBER2}) it is sufficient to impose the following relation between the constants $c^{L}$ and $c^{\tilde{L}}$
\begin{eqnarray}\label{relation-c-cLtilde}
c^{\tilde{L}} = -c^{L} + \pi \big( s + \tfrac{1}{2} \big) \qquad \text{with} \qquad s\in \mathbbmss{Z}.    
\end{eqnarray}
Throughout the rest of this section we shall assume this relation holds true, and work with the following rewriting of the above $K$-matrices \footnote{While such rewriting is always possible for $K^{L}$, since $\tan{(c^{L}+\chi(\gamma_{i}))}=f^{L}(\gamma_{i}) \,\,\,\, \forall \, c^{L}$, for $K^{\tilde{L}}$ one needs to impose the relation \eqref{relation-c-cLtilde} to make sure that $\cot{(c^{L}+\chi(\gamma_{i}))}=f^{\tilde{L}}(\gamma_{i})$.}
\begin{equation}\label{K-matrices-solving-BYBE}
\begin{aligned}
& K^{L}(\gamma_{i}) = 
\begin{pmatrix}
1 & 0 \\
0 & \tan{(c^{L}+\chi(\gamma_{i}))}
\end{pmatrix}
\qquad \qquad  
K^{\tilde{L}}(\gamma_{i}) = 
\begin{pmatrix}
1 & 0 \\
0 & \cot{( c^{L}+\chi(\gamma_{i}))}
\end{pmatrix}
\\
& \quad \text{with}  \qquad \qquad 
\chi(\gamma_{i}):= \frac{i}{2}\Big(i\, (\pi+2\pi k) + \log{\tanh{\frac{-\gamma_{i}}{2}}}\Big) \qquad \qquad \gamma_{i} < 0, \, k \in \mathbbmss{Z} \, \, .
\end{aligned}
\end{equation}
So far we have solved the boundary Yang-Baxter equations disregarding the bulk dressing factors. This means that the dressing factors will have to satisfy the properties which effectively allow them to cancel out in all the boundary Yang-Baxter equations. These conditions are:
\begin{eqnarray}
\Sigma_{++}^{LL}(\gamma)=\Sigma_{--}^{LL}(\gamma), \qquad \Sigma^{\tilde{L}\tilde{L}}_{++}(\gamma)=\Sigma^{\tilde{L}\tilde{L}}_{--}(\gamma),  \label{think1}
\end{eqnarray}
and 
\begin{eqnarray}
\Sigma^{L\tilde{L}}_{++}(\gamma)=\Sigma^{\tilde{L}L}_{--}(\gamma), \qquad \Sigma^{L\tilde{L}}_{--}(\gamma)=\Sigma^{\tilde{L}L}_{++}(\gamma)\qquad \Sigma_{+-}^{\tilde{L}L}(\gamma)=\Sigma^{L\tilde{L}}_{+-}(\gamma).\label{think2}
\end{eqnarray}
The conditions (\ref{think1}) should be compared with what is known in the literature \cite{AleSSergey}. In particular, we ought to  consider that we should construct physical excitations. As explained in \cite{Borsato:2014hja}, the massless physical excitations are obtained by taking tensor product of a single particle state in the $L$-representation with another single particle state in $\Tilde{L}$-representation, i.e. the physical massless state is of the form $\pi^L \otimes \pi^{\Tilde{L}}$. As a consequence, in the massless sector we must only allow boundary scattering which does not alter the $L$, $\tilde{L}$ flavour, in such a way that the physical excitations are maintained after reflection. This means that even for the singlet boundary, achiral reflections which change the representation of the reflecting bulk particle are prohibited, which is in contrast with the massive case considered in~\cite{Prinsloo:2015apa}.    
This should also make the difference in minus sign mentioned below (\ref{compa}) - which allows us to solve (\ref{BYBER}) and (\ref{BYBER2}) - inconsequential. This means that (\ref{BYBER}) and (\ref{BYBER2}) would actually never appear, and we would only need the processes analogue to formula (N.5) in \cite{Borsato:2014hja}.\\

\noindent
\textbf{Boundary subalgebras.} \quad Having solved the BYBE for the scattering of particles in the $L$ and $\tilde{L}$ representations off a right wall, it is natural to wonder about the number of symmetries that can be preserved by such a boundary. This means finding which generators (if any) of the bulk symmetry algebra $\mathcal{A}$ are not broken by the presence of the boundary, and are also symmetries of the $K$-matrix. These generators will constitute the boundary subalgebra $\mathcal{B}$. This amounts to searching for solutions of the intertwining equations (4.17) in \cite{Prinsloo:2015apa},
\begin{equation}\label{right-wall-singlet-intertwining-equation}
\pi_{-p}^{a-}(\mathfrak{b}) \, K^{a}\bigl(\gamma(p)\bigr)=K^{a}\bigl(\gamma(p)\bigr) \, \pi_{p}^{a+}(\mathfrak{b}) \quad \forall \, \mathfrak{b} \in \mathcal{B} \quad \text{and} \quad a \in \{ L,\tilde{L} \}, \quad \gamma<0, \quad p\in (0,\pi) \,\, .
\end{equation}
In the above equation, superscript $a$ refers to $L$ and $\tilde{L}$ representations, while $\pm$ following it refers to either right (+) or left (-) movers respectively. The subscripts $p$ refers to the momentum of the bulk particle which is taken to be positive i.e. $p\in (0,\pi)$, and hence the momentum for the left movers, which is negative and between $(-\pi,0)$, is denoted by $-p$.\\

\noindent
A brute force approach to the above task leads to the following results:
\begin{itemize}
\item The intertwining equations for the generators $\mathfrak{H}_L,\mathfrak{H}_R$ in the $L$ and $\tilde{L}$ representations are solved by the $K$-matrices $K^{L}(\gamma)$ and $K^{\tilde{L}}(\gamma)$, given in \eqref{K-matrices-solving-BYBE}, for any choice of constants $c^{L}$.
\item The intertwining equations for the generators $\mathfrak{P},\mathfrak{K}$ in the $L$ and $\tilde{L}$ representations cannot be solved by the $K$-matrices $K^{L}(\gamma)$ and $K^{\tilde{L}}(\gamma)$ for any choice of $c^{L}$.
\item The intertwining equations for the generators $\mathfrak{Q}_{L},\mathfrak{G}_{L}$ admit simultaneous solutions in either the $L$ or $\tilde{L}$ representations only if the constant $c^{L}$ is allowed to depend on the momenta. This implies that no allowed solution exists.
\item The intertwining equations for the generators $\mathfrak{Q}_{R},\mathfrak{G}_{R}$ admit simultaneous solutions in either the $L$ and $\tilde{L}$ representations provided that the constant $c^{L}$ is allowed to depend on the momenta. Hence, also in this case one finds no consistent solutions.
\end{itemize}
These findings seem to be in contrast with the results of \cite{Prinsloo:2015apa} for the massive case: it was there shown that a right boundary always preserves $\mathfrak{H}_L,\mathfrak{H}_R$ and that this boundary symmetry can be enhanced to two possible supersymmetric subalgebras, namely 
\begin{equation}\label{massive-boundary-susy-subalgebras}
\mathcal{B}_{L}= \langle \mathfrak{H}_L,\mathfrak{H}_R,\mathfrak{Q}_{L},\mathfrak{G}_{L} \rangle \qquad \text{and} \qquad \mathcal{B}_{R}= \langle \mathfrak{H}_L,\mathfrak{H}_R,\mathfrak{Q}_{R},\mathfrak{G}_{R} \rangle.
\end{equation}
Given this premise, to get a better grasp on the massless case it is then useful to take a step back and look at the problem from an algebraic perspective \cite{Niall}. A \textit{right boundary symmetry algebra} $\mathcal{B}$ with elements $\mathfrak{b}\in \mathcal{B}$ should be a subalgebra $\mathcal{B} \subset \mathcal{A}$ of the full bulk symmetry Hopf algebra $\mathcal{A}$, but not a Hopf subalgebra of the latter. Integrability rather requires it to be a \textit{coideal subalgebra}, which translates into the condition 
\begin{equation}\label{right-coideal-subalgebra}
\Delta(\mathfrak{b}) \in \mathcal{A}\otimes \mathcal{B} \qquad \forall \, \mathfrak{b} \in \mathcal{B},
\end{equation}
and, at the Lie algebra level, bulk and boundary should form a symmetric pair $(\mathfrak{g},\mathfrak{h})$ - $\mathfrak{g}$ corresponding to the bulk $\mathcal{A}$ and $\mathfrak{h}$ to the boundary $\mathcal{B}$, such that $\mathfrak{g}\simeq \mathfrak{h}\oplus \mathfrak{m}$, with $\mathfrak{h}$ a subalgebra of $\mathfrak{g}$ and commutation relations restricted to 
\begin{equation}\label{symmetric-space}
[\mathfrak{h},\mathfrak{h}] \subset \mathfrak{h}, \qquad \quad [\mathfrak{h},\mathfrak{m}] \subset \mathfrak{m}, \qquad \quad [\mathfrak{m},\mathfrak{m}] \subset \mathfrak{h}.
\end{equation}
In light of these requirements it is not hard to realise that while the conditions \eqref{right-coideal-subalgebra} and \eqref{symmetric-space} are satisfied for the above two subalgebras $\mathcal{B}_{L}$ and $\mathcal{B}_{R}$ given the choice of coproducts in \cite{Prinsloo:2015apa}, the same is not true for the case under consideration: the right-coideal subalgebra condition is indeed spoiled by the choice of a symmetric coproduct \eqref{simo}, due to the presence of the central group-like generator $\mathfrak{U}=e^{i\tfrac{\mathfrak{p}}{2}}$, introduced in \cite{Prinsloo:2015apa} via its action on single-magnon states
\begin{equation}
\mathfrak{U} \,| \varphi^{r}_{p}\rangle = e^{i\tfrac{p}{2}} \, | \varphi^{r}_{p}\rangle \qquad \qquad \text{and} \qquad \qquad
\mathfrak{U} \, | \bar{\varphi}^{r}_{p}\rangle = e^{i\tfrac{p}{2}} \, | \bar{\varphi}^{r}_{p}\rangle.
\end{equation}
Rewriting \eqref{simo} as
\begin{equation}\label{symm-coproduc-U}
\Delta(\mathfrak{Q}_{a})=\mathfrak{Q}_{a}\otimes \mathfrak{U}^{-\tfrac{1}{2}}+\mathfrak{U}^{\tfrac{1}{2}}\otimes \mathfrak{Q}_{a} \qquad \text{and} \qquad \Delta(\mathfrak{G}_{a})=\mathfrak{G}_{a}\otimes \mathfrak{U}^{\tfrac{1}{2}}+\mathfrak{U}^{-\tfrac{1}{2}}\otimes \mathfrak{G}_{a} \qquad \text{with} \qquad a \in \{ L,R \},
\end{equation}
and adopting from \cite{Prinsloo:2015apa} the coproduct 
\begin{equation}\label{U-coproduct}
\Delta({\mathfrak{U}^{\alpha}}) = \mathfrak{U}^{\alpha} \otimes \mathfrak{U}^{\alpha}, \qquad \qquad \forall \,\, \alpha \in \mathbb{R}, 
\end{equation}
it is then possible to realise that a more convenient choice of basis for the bulk symmetry algebra $\mathcal{A}$ is \footnote{The Hopf algebra $\mathcal{A}^{\mathfrak{U}}=\mathcal{A}$ but, due to the different basis choice, there are different (isomorphic) Lie algebras associated with $\mathcal{A}^{\mathfrak{U}}$ and $\mathcal{A}$ respectively. The same will be true for $\tilde{\mathcal{A}}^{\mathfrak{U}}$ in \eqref{new-algebra-basis-left-wall}. 
} 
\begin{equation}\label{new-algebra-basis-right-wall}
\mathcal{A}^{\mathfrak{U}} = \langle \mathfrak{H}_{L}, \, \mathfrak{H}_{R}, \, \mathfrak{U}^{\tfrac{1}{2}}\mathfrak{Q}_{L}, \, \mathfrak{U}^{\tfrac{1}{2}}\mathfrak{Q}_{R}, \, \mathfrak{U}^{-\tfrac{1}{2}}\mathfrak{G}_{L}, \, \mathfrak{U}^{-\tfrac{1}{2}}\mathfrak{G}_{R}, \, \mathfrak{U}\mathfrak{P}, \, \mathfrak{U}^{-1}\mathfrak{K}, \,\mathfrak{p}\rangle \quad \text{with} \quad \mathfrak{U}^{\alpha} = e^{i \frac{\alpha}{2} \mathfrak{p}} \quad \forall \, \alpha \in \mathbb{R}.
\end{equation}
In this basis, one can find two subalgebras equivalent to \eqref{massive-boundary-susy-subalgebras} satisfying conditions \eqref{right-coideal-subalgebra} and \eqref{symmetric-space}
\begin{equation}\label{rescaled-susy-boundary-algebras}
\mathcal{B}_{L}^{\mathfrak{U}} = \langle \mathfrak{H}_{L}, \, \mathfrak{H}_{R}, \, \mathfrak{U}^{\tfrac{1}{2}}\mathfrak{Q}_{L}, \, \mathfrak{U}^{-\tfrac{1}{2}}\mathfrak{G}_{L}  \rangle \qquad \text{and} \qquad
\mathcal{B}_{R}^{\mathfrak{U}} = \langle \mathfrak{H}_{L}, \, \mathfrak{H}_{R}, \, \mathfrak{U}^{\tfrac{1}{2}}\mathfrak{Q}_{R}, \, \mathfrak{U}^{-\tfrac{1}{2}}\mathfrak{G}_{R}  \rangle.
\end{equation}
One can then go back to the search for solutions to the intertwining equations \eqref{right-wall-singlet-intertwining-equation}, which in the new basis leads to the following results:
\begin{itemize}
\item Nothing changes for $\mathfrak{H}_L,\mathfrak{H}_R$, which are always preserved by the right wall in both the $L$ and $\tilde{L}$ representations for any choice of constant $c^{L}$ appearing in the $K$-matrices \eqref{refa} and \eqref{refb}.
\item The rescaled generators $\mathfrak{U}\mathfrak{P}, \, \mathfrak{U}^{-1}\mathfrak{K}$ are still never preserved in either the $L$ or $\tilde{L}$ representation.
\item The intertwining equations for $\mathfrak{U}^{\tfrac{1}{2}}\mathfrak{Q}_{L},\mathfrak{U}^{-\tfrac{1}{2}}\mathfrak{G}_{L}$ admit simultaneous solutions in both the $L$ and $\tilde{L}$ representations provided one chooses $c^{L}=c^{L}_{-}=\pi \big( m - \frac{1}{4} \big)$ with $m\in \mathbbmss{Z}$. For these values of $c^{L}$, the intertwining equations for $\mathfrak{U}^{\tfrac{1}{2}}\mathfrak{Q}_{R},\mathfrak{U}^{-\tfrac{1}{2}}\mathfrak{G}_{R}$ are not satisfied and preservation of $\mathfrak{H}_L,\mathfrak{H}_R$ is enhanced to the supersymmetric boundary subalgebra $\mathcal{B}_{L}^{\mathfrak{U}}$ defined in \eqref{rescaled-susy-boundary-algebras}.
\item The intertwining equations for $\mathfrak{U}^{\tfrac{1}{2}}\mathfrak{Q}_{R},\mathfrak{U}^{-\tfrac{1}{2}}\mathfrak{G}_{R}$ admit simultaneous solutions in both the $L$ and $\tilde{L}$ representations provided one chooses $c^{L}=c^{L}_{+}=\pi \big( m + \frac{1}{4} \big)$ with $m\in \mathbbmss{Z}$. For these values of $c^{L}$, the intertwining equations for $\mathfrak{U}^{\tfrac{1}{2}}\mathfrak{Q}_{L},\mathfrak{U}^{-\tfrac{1}{2}}\mathfrak{G}_{L}$ are not satisfied and preservation of $\mathfrak{H}_L,\mathfrak{H}_R$ is enhanced to the supersymmetric boundary subalgebra $\mathcal{B}_{R}^{\mathfrak{U}}$ defined in \eqref{rescaled-susy-boundary-algebras}.
\end{itemize}
Notice that in obtaining the above results we represented, for any $\mathfrak{a} \in \mathcal{A}^{\mathfrak{U}}$, the generator $\mathfrak{U}^{\alpha}\,\, \forall \alpha \in \mathbbmss{R}$ as
\begin{equation}\label{U-generator-rep.}
\pi^{a\pm}_{\pm p}(\mathfrak{U}^{\alpha} \mathfrak{a}) = \pi^{a\pm}_{\pm p}(\mathfrak{U}^{\alpha})  \pi^{a\pm}_{\pm p}(\mathfrak{a})  \qquad \text{with} \qquad \pi^{a\pm}_{\pm p}(\mathfrak{U}^{\alpha}) = e^{\pm i\alpha\tfrac{p}{2}}\mathbbmss{1} \qquad \text{for} \qquad p\in (0,\pi) \,\, , \,\, a\in\{ L,\tilde{L} \} 
\end{equation}
We must at this stage make some further comparison with the results obtained for the massive case in \cite{Prinsloo:2015apa}. The first thing to notice is that the massive $K$-matrices found to solve the BYBE and preserving the Abelian subalgebra $\langle \mathfrak{H}_{L}, \mathfrak{H}_{R} \rangle$ (see eq. 4.22 in \cite{Prinsloo:2015apa}) exhibit dependence on the momenta $p$ and a constant $c$ which is different from the one found in the massless case \eqref{refa} and \eqref{refb} (for the different constant $c^{L}$) and there exists no choice of $c^{L}$ which allows one to recast \eqref{K-matrices-solving-BYBE} in the form found for the massive case in terms of the constant $c$. A second important difference lies then in the $K$-matrices preserving the enhanced symmetry superalgebras $\mathcal{B}_{L}^{\mathfrak{U}}$ and $\mathcal{B}_{R}^{\mathfrak{U}}$ at the special values $c^{L}=c^{L}_{\pm}$ in \eqref{K-matrices-solving-BYBE}.
While the massive $K$-matrices exhibiting symmetry enhancement are characterised by the presence of $-e^{ip}$ factors only (see eq. 4.18, 4.19 in \cite{Prinsloo:2015apa}), the respective massless $K$-matrices \eqref{K-matrices-solving-BYBE} reduce to the form \eqref{KL-matrix-c+-} and \eqref{KLtilde-matrix-c+-} at $c=c_{\pm}$, and hence depend on $\pm e^{\pm i \tfrac{p}{2}}$. These differences should be expected, as a result of the representations involved in the analysis: while in the massive case the relevant representations are $L$ and $R$, in the massless one are $L$ and $\tilde{L}$, which then further split depending on the chirality of the excitations, leading to the four cases described above. Additionally, one should take into account that the $L$ representation considered in this work is not the same as the massless limit of the one considered in \cite{Prinsloo:2015apa} - which we briefly discuss in Appendix \ref{App:A}. Altogether, this accounts for the differences encountered above and the ones that we shall find in the next paragraph.\\

\noindent
\textbf{Hidden symmetry.} \quad 
In further analogy with the massive scenario, it is natural to wonder about the existence of some (hidden) larger supersymmetric subalgebra of $\mathcal{A}^{\mathfrak{U}}$ solving the intertwining equations for the $K$-matrices \eqref{K-matrices-solving-BYBE}, preserving $\langle \mathfrak{H}_{L},\mathfrak{H}_{R} \rangle$ for generic $c^L$. In \cite{Prinsloo:2015apa}, the authors managed to provide a positive answer to such question (see around equation 4.23) upon identifying a new set of generators, constructed as non-linear combinations of the original ones and satisfying an isomorphic set of commutation relations as well as various symmetric pair decompositions \eqref{symmetric-space}. Satisfaction of the right coideal subalgebra condition \eqref{right-coideal-subalgebra} remained more misterious, for technical reasons we shall comment on at the end of this paragraph.

For the massless excitations considered in this work, the choice of new basis \eqref{new-algebra-basis-right-wall}  turned out to be fundamental in showing symmetry preservation of the right wall. Hence one should reasonably expect this to play a role in the existence of a hidden symmetry algebra and the intuitive guess would be defining the new generators as in \cite{Prinsloo:2015apa}, after taking into account the modified basis \eqref{new-algebra-basis-right-wall}. Due to the different dependences exhibited by the massive and massless $K$-matrices on the respective constants $c$ and $c^{L}$, such guess would however not be sufficient to show satisfaction of the intertwining equations. This means one should not expect to be able to solve them by simply replacing $c \rightarrow c^{L}$ in the definition of hidden generators given in \cite{Prinsloo:2015apa}: some more general dependence on $c^{L}$ should be considered. 

A natural way of taking into account these aspects is looking for solutions of the intertwining equations \eqref{right-wall-singlet-intertwining-equation} for the following set of hidden generators
\begin{equation}\label{rescaled-hidden-generators}
\begin{aligned}
\mathfrak{q}_{+} &:= \mathfrak{U}^{-1/2}\mathfrak{K}\mathfrak{Q}_{L}+ie(c^{L}) \,  \mathfrak{U}^{1/2}\mathfrak{P}\mathfrak{G}_{R} \qquad \qquad \mathfrak{d} := (\mathfrak{H}_{L}-e(c^{L})^2 \, \mathfrak{H}_{R}+ie(c^{L}) \, (\mathfrak{U}\mathfrak{P}+\mathfrak{U}^{-1}\mathfrak{K}))\mathfrak{P}\mathfrak{K}
\\
\mathfrak{q}_{-} &:= \mathfrak{U}^{1/2}\mathfrak{P}\mathfrak{G}_{L}+ie(c^{L}) \, \mathfrak{U}^{-1/2}\mathfrak{K}\mathfrak{Q}_{R} \qquad \qquad \tilde{\mathfrak{d}} := (\mathfrak{H}_{L}-e(c^{L})^2 \, \mathfrak{H}_{R}-ie(c^{L}) \, (\mathfrak{U}\mathfrak{P}+\mathfrak{U}^{-1}\mathfrak{K}))\mathfrak{P}\mathfrak{K}
\\
\mathfrak{s}_{+} &:= \mathfrak{U}^{-1/2}\mathfrak{K}\mathfrak{Q}_{L}-ie(c^{L}) \, 
\mathfrak{U}^{1/2}\mathfrak{P}\mathfrak{G}_{R} \qquad \qquad \mathfrak{n} := (\mathfrak{H}_{L}+e(c^{L})^2 \, \mathfrak{H}_{R}+ie(c^{L}) \, (\mathfrak{U}\mathfrak{P}-\mathfrak{U}^{-1}\mathfrak{K}))\mathfrak{P}\mathfrak{K}
\\
\mathfrak{s}_{-} &:= \mathfrak{U}^{1/2}\mathfrak{P}\mathfrak{G}_{L}-ie(c^{L}) \, \mathfrak{U}^{-1/2}\mathfrak{K}\mathfrak{Q}_{R} \qquad \qquad \tilde{\mathfrak{n}} := (\mathfrak{H}_{L}+e(c^{L})^2 \, \mathfrak{H}_{R}-ie(c^{L}) \, (\mathfrak{U}\mathfrak{P}-\mathfrak{U}^{-1}\mathfrak{K}))\mathfrak{P}\mathfrak{K} \, \, .
\end{aligned}
\end{equation}
which closely resemble the ones defined in \cite{Prinsloo:2015apa} while accounting for the new basis \eqref{new-algebra-basis-right-wall} and the different dependence on the integration constant. These satisfy the commutation relations
\begin{equation}\label{hidden_generators_commutators}
\begin{aligned}
\{ \mathfrak{q}_{+},\mathfrak{q}_{-} \} &= \mathfrak{d} \qquad \qquad 
\{ \mathfrak{q}_{+},\mathfrak{s}_{-} \} = \mathfrak{n} \qquad \qquad 
\{ \mathfrak{q}_{+},\mathfrak{s}_{+} \}=0  
\\
\{ \mathfrak{s}_{+},\mathfrak{s}_{-} \} &= \tilde{\mathfrak{d}} \qquad \qquad 
\{ \mathfrak{q}_{-},\mathfrak{s}_{+} \} = \tilde{\mathfrak{n}} \qquad \qquad 
\{ \mathfrak{q}_{-},\mathfrak{s}_{-} \}=0 \,\, ,
\end{aligned}
\end{equation}
and studying the intertwining conditions one can observe the followings:
\begin{itemize}
\item The hidden generators $\mathfrak{n},\tilde{\mathfrak{n}}$ are always preserved by the right wall in both the $L$ and $\tilde{L}$ representations, for any choice of function $e(c^{L})\neq 0$.
\item Generators $\mathfrak{d},\tilde{\mathfrak{d}}$ are never preserved by the right wall, in either the $L$ or $\tilde{L}$ representation, for any non-zero choice of $e(c^{L})$.
\item The intertwining equations for $\mathfrak{q}_{+}, \mathfrak{s}_{-}$ admit simultaneous solutions in both the $L$ and $\tilde{L}$ representations provided one chooses $e(c^{L})= e_{-}(c^{L}) = -i \, \frac{\cos{(c^{L})}-\sin{(c^{L})}}{\cos{(c^{L})}+\sin{(c^{L})}}$. For such choice of $e(c^{L})$ the intertwining equations for $\mathfrak{q}_{-},\mathfrak{s}_{+}$ are never satisfied, unless one further specifies $c^{L}$ such that $\cos{(c^{L})}=\sin{(c^{L})}$. This would precisely correspond to choosing $c^{L}=c_{+}^{L}=\pi(m +\tfrac{1}{4})$ with $m \in \mathbb{Z}$, namely the $c^{L}$ leading to preservation of $\mathcal{B}_{R}^{\mathfrak{U}}$ defined in \eqref{rescaled-susy-boundary-algebras}. However, this is obviously forbidden by the fact that would imply $e(c^{L})=0$ and thus trivialisation of the hidden generators. The value $c^L = c^L_{-}$ is also forbidden, as it leads to vanishing denominator in $e_{-}(c^L)$.
\item The intertwining equations for $\mathfrak{q}_{-}, \mathfrak{s}_{+}$ admit simultaneous solutions in both the $L$ and $\tilde{L}$ representations provided one chooses $e(c^{L})=e_{+}(c^{L}) = i \, \frac{\cos{(c^{L})}-\sin{(c^{L})}}{\cos{(c^{L})}+\sin{(c^{L})}}$. For such choice of $e(c^{L})$ the intertwining equations for $\mathfrak{q}_{+},\mathfrak{s}_{-}$ are never satisfied, unless one further specifies $c^{L}$ such that $\cos{(c^{L})}=\sin{(c^{L})}$. Again, this would correspond to choosing $c^{L}=c_{+}^{L}=\pi(m +\tfrac{1}{4})$ with $m \in \mathbb{Z}$, which is forbidden by the fact that $e(c^{L})=0$ for such values. Also in this case one needs $c^L \neq c^L_{-}$.
\end{itemize}
We further notice that, as expected, defining the hidden symmetry generators without making use of the basis \eqref{new-algebra-basis-right-wall} leads to no solution of the intertwining equations, i.e. no hidden symmetry being preserved by the boundary. Moreover, one may wonder whether the dependence on $c^L$ might be generalised by including arbitrary rescaling functions in each of the terms which define the generators \eqref{rescaled-hidden-generators} and whether this may lead to different sets of preserved symmetries: it seems that a more general dependence on $c^{L}$ may always be brought back to the form \eqref{rescaled-hidden-generators} upon requiring the generators and their coproducts to satisfy the commutation relations \eqref{hidden_generators_commutators}.

It is then necessary to highlight that as a result of \eqref{hidden_generators_commutators}, the symmetric pair condition \eqref{symmetric-space} are still respected by the sets of generators preserved by the right boundary for any choice of $c^{L}\neq c_{\pm}^{L}$, namely 
\begin{equation}\label{preserved_hidden_symmetric_pairs}
\mathfrak{h}_{\pm}=\langle \mathfrak{q}_{\pm},\mathfrak{s}_{\mp},\mathfrak{n},\tilde{\mathfrak{n}} \rangle \qquad \text{with} \qquad \mathfrak{m}_{\pm}=\langle \mathfrak{q}_{\mp},\mathfrak{s}_{\pm},\mathfrak{d},\tilde{\mathfrak{d}} \rangle \,\, .
\end{equation}
This translates into preservation of the following hidden symmetry subalgebras
\begin{equation}
\mathcal{B}_{D}^{\mathfrak{U}\pm} = \langle \mathfrak{q}_{\pm}, \mathfrak{s}_{\mp}, \mathfrak{n}, \tilde{\mathfrak{n}} \rangle \qquad \text{for} \qquad e(c^L) = e_{\mp}(c^L)\, .
\end{equation}
At this point, the right coideal subalgebra condition \eqref{right-coideal-subalgebra} remains  the final ingredient to be considered, and in this respect an important comment is in order. An explicit check of the latter property would require computing the coproduct of the hidden symmetry generators \eqref{rescaled-hidden-generators} by using the coproduct homomorphism property and applying the relations \eqref{symm-coproduc-U} and \eqref{U-coproduct}. This way, the coproducts for the hidden symmetry generators would be written as tensor products of the original generators and the non-trivial step would correspond to re-expressing such results into the form
\begin{equation}\label{desired_hidden_coproduct}
\Delta(\mathfrak{b}) = \sum_{\mathfrak{a}^{i}\in \mathcal{A}, \,\mathfrak{b}^{j}\in \mathcal{B}} c_{ij} \, \mathfrak{a}^{i}\otimes \mathfrak{b}^{j} \quad \in \mathcal{A}\otimes \mathcal{B} \qquad \forall \, \mathfrak{b} \, \in \mathcal{B} \, \, ,
\end{equation}
by inverting the relations \eqref{rescaled-hidden-generators}. In the latter expression $c_{ij}$ denote some arbitrary coefficients to be determined, $\mathcal{A}=\langle \mathfrak{q}_{+},\mathfrak{q}_{-},\mathfrak{s}_{+},\mathfrak{s}_{-}, \mathfrak{d},\tilde{\mathfrak{d}},\mathfrak{n},\tilde{\mathfrak{n}}, \mathfrak{p} \rangle$ and $\mathcal{B}$ is either  $\langle \mathfrak{q}_{-},\mathfrak{s}_{+},\mathfrak{n},\tilde{\mathfrak{n}} \rangle$ or $\langle \mathfrak{q}_{+},\mathfrak{s}_{-},\mathfrak{n},\tilde{\mathfrak{n}} \rangle$.
Performing such inverse computation turns out to be impossible in both the massless case considered here and in the massive case considered in \cite{Prinsloo:2015apa}, but the possibility of constructing a coproduct of hidden generators compatible with the coideal subalgebra condition should be ensured by the symmetric pair decompositons \eqref{preserved_hidden_symmetric_pairs}, and their massive analogue, as discussed in \cite{Belliard:2014uja}\footnote{We thank Vidas Regelskis for very helpful discussion on this point.}.
Strictly speaking, the construction of \cite{Belliard:2014uja} applies to complex simple Lie algebras, but the very non-trivial fact of being able to determine a $K$-matrix preserving (part of) the hidden generators \eqref{rescaled-hidden-generators} for the massless case - and similarly for the massive case of \cite{Prinsloo:2015apa} - suggests it should be extendable to the case under consideration. Applying this, would ensure the possibility of finding coproducts for the hidden symmetry generators with the right structure \eqref{desired_hidden_coproduct}, and would in turn induce a set of coproducts on the original generators by direct application of the relations \eqref{rescaled-hidden-generators}. The latter coproducts would however be not guaranteed to take the original form \eqref{symm-coproduc-U} and \eqref{U-coproduct}, as the procedure would generally lead to more complicated expressions.
\\

\noindent
\textbf{Short summary of the right wall singlet boundary.} \quad  Solving the four BYB equations \eqref{BYBE},\eqref{BYBERR}, \eqref{BYBER},\eqref{BYBER2} leads to the $K$-matrices \eqref{K-matrices-solving-BYBE}, which we report here for self consistency of the recap
\begin{equation}
\begin{aligned}
& K^{L}(\gamma_{i}) = 
\begin{pmatrix}
1 & 0 \\
0 & \tan{(c^{L}+\chi(\gamma_{i}))}
\end{pmatrix}
\qquad \qquad  
K^{\tilde{L}}(\gamma_{i}) = 
\begin{pmatrix}
1 & 0 \\
0 & \cot{( c^{L}+\chi(\gamma_{i}))}
\end{pmatrix}
\\
& \,\, \text{with}  \qquad 
\chi(\gamma_{i}):= \frac{i}{2}\Big(i \,(\pi+2\pi k) + \log{\tanh{\frac{-\gamma_{i}}{2}}}\Big) \qquad  \text{and} \qquad \gamma_{i} < 0, \, k \in \mathbbmss{Z},\,   c^{L} \in \mathbb{C} \,\, .
\end{aligned}
\end{equation}
Satisfying the right coideal subalgebra condition \eqref{right-coideal-subalgebra} without breaking the symmetric pair decomposition \eqref{symmetric-space} then requires choosing the following basis for the bulk symmetry algebra
\begin{equation}
\mathcal{A}^{\mathfrak{U}}=
\langle \mathfrak{H}_{L}, \, \mathfrak{H}_{R}, \, \mathfrak{U}^{\tfrac{1}{2}}\mathfrak{Q}_{L}, \, \mathfrak{U}^{\tfrac{1}{2}}\mathfrak{Q}_{R}, \, \mathfrak{U}^{-\tfrac{1}{2}}\mathfrak{G}_{L}, \, \mathfrak{U}^{-\tfrac{1}{2}}\mathfrak{G}_{R}, \, \mathfrak{U}\mathfrak{P}, \, \mathfrak{U}^{-1}\mathfrak{K}, \,\mathfrak{p}\rangle \quad \text{with} \quad \mathfrak{U}^{\alpha} = e^{i \frac{\alpha}{2} \mathfrak{p}} \quad \forall \, \alpha \in \mathbb{R}.
\end{equation}
The central generators $\mathfrak{H}_{L},\mathfrak{H}_{R},\mathfrak{p}$ are then preserved by the right wall in both the $L$ and $\tilde{L}$ representations, i.e. they satisfy the intertwining equations \eqref{right-wall-singlet-intertwining-equation}, for any choice of the constant $c^{L}$, while the generators $\mathfrak{U}\mathfrak{P}$ and $\mathfrak{U}^{-1}\mathfrak{K}$ are never preserved in either the $L$ or $\tilde{L}$ representation. The fermionic generators $\mathfrak{U}^{\frac{1}{2}}\mathfrak{Q}_{a}$ and $\mathfrak{U}^{-\frac{1}{2}}\mathfrak{G}_{a}$ are generally not preserved, unless one sits at the special values $c^{L} = c^{L}_{\pm} = \pi(m\pm\frac{1}{4})$, with $m\in \mathbb{Z}$, where the following supersymmetry enhanced subalgebras are preserved
\begin{equation}
c^{L}_{+}: \quad \mathcal{B}_{R}^{\mathfrak{U}} = \langle \mathfrak{H}_{L},\mathfrak{H}_{R},\mathfrak{U}^{\frac{1}{2}}\mathfrak{Q}_{R},\mathfrak{U}^{-\frac{1}{2}}\mathfrak{G}_{R}  \rangle \qquad \quad \text{and}  \qquad \quad c^{L}_{-}: \quad \mathcal{B}_{L}^{\mathfrak{U}} = \langle \mathfrak{H}_{L},\mathfrak{H}_{R},\mathfrak{U}^{\frac{1}{2}}\mathfrak{Q}_{L},\mathfrak{U}^{-\frac{1}{2}}\mathfrak{G}_{L}  \rangle \,\,.
\end{equation}
The two $K$-matrices then take the following very simple form
\begin{equation}\label{summary-right-wall-singlet-K-matrices}
\begin{aligned}
&c^{L}_{+}: \qquad K^{L}_{\mathcal{B}_{R}^{\mathfrak{U}}}(p) = \begin{pmatrix}1&0\\0&-e^{i\frac{p}{2}}\end{pmatrix}
\qquad \qquad &&\text{and} \qquad \qquad
K^{\tilde{L}}_{\mathcal{B}_{R}^{\mathfrak{U}}}(p) = \begin{pmatrix}1&0\\0&-e^{-i\frac{ p}{2}}\end{pmatrix}
\\
&c^{L}_{-}: \qquad K^{L}_{\mathcal{B}_{L}^{\mathfrak{U}}}(p) = \begin{pmatrix}1&0\\0&e^{-i\frac{p}{2}}\end{pmatrix}
\qquad \qquad  &&\text{and} \qquad \qquad 
K^{\tilde{L}}_{\mathcal{B}_{L}^{\mathfrak{U}}}(p) = \begin{pmatrix}1&0\\0&e^{i\frac{p}{2}}\end{pmatrix}.
\end{aligned}
\end{equation}
It is finally possible to define, in terms of $\mathfrak{a} \in \mathcal{A}^{\mathfrak{U}}$, a set of new generators \eqref{rescaled-hidden-generators} depending on some arbitrary function $e(c^{L})$ of the constant which appears in the $K$-matrices, and exhibiting commutation relations \eqref{hidden_generators_commutators} with the same structure as \eqref{comma}. This complicated and non-linear basis redefinition allows to uncover the presence of two hidden symmetry boundary superalgebras $\mathcal{B}_{D}^{\mathfrak{U}\pm} = \langle \mathfrak{q}_{\pm}, \mathfrak{s}_{\mp}, \mathfrak{n}, \tilde{\mathfrak{n}} \rangle$, which respectively solve the boundary intertwining equations \eqref{right-wall-singlet-intertwining-equation} for any $c^{L} \neq c^{L}_{\pm}$ upon choosing
\begin{equation}
e_{\mp}(c^{L}) = \mp i \,\, \frac{\cos{(c^{L})}-\sin{(c^{L})}}{\cos{(c^{L})}+\sin{(c^{L})}} \,\, .
\end{equation}\\

\noindent
\textbf{Unitarity.} \quad
\begin{figure}
\centerline{\includegraphics[width=10cm]{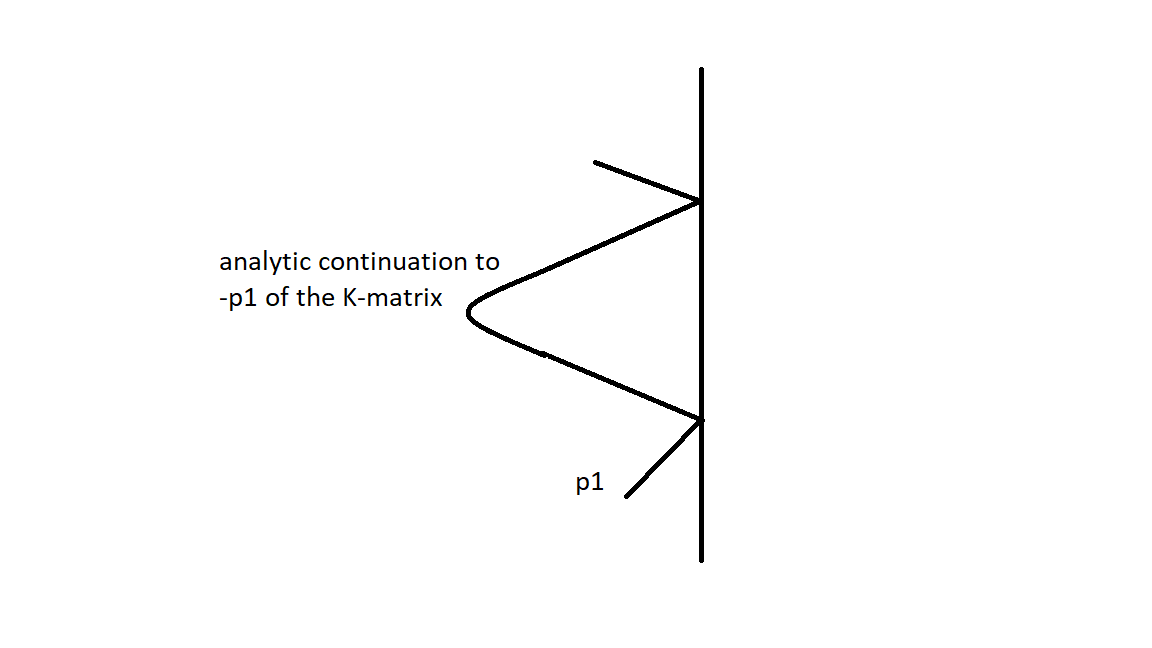}}
\caption{\label{3} The unitarity condition for the right wall. One has to analytically continue the expression for the reflection matrix (which is defined for positive incoming momentum) to negative incoming momentum (a left mover is fed back into the right wall).}
\end{figure}
In analogy with $S$-matrices, physically relevant $K$-matrices should satisfy the following unitarity relations 
\begin{equation}\label{Unitarity-relations-K-matrices}
\begin{aligned}
\text{Physical Unitarity} \qquad\quad  \qquad K(\gamma)K^{\dagger}(\gamma) &= \mathbbmss{1}
\\
\text{Braiding Unitarity} \qquad \qquad \,\,\,\,
K(\gamma)K(-\gamma) &= \mathbbmss{1}
\end{aligned}
\end{equation}
with $^\dagger$ denoting Hermitean conjugation (See figure \ref{3} for a pictorial view on braiding unitarity). It is straightforward to check that the $K$-matrices summarised in \eqref{summary-right-wall-singlet-K-matrices} for the case of a right wall singlet boundary indeed satisfy the physical unitarity relations, provided the associated dressing factors also satisfy
\begin{equation}
k^{L}(\gamma)\big[ k^{L}(\gamma)\big]^{*}=1=
k^{\tilde{L}}(\gamma)\big[ k^{\tilde{L}}(\gamma)\big]^{*} \qquad \text{with} \qquad \gamma <0
\end{equation}

However, due to the presence of different representations for left and right moving particles in the massless limit, the naive braiding unitarity relation (\ref{Unitarity-relations-K-matrices}) - which holds in the massive case discussed in Sec.~\ref{massive-case} - is not satisfied by the above $K$-matrices. As indicated in figure \ref{3}, taking the analytical continuation of the $K$-matrix to $-p$ turns a right moving particle into a left moving particle, which in the massless limit has a different bulk representation.  Hence, one needs to introduce a new matrix, denoted $K'$, which is obtained by solving  the boundary intertwining equation (\ref{BIE_sing_unitarity}) for the opposite particle momentum, i.e. symbolically a left moving particle reflecting off a right wall,
\begin{eqnarray}\label{BIE_sing_unitarity}
\pi^{a+}_{p}(\mathfrak{b}) \, K'(-\gamma) = K'(-\gamma)  \, \pi^{a-}_{-p}(\mathfrak{b}) \,\, \qquad \forall \, \mathfrak{b} \in \mathcal{B},
\end{eqnarray}
where $\mathcal{B}$ is the corresponding boundary subalgebra. Coincidentally, solving \eqref{BIE_sing_unitarity} in the singlet boundary case gives us the same $K$-matrices as before i.e. $K'(\gamma) = K(\gamma)$. This is however not always the case as we shall see for the vector boundary in the next subsection. The braiding unitarity leads to the following constraint on the dressing factors,
\begin{equation}
k^{L}(\gamma)\big[ k'^{L}(-\gamma)\big]=1=
k^{\tilde{L}}(\gamma)\big[ k'^{\tilde{L}}(-\gamma)\big]
\end{equation}
Note, however, that the matrix $K'(\gamma)$ can also be obtained by taking the massless limit of the massive BYBE solutions carefully. So while a priori there is no reason why the dressing factors for $K(\gamma)$ and $K'(\gamma)$ must be the same, they are not entirely independent either.

\subsection{Vector boundary\label{subsection-right-wall-vector-boundary}}
Let us consider now a boundary which itself carries a two-dimensional representation\footnote{This akin to \cite{Fring}, see also \cite{Muss}.}. In this case the reflection matrix is a $4 \times 4$ matrix acting on the tensor product of a bulk and a boundary representation. The boundary Yang-Baxter equation still reads as in (\ref{BYBE}), without the $L$ superscripts, however it is now an equation on three two-dimensional spaces $V_1 \otimes V_2 \otimes V_B$, where ${}_B$ denotes the boundary space:
\begin{eqnarray}
&&K_{2B}(\gamma_2,\gamma_B) R^{op}_{+-}\Big(\gamma_2 - (-\gamma_1)\Big)K_{1B}(\gamma_1,\gamma_B)R_{++}\Big(\gamma_1-\gamma_2\Big) \nonumber\\
&&\qquad \qquad = R^{op}_{--}\Big((-\gamma_2)-(-\gamma_1)\Big)K_{1B}(\gamma_1,\gamma_B)R_{+-}\Big(\gamma_1 - (-\gamma_2)\Big)K_{2B}(\gamma_2,\gamma_B).    \label{BYBEv}
\end{eqnarray}
The $R$-matrices always act in spaces $V_1 \otimes V_2$ which represent the two bulk particles. Here $K_{1B}(\gamma,\gamma_B)=(\mathds{1}\otimes\mathds{P})(K(\gamma,\gamma_B)\otimes \mathds{1})(\mathds{1}\otimes\mathds{P})$ and $K_{2B}(\gamma,\gamma_B)= \mathds{1} \otimes K(\gamma,\gamma_B)$ denote the reflection of the first and second particle with the boundary, with $\mathds{P}$ being the graded permutation and $K(\gamma,\gamma_B)$ being the complete vector boundary reflection matrix. In this setting it is clear that, contrarily to the singlet case, the presence of a non-trivial boundary representation increases considerably the complexity of the boundary Yang-Baxter equation. It is thus useful to recall the result of \cite{Prinsloo:2015apa}, stating that for vector boundaries the $K$-matrix must  factorise as
\begin{equation}\label{Vector-K-matrix-factorisation}
K(\gamma,\gamma_{B})=K^{LL}(\gamma,\gamma_{B})\oplus K^{\tilde{L}\tilde{L}}(\gamma,\gamma_{B})\oplus K^{L\tilde{L}}(\gamma,\gamma_{B})\oplus K^{\tilde{L}L}(\gamma,\gamma_{B})
\end{equation}
and consequently the BYBE \eqref{BYBEv} can be rewritten as a set of eight equations of the form
\begin{eqnarray}\label{right-wall-BYBE-rep-indices}
&&K_{2B}^{bc}(\gamma_2,\gamma_B) [R_{+-}^{ba}]^{op}\Big(\gamma_2 - (-\gamma_1)\Big)K_{1B}^{ac}(\gamma_1,\gamma_B)R^{ab}_{++}\Big(\gamma_1-\gamma_2 \Big) \nonumber\\
&&\qquad \qquad = [R_{--}^{ba}]^{op}\Big((-\gamma_2)-(-\gamma_1)\Big)K_{1B}^{ac}(\gamma_1,\gamma_B)R^{ab}_{+-}\Big(\gamma_1 - (-\gamma_2)\Big)K_{2B}^{bc}(\gamma_2,\gamma_B)
\end{eqnarray}
with $a,b,c \in \{ L,\tilde{L} \}$. This certainly helps in decreasing the complexity of BYBE, but unfortunately is still not enough to ensure the possibility of solving it in full generality, as we have done in the singlet case. For this reason, we shall start by following a slightly different logic, i.e. first analysing the intertwining equations, which are equations on the tensor product of two vector spaces rather than three, and successively checking the BYBE is satisfied. As for the case of the singlet boundary, the use of a symmetric coproduct forces us to consider the new choice of basis \eqref{new-algebra-basis-right-wall} for the bulk symmetry algebra, while searching for a subalgebra $\mathcal{B}$ preserved by the vector boundary. In this setting, the boundary intertwining equations are given by
\begin{eqnarray}\label{BIE_vec}
(\pi^{a-}_{-p}\otimes \pi^b_B)\Delta(\mathfrak{b}) \, K^{ab}(\gamma(p)) = K^{ab}(\gamma(p))  \, (\pi^{a+}_{p}\otimes \pi^b_B)\Delta(\mathfrak{b}) \quad \quad \forall \, \mathfrak{b} \in \mathcal{B},
\end{eqnarray}
where $a,b \in \{L,\Tilde{L} \}$ denote the representations, $\pm$ in the superscript denote right and left mover respectively, and $K^{ab}$ denotes the partial reflection matrix with $a$ (the first index) being the representation of the particle and $b$ (the second index) being the representation of the boundary. $p \in (0,\pi)$ in the subscript is the momentum of the particle and $B$ denotes the boundary representation. 

We shall begin by studying the above intertwining equation using the representations (\ref{Lrep+}, \ref{Lrep-}) for $L$ and (\ref{Ltilrep+}, \ref{Ltilrep-}) for $\Tilde{L}$, not only for the bulk excitations but also for the boundary itself. The boundary representation will simply be chosen to be $\pi_B^a(\mathfrak{b}) = \pi^{a+}_{p}(\mathfrak{b})|_{p=\pi,h\to h/2}$ for all generators $\mathfrak{b}$ of the boundary subalgebra $\mathcal{B}$ and $a \in \{ L,\Tilde{L} \}$. Such a choice of vector boundary representation allows us to keep contact with \cite{Prinsloo:2015apa} and enjoys a simple physical interpretation in terms of the bulk representations, since it just requires a fixing of its parameters. To further motivate this choice of boundary representation, if we focus on non-vanishing momentum, we note that the modulus of the group velocity of the wave packet $\vert \partial E / \partial p \vert = h \cos{\frac{p}{2}}$ vanishes at $p = \pm \pi$, hence giving us a stationary wave packet corresponding to the boundary. The rescaling $h\to h/2$ can be interpreted as arising from the $\mathbb{Z}_2$ orbifolding of the bulk line to the half-line with boundary. This does not represent the only possible choice of vector boundary representation. We can also consider other more generic choices for this boundary representation, albeit without clear physical interpretations. We refer to Sec. \ref{Gen Vec Bound} for an attempt at such a generalisation.

Analysing \eqref{BIE_vec} starting from a general unconstrained $K$-matrix, it is possible to fix its free coefficients in such a way that the following boundary algebra is preserved
\begin{eqnarray}\label{vec_bound_alg}
\mathcal{B}^{\mathfrak{U}}_{V}= \langle \mathfrak{H}_{L}, \, \mathfrak{H}_{R}, \, \mathfrak{U}^{\tfrac{1}{2}}\mathfrak{Q}_{L}, \, \mathfrak{U}^{\tfrac{1}{2}}\mathfrak{Q}_{R}, \, \mathfrak{U}^{-\tfrac{1}{2}}\mathfrak{G}_{L}, \, \mathfrak{U}^{-\tfrac{1}{2}}\mathfrak{G}_{R}, \, \mathfrak{U}\mathfrak{P}, \, \mathfrak{U}^{-1}\mathfrak{K}\rangle.
\end{eqnarray}
It is not hard to realise that it correctly satisfies both the \textit{coideal subalgebra} condition (\ref{right-coideal-subalgebra}) and the symmetric pair condition (\ref{symmetric-space}) together with $\mathcal{A}^{\mathfrak{U}}$ given in \eqref{new-algebra-basis-right-wall}. Notably, in this case, the amount of supersymmetry in the boundary algebra $\mathcal{B}^{\mathfrak{U}}_{V}$ is enough to fully constrain all the coefficients of the reflection matrix in the boundary intertwining equations, which in terms of the rapidity $\gamma$ takes the form
\begin{eqnarray}\label{special-sol-right-wall-vector-K-matrix}
    K^{LL}(\gamma,\gamma_B) &=& k^{LL}(\gamma,\gamma_B)\, \Big[ E_{11} \otimes E_{11} + \left( 1 - 2 \sech{\gamma} \right) \left(E_{11} \otimes E_{22} + i E_{22} \otimes E_{11}\right) + i E_{22} \otimes E_{22} \nonumber \\ 
    && + 2 \sqrt{2}\sinh{\frac{\gamma}{2}}\sech{\gamma} \left( E_{12} \otimes E_{21} + i \, E_{21} \otimes E_{12} \right) \Big]  
\end{eqnarray}

\begin{eqnarray}
    K^{\Tilde{L}\Tilde{L}}(\gamma,\gamma_B) &=& k^{\Tilde{L}\Tilde{L}}(\gamma,\gamma_B)\, \Big[ E_{11} \otimes E_{11} + \left( 1 - 2 \sech{\gamma} \right) \left(E_{11} \otimes E_{22} -i  E_{22} \otimes E_{11} \right) - i E_{22} \otimes E_{22} \nonumber \\ 
    && + 2 \sqrt{2}\sinh{\frac{\gamma}{2}}\sech{\gamma} \left(   E_{12} \otimes E_{21} -i \, E_{21} \otimes E_{12} \right) \Big]  
\end{eqnarray}

\begin{eqnarray}
    K^{L\Tilde{L}}(\gamma,\gamma_B) &=& k^{L\Tilde{L}}(\gamma,\gamma_B)\, \Big[ E_{11} \otimes E_{11} +  \left( \frac{1}{1 - 2 \sech{\gamma}} \right) \left( E_{11} \otimes E_{22} +i\, E_{22} \otimes E_{11} \right) + i\, E_{22} \otimes E_{22} \nonumber \\ 
    && + \left( \frac{2\sqrt{2} \sinh{\frac{\gamma}{2}} }{\cosh{\gamma} - 2 } \right) \left( \, E_{12} \otimes E_{12} +  iE_{21} \otimes E_{21} \right) \Big]  
\end{eqnarray}

\begin{eqnarray}
    K^{\Tilde{L}L}(\gamma,\gamma_B) &=& k^{\Tilde{L}L}(\gamma,\gamma_B)\, \Big[ E_{11} \otimes E_{11} +  \left( \frac{1}{1- 2 \sech{\gamma}} \right) \left( E_{11} \otimes E_{22} -i\, E_{22} \otimes E_{11} \right) - i\, E_{22} \otimes E_{22} \nonumber \\ 
    && + \left( \frac{2\sqrt{2} \sinh{\frac{\gamma}{2}} }{\cosh{\gamma} - 2 } \right) \left( \, E_{12} \otimes E_{12} - i E_{21} \otimes E_{21} \right) \Big]  
\end{eqnarray}
Here $\gamma_B = \gamma|_{p \to \pi}$.

The partial reflection matrices stated above correctly satisfy the BYBE and hence $\mathcal{B}^{\mathfrak{U}}_{V}$ represents a consistent reflection algebra for the vector boundary. The dressing phases $k^{ab}(\gamma,\gamma_B)$, for $a,b \in \{ L,\tilde{L} \}$, can be determined by the unitarity relations \eqref{Unitarity-relations-K-matrices} as well as the boundary crossing equations.\\

\noindent
\textbf{Unitarity.} \quad It is finally important to look again at the unitarity conditions \eqref{Unitarity-relations-K-matrices}, that physically relevant $K$-matrices should satisfy. For the case at hand it is useful to rewrite the physical unitarity condition as
\begin{equation}\label{unitarity-vector-boundary}
K^{ab}(\gamma,\gamma_{B})\bigl[K^{ab}(\gamma,\gamma_{B})\bigr]^{\dagger}=\mathbbmss{1} \qquad \forall \,\, a,b \in \{ L,\tilde{L} \} \, \, .
\end{equation}
Imposition of \eqref{unitarity-vector-boundary} on the $K$-matrices leads to the following constraints on the dressing factors $k^{ab}(\gamma,\gamma_B)$,  
\begin{equation}\label{right-wall-vector-boundary-dressing-factors-conditions}
\begin{aligned}
k^{ab}(\gamma,\gamma_{B})\bigl[ k^{ab}(\gamma,\gamma_{B}) \bigr]^{*} & = 1 \qquad &&\text{for} \qquad ab=\{ LL,\tilde{L}\tilde{L} \}
\\
k^{ab}(\gamma,\gamma_{B})\bigl[ k^{ab}(\gamma,\gamma_{B}) \bigr]^{*} & = \frac{(\cosh{\gamma}-2)^2}{(\cosh{\gamma})^2} 
\qquad &&\text{for} \qquad ab=\{ L\tilde{L},\tilde{L}L \}
\end{aligned}
\end{equation}

\indent

As for the singlet case, the naive braiding unitarity relation (\ref{Unitarity-relations-K-matrices}) needs to be modified. Taking the analytical continuation of the $K$-matrix to $-p$ gives us the following boundary intertwining equation.
\begin{eqnarray}\label{BIE_vec_unitarity}
(\pi^{a+}_{p}\otimes \pi^b_B)\Delta(\mathfrak{a}) \, K'^{ab}(-\gamma,\gamma_B) = K'^{ab}(-\gamma,\gamma_B)  \, (\pi^{a-}_{-p}\otimes \pi^b_B)\Delta(\mathfrak{a}),
\end{eqnarray}
Braiding unitarity, in case of the vector boundary, then takes the following form,
\begin{eqnarray}\label{Braiding_Unitarity_Vector}
K(\gamma,\gamma_B)K'(-\gamma,\gamma_B) = \mathbbmss{1} \, \, ,
\end{eqnarray}
and constrains the dressing factors of the $K$-matrices to satisfy,
\begin{equation}\label{right-wall-vec-bdry-braidU-dressing-factors-conditions}
\begin{aligned}
k^{ab}(\gamma,\gamma_{B})\bigl[ k'^{ab}(-\gamma,\gamma_{B}) \bigr] & = 1 \qquad &&\text{for} \qquad ab=\{ LL,\tilde{L}\tilde{L} \}
\\
k^{ab}(\gamma,\gamma_{B})\bigl[ k'^{ab}(-\gamma,\gamma_{B}) \bigr] & = \frac{(\cosh{\gamma}-2)^2}{(\cosh{\gamma})^2} 
\qquad &&\text{for} \qquad ab=\{ L\tilde{L},\tilde{L}L \}
\end{aligned}
\end{equation}


\section{Left wall reflection matrices} \label{sec::left wall}

\subsection{Singlet boundary}
We shall now repeat the analysis for a left wall.
\begin{figure}
\centerline{\includegraphics[width=10cm]{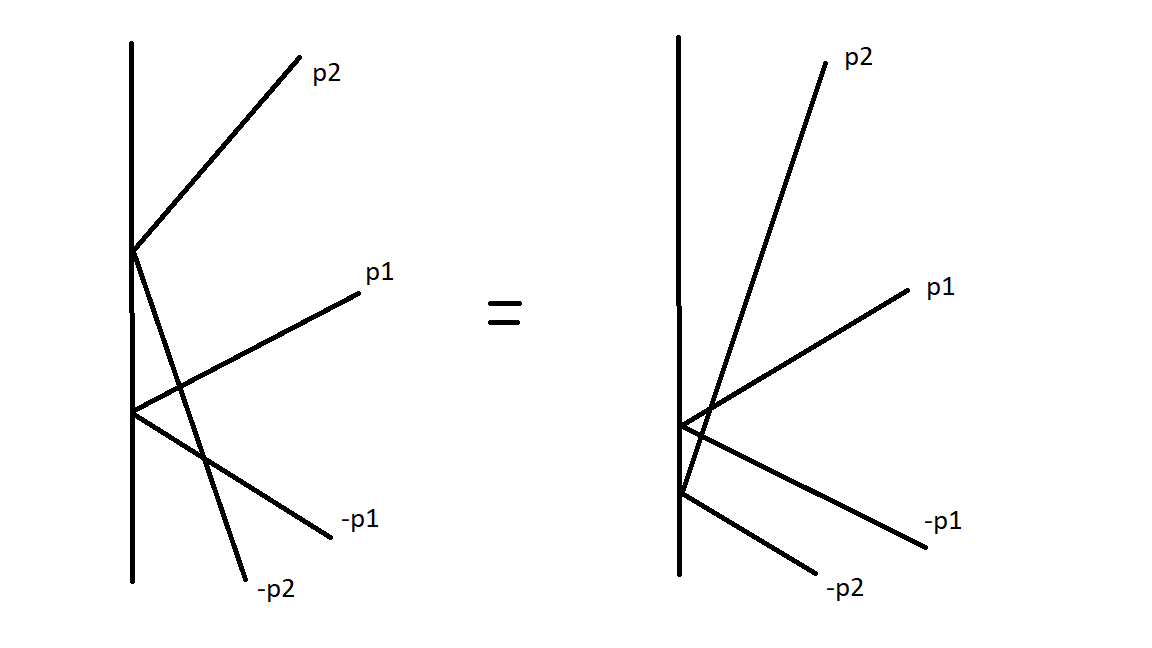}}
\caption{\label{2} The Boundary Yang-Baxter Equation (BYBE) for a left wall (hence all the incoming particles are left movers, and all the outgoing particles after reflection are right movers).}
\end{figure}
By observing figure \ref{2} we can again write
\begin{eqnarray}\label{BYBEl}
&&\tilde{K}^{L}_2(-\gamma_2) R^{LL}_{+-}\Big(\gamma_1 - (-\gamma_2)\Big)\tilde{K}^{L}_1(-\gamma_1)[R^{LL}_{--}]^{op}\Big(-\gamma_2-(-\gamma_1)\Big) = \nonumber\\
&&\qquad \qquad \qquad \qquad R^{LL}_{++}\Big(\gamma_1-\gamma_2 \Big)\tilde{K}^{L}_1(-\gamma_1)[R^{LL}_{+-}]^{op}\Big(\gamma_2 - (-\gamma_1)\Big)\tilde{K}^{L}_2(-\gamma_2),
\end{eqnarray}
where $\tilde{K}^{L}(x)$ represents, for $x>0$, the left wall $K$-matrix for particles in the $L$ representation and again we pull out minus signs from negative momenta and make use of
\begin{eqnarray}
\gamma_1 = \log \tan \frac{p_1}{4}, \qquad  \gamma_2 = \log \tan \frac{p_2}{4}, \qquad p_1, p_2 \in (0,\pi), \qquad \text{such that} \qquad \gamma_{1},\gamma_{2} < 0
\end{eqnarray}
It turns out that the equation which this produces for the singlet boundary is the exact same equation as we encountered for the right wall, except that now we consider the left wall reflection matrix $\tilde{K}^{L}$. In turn, the equation has the same solution as for the right wall, while exhibiting dependence on some new integration constant \footnote{The integration constants in the left and right wall cases must be related to each other via parity and crossing symmetries. However, we treat both walls independently and deal with the relation between left and right wall reflection matrices and fixing the dressing phases in future work.},
\begin{eqnarray}\label{left-wall-KL-matrix}
\tilde{K}^{L}(\gamma_{i})=\begin{pmatrix}1&0\\0&\tilde{f}^{L}(\gamma_i)\end{pmatrix},
\qquad 
\tilde{f}^{L}(\gamma_i) = -i \tanh \Big( \mbox{arccoth} (e^{\gamma_i}) + i \tilde{c}^{L}\Big), 
\qquad 
\gamma_{i} > 0 \,\, ,
\end{eqnarray}
with $\tilde{c}^{L}$ any constant. This similarity is not really a surprise, once we consider that the bulk $R$-matrices also coincide term by term with the previous case, thanks to
\begin{eqnarray}\label{thn} 
R^{aa}_{+-}(\gamma) =[R^{aa}_{+-}]^{op}(\gamma), \qquad R^{aa}_{++}(\gamma) = [R^{aa}_{--}]^{op}(\gamma), \qquad \text{for} \qquad a=\{L,\tilde{L}\}
\end{eqnarray}
 which is consistent with the requirements (\ref{think1}) and (\ref{think2}), and which makes the BYBE to exactly coincide for right and left wall. Let us point out that this is another convenient feature of the most symmetric frame, and it will not be true in generic frames (such as the frame discussed in Appendix \ref{App:A}).  

As for the right wall, the above $K$-matrix enjoys two particularly interesting choices of $\tilde{c}^{L}$, namely $\tilde{c}^{L}_{\pm} = \pi \big( \tilde{m} \pm \frac{1}{4} \big)$, with $\tilde{m}\in \mathbbmss{Z}$, which produce, upon using $\gamma_i(p_i) = \log \cot \frac{-p_i}{4}$, the results
\begin{eqnarray}
\tilde{K}^{L}\Big(\gamma_i(p_i)\Big)\Big\vert_{\tilde{c}_{+}^{L}} = \begin{pmatrix}1&0\\0& e^{\frac{i p_i}{2}}\end{pmatrix} \qquad \text{and} \qquad \tilde{K}^{L}\Big(\gamma_i(p_i)\Big)\Big\vert_{\tilde{c}_{-}^{L}} = \begin{pmatrix}1&0\\0&-e^{-\frac{i p_i}{2}}\end{pmatrix}\qquad p\in(-\pi,0).
\end{eqnarray}

If we then scatter two particles in the $\tilde{L}$ representation against the left wall, we can write the associated boundary Yang-Baxter equation as
\begin{eqnarray}
&&\tilde{K}^{\tilde{L}}_2(-\gamma_2) R^{\tilde{L}\tilde{L}}_{+-}\Big(\gamma_1 - (-\gamma_2)\Big)\tilde{K}^{\tilde{L}}_1(-\gamma_1)\big[R_{--}^{\tilde{L}\tilde{L}}\big]^{op}\Big(-\gamma_2-(-\gamma_1)\Big) = \nonumber\\
&&\qquad \qquad \qquad \qquad R^{\tilde{L}\tilde{L}}_{++}\Big( \gamma_1-\gamma_2\Big) \tilde{K}^{\tilde{L}}_1(-\gamma_1)\big[R_{+-}^{\tilde{L}\tilde{L}}\big]^{op}\Big(\gamma_2 - (-\gamma_1)\Big)\tilde{K}^{\tilde{L}}_2(-\gamma_2)\, .
\label{BYBElR}
\end{eqnarray}
Again, this has the same solution as the right wall,
\begin{eqnarray}\label{left-wall-KLtilde-matrix}
\tilde{K}^{\tilde{L}}(\gamma_i) = \begin{pmatrix}1&0\\0&\tilde{f}^{\tilde{L}}(\gamma_i)\end{pmatrix},
\qquad
\tilde{f}^{\tilde{L}}(\gamma_i) = i \tanh \Big( \mbox{arccoth} (e^{\gamma_i}) - i \tilde{c}^{\tilde{L}}\Big)\, ,
\qquad
\gamma > 0 \,\, ,
\end{eqnarray}
with $\tilde{c}^{\tilde{L}}$ a new constant, allowing the two interesting choices $\tilde{c}^{\tilde{L}}_{\pm} = \pi \big( \tilde{n} \pm \frac{1}{4} \big)$, with $\tilde{n}\in \mathbbmss{Z}$. These lead to 
\begin{eqnarray}
\tilde{K}^{\tilde{L}}\Big(\gamma_i(p_i)\Big)\Big\vert_{\tilde{c}^{\tilde{L}}_{+}} = \begin{pmatrix}1&0\\0&e^{-\frac{i p_i}{2}}\end{pmatrix} \qquad \text{and} \qquad \tilde{K}^{\tilde{L}}\Big(\gamma_i(p_i)\Big)\Big\vert_{\tilde{c}^{\tilde{L}}_{-}} = \begin{pmatrix}1&0\\0&-e^{\frac{i p_i}{2}}\end{pmatrix}\qquad p\in(-\pi,0).
\end{eqnarray}
The mixed boundary Yang-Baxter equations for one $L$ and one $\tilde{L}$ (resp., one $\tilde{L}$ and one $L$) particle read
\begin{eqnarray}
&&\tilde{K}^{\tilde{L}}_2(-\gamma_2) R^{L\tilde{L}}_{+-}\Big(\gamma_1 - (-\gamma_2)\Big)\tilde{K}^{L}_1(-\gamma_1)\big[R_{--}^{\tilde{L}L}\big]^{op}\Big(-\gamma_2-(-\gamma_1)\Big) = \nonumber\\
&&\qquad \qquad \qquad \qquad R^{L\tilde{L}}_{++}\Big(\gamma_1-\gamma_2\Big)\tilde{K}^{L}_1(-\gamma_1)\big[R_{+-}^{\tilde{L}L}\big]^{op}\Big(\gamma_2 - (-\gamma_1)\Big)\tilde{K}^{\tilde{L}}_2(-\gamma_2)\, ,
\label{BYBElR2}
\end{eqnarray}
and
\begin{eqnarray}
&&\tilde{K}^{L}_2(-\gamma_2) R^{\tilde{L}L}_{+-}\Big(\gamma_1 - (-\gamma_2)\Big)\tilde{K}^{\tilde{L}}_1(-\gamma_1)\big[R_{--}^{L\tilde{L}}\big]^{op}\Big(-\gamma_2-(-\gamma_1)\Big) = \nonumber\\
&&\qquad \qquad \qquad \qquad R^{\tilde{L}L}_{++}\Big(\gamma_1-\gamma_2\Big)\tilde{K}^{\tilde{L}}_1(-\gamma_1)\big[R_{+-}^{L\tilde{L}}\big]^{op}\Big(\gamma_2 - (-\gamma_1)\Big)\tilde{K}^{L}_2(-\gamma_2)\, .
\label{BYBElR3}
\end{eqnarray}
Both are solved provided that
\begin{equation}
\tilde{c}^{\tilde{L}}=-\tilde{c}^{L}+2\pi \tilde{s} \pm \tfrac{\pi}{2} \qquad \text{with} \qquad \tilde{s}\in \mathbbmss{Z}.    
\end{equation}
As for the right wall, we shall proceed assuming the latter relation to hold true, and work with
\begin{equation}\label{Ktilde-matrices-solving-BYBE}
\begin{aligned}
& \tilde{K}^{L}(\gamma_{i}) = 
\begin{pmatrix}
1 & 0 \\
0 & \tan{(\tilde{c}^{L}+\tilde{\chi}(\gamma_{i}))}
\end{pmatrix}
\qquad \qquad  
\tilde{K}^{\tilde{L}}(\gamma_{i}) = 
\begin{pmatrix}
1 & 0 \\
0 & \cot{( \tilde{c}^{L}+\tilde{\chi}(\gamma_{i}))}
\end{pmatrix}
\\
& \quad \text{with}  \qquad \qquad 
\tilde{\chi}(\gamma_{i}):=- \frac{i}{2}\Big(i\, (2\pi+4\pi k) + \log{\coth{\frac{\gamma_{i}}{2}}}\Big) \qquad \qquad \gamma_{i} > 0, \, k \in \mathbbmss{Z}.
\end{aligned}
\end{equation}

\noindent
\textbf{Boundary subalgebras.} \quad The next step is understanding which bulk symmetries are preserved by the left wall for a singlet boundary. With this aim, the intertwining equations to be solved read
\begin{equation}\label{left-wall-singlet-intertwining-equation}
\pi_{p}^{a+}(\mathfrak{b}) \, \tilde{K}^{a}\bigl(-\gamma(p)\bigr)=\tilde{K}^{a}\bigl(-\gamma(p)\bigr) \, \pi_{-p}^{a-}(\mathfrak{b}) \quad \forall \, \mathfrak{b}\in \mathcal{B} \qquad \text{and} \quad a\in \{L,\tilde{L}\}\, , \quad \gamma < 0, \quad p\in (0,\pi)
\end{equation}
with subscripts and superscripts following the same logic as \eqref{right-wall-singlet-intertwining-equation} and $\mathcal{B}$ the boundary subalgebra.
As for the right wall, trying to solve it brute force in the original bulk symmetry basis produces incorrect results, due to the incompatibility of the symmetric coproduct \eqref{simo} and the boundary coideal subalgebra requirement. Since for a left wall the condition \eqref{right-coideal-subalgebra} becomes
\begin{equation}\label{left-coideal-subalgebra}
\Delta(\mathfrak{b}) \in \mathcal{B}\otimes \mathcal{A} \qquad \forall \, \mathfrak{b} \in \mathcal{B} \,\, ,
\end{equation}
it is clear that the two boundary subalgebras \eqref{massive-boundary-susy-subalgebras}, identified in the massive case with the non-symmetric coproduct, do not satisfy the latter condition and a more suitable choice of basis is needed. However, this cannot be the same as for the right wall, since the exchange of $\mathcal{A}$ and $\mathcal{B}$ in the tensor product \eqref{left-coideal-subalgebra}, with respect to \eqref{right-coideal-subalgebra}, would not be respected. Inspecting \eqref{symm-coproduc-U} and \eqref{U-coproduct} it is not hard to recognise that a natural left-analogue of \eqref{new-algebra-basis-right-wall} is given by 
\begin{equation}\label{new-algebra-basis-left-wall}
\tilde{\mathcal{A}}^{\mathfrak{U}}=\langle \mathfrak{H}_{L}, \, \mathfrak{H}_{R}, \, \mathfrak{U}^{-\tfrac{1}{2}}\mathfrak{Q}_{L}, \, \mathfrak{U}^{-\tfrac{1}{2}}\mathfrak{Q}_{R}, \, \mathfrak{U}^{\tfrac{1}{2}}\mathfrak{G}_{L}, \, \mathfrak{U}^{\tfrac{1}{2}}\mathfrak{G}_{R}, \, \mathfrak{U}^{-1}\mathfrak{P}, \, \mathfrak{U}\mathfrak{K}, \, \mathfrak{p} \rangle \, \quad \text{with} \quad \mathfrak{U}^{\alpha} = e^{i \frac{\alpha}{2} \mathfrak{p}} \quad \forall \, \alpha \in \mathbb{R},
\end{equation}
which allows us to identify two subalgebras satisfying conditions \eqref{left-coideal-subalgebra} and \eqref{symmetric-space}
\begin{equation}\label{left-wall-rescaled-susy-boundary-algebras}
\tilde{\mathcal{B}}_{L}^{\mathfrak{U}} = \langle \mathfrak{H}_{L}, \, \mathfrak{H}_{R}, \, \mathfrak{U}^{-\tfrac{1}{2}}\mathfrak{Q}_{L}, \, \mathfrak{U}^{\tfrac{1}{2}}\mathfrak{G}_{L}  \rangle \qquad 
\text{and}
\qquad
\tilde{\mathcal{B}}_{R}^{\mathfrak{U}} = \langle \mathfrak{H}_{L}, \, \mathfrak{H}_{R}, \, \mathfrak{U}^{-\tfrac{1}{2}}\mathfrak{Q}_{R}, \, \mathfrak{U}^{\tfrac{1}{2}}\mathfrak{G}_{R}  \rangle \, .
\end{equation}
From this it is clear that the left wall rescaling factors can be obtained by letting $\mathfrak{p}\rightarrow - \mathfrak{p}$ in those found for the right wall. One can now go back to the intertwining equation \eqref{left-wall-singlet-intertwining-equation} and using \eqref{U-generator-rep.} check that
\begin{itemize}
\item The generators $\mathfrak{H}_L,\mathfrak{H}_R$ are preserved by the left wall in both $L$ and $\tilde{L}$ representations for any choice of constant $\tilde{c}^{L}$ appearing in the $K$-matrices \eqref{Ktilde-matrices-solving-BYBE}.
\item The rescaled generators $\mathfrak{U}^{-1}\mathfrak{P}, \, \mathfrak{U}\mathfrak{K}$ are never preserved in either the $L$ or $\tilde{L}$ representation.
\item The intertwining equations for $\mathfrak{U}^{-\tfrac{1}{2}}\mathfrak{Q}_{L},\mathfrak{U}^{\tfrac{1}{2}}\mathfrak{G}_{L}$ admit simultaneous solutions in both the $L$ and $\tilde{L}$ representations provided one chooses $\tilde{c}^{L}=\tilde{c}^{L}_{+}=\pi \big( \tilde{m} + \frac{1}{4} \big)$ with $\tilde{m}\in \mathbbmss{Z}$. For these values of $\tilde{c}^{L}$, the intertwining equations for $\mathfrak{U}^{-\tfrac{1}{2}}\mathfrak{Q}_{R},\mathfrak{U}^{\tfrac{1}{2}}\mathfrak{G}_{R}$ are not satisfied and preservation of $\mathfrak{H}_L,\mathfrak{H}_R$ is enhanced to the supersymmetric boundary subalgebra $\tilde{\mathcal{B}}^{\mathfrak{U}}_{L}$ defined in \eqref{left-wall-rescaled-susy-boundary-algebras}.
\item The intertwining equations for $\mathfrak{U}^{-\tfrac{1}{2}}\mathfrak{Q}_{R},\mathfrak{U}^{\tfrac{1}{2}}\mathfrak{G}_{R}$ admit simultaneous solutions in both the $L$ and $\tilde{L}$ representations provided one chooses $\tilde{c}^{L}=\tilde{c}^{L}_{-}=\pi \big( \tilde{m} - \frac{1}{4} \big)$ with $\tilde{m}\in \mathbbmss{Z}$. For these values of $\tilde{c}^{L}$, the intertwining equations for $\mathfrak{U}^{-\tfrac{1}{2}}\mathfrak{Q}_{L},\mathfrak{U}^{\tfrac{1}{2}}\mathfrak{G}_{L}$ are not satisfied and preservation of $\mathfrak{H}_L,\mathfrak{H}_R$ is enhanced to the supersymmetric boundary subalgebra $\tilde{\mathcal{B}}^{\mathfrak{U}}_{R}$ defined in \eqref{left-wall-rescaled-susy-boundary-algebras}.
\end{itemize}

\noindent
\textbf{Hidden symmetry.} \quad  To conclude our analysis of the singlet left wall, we should now look for the existence of a hidden symmetry algebra, in analogy with the results obtained for the right wall. The latter taught us that we should keep using the convenient basis \eqref{new-algebra-basis-left-wall} while including a generic dependence of the hidden generators on the constant $\tilde{c}^{L}$, which appears in the $K$-matrices \eqref{Ktilde-matrices-solving-BYBE}. Taking this into account, the hidden symmetry generators defined in \cite{Prinsloo:2015apa} for the massive case generalise to
\begin{equation}\label{rescaled-hidden-generators-left-wall}
\begin{aligned}
\mathfrak{q}_{+} &:= \mathfrak{U}^{1/2}\mathfrak{K}\mathfrak{Q}_{L}+i\tilde{e}(\tilde{c}^{L}) \,  \mathfrak{U}^{-1/2}\mathfrak{P}\mathfrak{G}_{R} \qquad \qquad \mathfrak{d} := (\mathfrak{H}_{L}-\tilde{e}(\tilde{c}^{L})^2 \, \mathfrak{H}_{R}+i\tilde{e}(\tilde{c}^{L}) \, (\mathfrak{U}^{-1}\mathfrak{P}+\mathfrak{U}\mathfrak{K}))\mathfrak{P}\mathfrak{K}
\\
\mathfrak{q}_{-} &:= \mathfrak{U}^{-1/2}\mathfrak{P}\mathfrak{G}_{L}+i\tilde{e}(\tilde{c}^{L}) \, \mathfrak{U}^{1/2}\mathfrak{K}\mathfrak{Q}_{R} \qquad \qquad \tilde{\mathfrak{d}} := (\mathfrak{H}_{L}-\tilde{e}(\tilde{c}^{L})^2 \, \mathfrak{H}_{R}-i\tilde{e}(\tilde{c}^{L}) \, (\mathfrak{U}^{-1}\mathfrak{P}+\mathfrak{U}\mathfrak{K}))\mathfrak{P}\mathfrak{K}
\\
\mathfrak{s}_{+} &:= \mathfrak{U}^{1/2}\mathfrak{K}\mathfrak{Q}_{L}-i\tilde{e}(\tilde{c}^{L}) \, 
\mathfrak{U}^{-1/2}\mathfrak{P}\mathfrak{G}_{R} \qquad \qquad \mathfrak{n} := (\mathfrak{H}_{L}+\tilde{e}(\tilde{c}^{L})^2 \, \mathfrak{H}_{R}+i\tilde{e}(\tilde{c}^{L}) \, (\mathfrak{U}^{-1}\mathfrak{P}-\mathfrak{U}\mathfrak{K}))\mathfrak{P}\mathfrak{K}
\\
\mathfrak{s}_{-} &:= \mathfrak{U}^{-1/2}\mathfrak{P}\mathfrak{G}_{L}-i\tilde{e}(\tilde{c}^{L}) \, \mathfrak{U}^{1/2}\mathfrak{K}\mathfrak{Q}_{R} \qquad \qquad \tilde{\mathfrak{n}} := (\mathfrak{H}_{L}+\tilde{e}(\tilde{c}^{L})^2 \, \mathfrak{H}_{R}-i\tilde{e}(\tilde{c}^{L}) \, (\mathfrak{U}^{-1}\mathfrak{P}-\mathfrak{U}\mathfrak{K}))\mathfrak{P}\mathfrak{K} \, \, ,
\end{aligned}
\end{equation}
with the function $\tilde{e}(\tilde{c}^{L})$ to be determined by solving the intertwining equations \eqref{left-wall-singlet-intertwining-equation} for the hidden-symmetry generators. This leads to the following results:
\begin{itemize}
\item The generators $\mathfrak{n},\tilde{\mathfrak{n}}$ are always preserved by the left wall in both the $L$ and $\tilde{L}$ representations, for any choice of function $\tilde{e}(\tilde{c}^{L})$.
\item The generators $\mathfrak{d},\tilde{\mathfrak{d}}$ are never preserved by the left wall in either the $L$ or $\tilde{L}$ representation for any non-zero choice of $\tilde{e}(\tilde{c}^{L})$.
\item The intertwining equations for $\mathfrak{q}_{+}, \mathfrak{s}_{-}$ admit simultaneous solutions in both the $L$ and $\tilde{L}$ representations provided one chooses $\tilde{e}(\tilde{c}^{L})=\tilde{e}_{-}(\tilde{c}^{L}) = -i \, \frac{\cos{(\tilde{c}^{L})}+\sin{(\tilde{c}^{L})}}{\cos{(\tilde{c}^{L})}-\sin{(\tilde{c}^{L})}}$. For such choice of $\tilde{e}(\tilde{c}^{L})$ the intertwining equations for $\mathfrak{q}_{-},\mathfrak{s}_{+}$ are never satisfied, unless one further specifies $\tilde{c}^{L}$ such that $\cos{(\tilde{c}^{L})}=-\sin{(\tilde{c}^{L})}$. This would precisely correspond to choosing $\tilde{c}^{L}=\tilde{c}_{-}^{L}=\pi(\tilde{m} -\tfrac{1}{4})$ with $\tilde{m} \in \mathbb{Z}$, namely the $\tilde{c}^{L}$ leading to preservation of $\tilde{\mathcal{B}}_{R}^{\mathfrak{U}}$ defined in \eqref{left-wall-rescaled-susy-boundary-algebras}. However, this is obviously forbidden by the fact that would imply $\tilde{e}(\tilde{c}^{L})=0$ and thus trivialisation of the hidden generators. The value $\tilde{c}^L = \tilde{c}^L_{+}$ is also forbidden, due to the vanishing of the denominator in $\tilde{e}_{-}(\tilde{c}^L)$.
\item The intertwining equations for $\mathfrak{q}_{-}, \mathfrak{s}_{+}$ admit simultaneous solutions in both the $L$ and $\tilde{L}$ representations provided one chooses $\tilde{e}(\tilde{c}^{L})=\tilde{e}_{+}(\tilde{c}^{L}) = i \, \frac{\cos{(\tilde{c}^{L})}+\sin{(\tilde{c}^{L})}}{\cos{(\tilde{c}^{L})}-\sin{(\tilde{c}^{L})}}$. For such choice of $\tilde{e}(\tilde{c}^{L})$ the intertwining equations for $\mathfrak{q}_{+},\mathfrak{s}_{-}$ are never satisfied, unless one further specifies $\tilde{c}^{L}$ such that $\cos{(\tilde{c}^{L})}=-\sin{(\tilde{c}^{L})}$. This corresponds to choosing $\tilde{c}^{L}=\tilde{c}_{-}^{L}=\pi(\tilde{m} -\tfrac{1}{4})$ with $\tilde{m} \in \mathbb{Z}$, which is forbidden by the fact that $\tilde{e}(\tilde{c}^{L})=0$ for such values. Once again one also needs $\tilde{c}^L\neq \tilde{c}^L_{+}$.
\end{itemize}
The hidden symmetry algebras  $\langle \mathfrak{q}_{-},\mathfrak{s}_{+},\mathfrak{n},\tilde{\mathfrak{n}} \rangle$ and $ \langle \mathfrak{q}_{+},\mathfrak{s}_{-},\mathfrak{n},\tilde{\mathfrak{n}} \rangle$, found to be preserved by the left wall for any choice of $\tilde{c}^{L}\neq \tilde{c}^{L}_{\mp}$, turn out to be of exactly the same form as the ones preserved by the right wall
\begin{equation}
\tilde{\mathcal{B}}_{D}^{\mathfrak{U}\pm} = \langle \mathfrak{q}_{\pm}, \mathfrak{s}_{\mp}, \mathfrak{n}, \tilde{\mathfrak{n}} \rangle \qquad \text{for} \qquad \tilde{e}(\tilde{c}^L) = \tilde{e}_{\mp}(\tilde{c}^L) \,\, ,
\end{equation}
and considerations similar to those highlighted after equation \eqref{preserved_hidden_symmetric_pairs} also apply to this case. \\

\noindent
\textbf{Short summary of the left wall singlet boundary.} \quad Solving the four BYB equations \eqref{BYBEl},\eqref{BYBElR}, \eqref{BYBElR2},\eqref{BYBElR3} leads to the $K$-matrices \eqref{Ktilde-matrices-solving-BYBE}, which we also report here
\begin{equation}
\begin{aligned}
& \tilde{K}(\gamma_{i}) = 
\begin{pmatrix}
1 & 0 \\
0 & \tan{(\tilde{c}^{L}+\tilde{\chi}(\gamma_{i}))}
\end{pmatrix}
\qquad \qquad  
\tilde{K}^{\tilde{L}}(\gamma_{i}) = 
\begin{pmatrix}
1 & 0 \\
0 & \cot{( \tilde{c}^{L}+\tilde{\chi}(\gamma_{i}))}
\end{pmatrix}
\\
& \,\, \text{with}  \qquad 
\tilde{\chi}(\gamma_{i}):=- \frac{i}{2}\Big(i\, (2\pi+4\pi k) + \log{\coth{\frac{\gamma_{i}}{2}}}\Big) \qquad  \text{and} \qquad \gamma_{i} > 0, \, k \in \mathbbmss{Z},\,   \tilde{c}^{L} \in \mathbb{C} \,\, .
\end{aligned}
\end{equation}
Satisfying the left coideal subalgebra condition \eqref{left-coideal-subalgebra} without breaking the symmetric pair decomposition \eqref{symmetric-space} then requires choosing the following basis for the bulk symmetry algebra
\begin{equation}
\tilde{\mathcal{A}}^{\mathfrak{U}}=
\langle \mathfrak{H}_{L}, \, \mathfrak{H}_{R}, \, \mathfrak{U}^{-\tfrac{1}{2}}\mathfrak{Q}_{L}, \, \mathfrak{U}^{-\tfrac{1}{2}}\mathfrak{Q}_{R}, \, \mathfrak{U}^{\tfrac{1}{2}}\mathfrak{G}_{L}, \, \mathfrak{U}^{\tfrac{1}{2}}\mathfrak{G}_{R}, \, \mathfrak{U}^{-1}\mathfrak{P}, \, \mathfrak{U}\mathfrak{K}, \, \mathfrak{p} \rangle \quad \text{with} \quad \mathfrak{U}^{\alpha} = e^{i \frac{\alpha}{2} \mathfrak{p}} \quad \forall \, \alpha \in \mathbb{R}.
\end{equation}
The central generators $\mathfrak{H}_{L},\mathfrak{H}_{R},\mathfrak{p}$ are preserved by the left wall in both the $L$ and $\tilde{L}$ representations, i.e. they satisfy the intertwining equations \eqref{left-wall-singlet-intertwining-equation}, for any choice of the constant $\tilde{c}^{L}$, while the generators $\mathfrak{U}^{-1}\mathfrak{P}$ and $\mathfrak{U}\mathfrak{K}$ are never preserved in either the $L$ or $\tilde{L}$ representation. The fermionic generators $\mathfrak{U}^{-\frac{1}{2}}\mathfrak{Q}_{a}$ and $\mathfrak{U}^{\frac{1}{2}}\mathfrak{G}_{a}$ are generally not preserved, unless one sits at the special values $\tilde{c}^{L}= \tilde{c}^{L}_{\pm} = \pi(\tilde{m}\pm\frac{1}{4})$, with $\tilde{m}\in \mathbb{Z}$, where the following supersymmetry enhanced subalgebras are preserved
\begin{equation}
\tilde{c}^{L}_{+}: \quad \tilde{\mathcal{B}}_{L}^{\mathfrak{U}} = \langle \mathfrak{H}_{L},\mathfrak{H}_{R},\mathfrak{U}^{-\frac{1}{2}}\mathfrak{Q}_{L},\mathfrak{U}^{\frac{1}{2}}\mathfrak{G}_{L}  \rangle \qquad \quad \text{and}  \qquad \quad \tilde{c}^{L}_{-}: \quad \tilde{\mathcal{B}}_{R}^{\mathfrak{U}} = \langle \mathfrak{H}_{L},\mathfrak{H}_{R},\mathfrak{U}^{-\frac{1}{2}}\mathfrak{Q}_{R},\mathfrak{U}^{\frac{1}{2}}\mathfrak{G}_{R}  \rangle\,\,.
\end{equation}
The two $K$-matrices then take the following very simple form
\begin{equation}\label{left-wall-singlet-K-matrices-summary}
\begin{aligned}
&\tilde{c}^{L}_{+}: \qquad \tilde{K}^{L}_{\tilde{\mathcal{B}}_{L}^{\mathfrak{U}}}(p) = \begin{pmatrix}1&0\\0&e^{i\frac{p}{2}}\end{pmatrix}
\qquad \qquad &&\text{and} \qquad \qquad
\tilde{K}^{\tilde{L}}_{\tilde{\mathcal{B}}_{L}^{\mathfrak{U}}}(p) = \begin{pmatrix}1&0\\0&e^{-i\frac{ p}{2}}\end{pmatrix}
\\
&\tilde{c}^{L}_{-}: \qquad \tilde{K}^{L}_{\tilde{\mathcal{B}}_{R}^{\mathfrak{U}}}(p) = \begin{pmatrix}1&0\\0&-e^{-i\frac{p}{2}}\end{pmatrix}
\qquad \qquad  &&\text{and} \qquad \qquad 
\tilde{K}^{\tilde{L}}_{\tilde{\mathcal{B}}_{R}^{\mathfrak{U}}}(p) = \begin{pmatrix}1&0\\0&-e^{i\frac{p}{2}}\end{pmatrix},
\end{aligned}
\end{equation}
where $p\in(-\pi,0)$. Once again one can finally define, in terms of $\mathfrak{a} \in \tilde{\mathcal{A}}^{\mathfrak{U}}$, a set of new generators \eqref{rescaled-hidden-generators-left-wall} depending on some arbitrary function $\tilde{e}(\tilde{c}^{L})$ of the constant which appears in the $K$-matrices, and exhibiting commutation relations \eqref{hidden_generators_commutators} with the same structure as \eqref{comma}. This change of basis allows to uncover the presence of two hidden symmetry boundary superalgebras $\tilde{\mathcal{B}}_{D}^{\mathfrak{U}\pm} = \langle \mathfrak{q}_{\pm}, \mathfrak{s}_{\mp}, \mathfrak{n}, \tilde{\mathfrak{n}} \rangle$, which respectively solve the boundary intertwining equations \eqref{right-wall-singlet-intertwining-equation} for any $\tilde{c}^{L} \neq \tilde{c}^{L}_{\mp}$ upon choosing
\begin{equation}
\tilde{e}_{\mp}(\tilde{c}^{L}) = \mp i \,\, \frac{\cos{(\tilde{c}^{L})}+\sin{(\tilde{c}^{L})}}{\cos{(\tilde{c}^{L})}-\sin{(\tilde{c}^{L})}} \,\, .
\end{equation}
We conclude by noting that going from the right to the left wall the form of the $K$-matrices formally remains unchanged and the preserved symmetry algebras are mapped onto each other by simply letting $\mathfrak{U} \leftrightarrow \mathfrak{U}^{-1}$, which corresponds to $\mathfrak{p} \leftrightarrow -\mathfrak{p}$, and swapping the labels $L \leftrightarrow R$ on the fermionic generators. For what concerns the hidden symmetry sector, the same subalgebras are preserved by the two walls (keeping in mind they differ by $\mathfrak{U} \leftrightarrow \mathfrak{U}^{-1}$ in the definition of hidden generators) and upon setting $c^{L}=\tilde{c}^{L}\neq c^{L}_{\pm}$ one can readily notice that $\tilde{e}_{\pm} = e_{\mp}^{-1} = -e_{\pm}^{-1}$.
This makes clear that $\tilde{\mathcal{B}}_{D}^{\mathfrak{U}\pm}$ and $\mathcal{B}_{D}^{\mathfrak{U}\pm}$ are mapped onto each other by the simple rules $\mathfrak{U}\leftrightarrow \mathfrak{U}^{-1}$ and $\tilde{e}_{\pm} \leftrightarrow -e_{\pm}^{-1}$.\\

\noindent
\textbf{Unitarity.} \quad
Similar to the right wall singlet boundary, given the simple structure \eqref{left-wall-singlet-K-matrices-summary} for the left wall $K$-matrices, it is straightforward to check that both physical and braiding unitarity relations \eqref{Unitarity-relations-K-matrices} are satisfied (after careful consideration of the subtlety involving massless representations), provided the associated dressing factors also respect the relations
\begin{equation}
\tilde{k}^{L}(\gamma)\big[ \tilde{k}^{L}(\gamma)\big]^{*}=1=\tilde{k}^{L}(\gamma) \tilde{k}'^{L}(-\gamma) 
\qquad \qquad
\tilde{k}^{\tilde{L}}(\gamma)\big[ \tilde{k}^{\tilde{L}}(\gamma)\big]^{*}=1=\tilde{k}^{\tilde{L}}(\gamma) \tilde{k}'^{\tilde{L}}(-\gamma)\,\, \qquad \gamma > 0 \, \, .
\end{equation}

\subsection{Vector boundary}\label{subsec:Left-Wall-Vector-Boundary}
The case in which the left boundary carries a two dimensional representation is very similar to the one for the right wall. The boundary Yang-Baxter equation can be read off figure \ref{2} and exhibits the same structure as \eqref{BYBEl}, with the main difference being that it is now an equation on three two-dimensional spaces $V_{B} \otimes V_1 \otimes V_2$. Once again, ${}_B$ denotes the boundary space and one should note the swap in its position, as compared to the right wall of section \ref{subsection-right-wall-vector-boundary}, where we had $V_{1}\otimes V_{2}\otimes V_{B}$. The BYBE then reads
\begin{eqnarray}\label{BYBE-left-wall-vector}
&&\tilde{K}_{B2}(-\gamma_2,\gamma_B) R_{+-}\Big(\gamma_1 - (-\gamma_2)\Big)\tilde{K}_{B1}(-\gamma_1,\gamma_B)R_{--}^{op}\Big(-\gamma_2-(-\gamma_1)\Big) \nonumber\\
&&\qquad \qquad = R_{++}\Big(\gamma_{1}-\gamma_2\Big)\tilde{K}_{B1}(-\gamma_1,\gamma_B)R_{+-}^{op}\Big(\gamma_2 - (-\gamma_1)\Big)\tilde{K}_{B2}(-\gamma_2,\gamma_B).    
\end{eqnarray}
The $R$-matrices always act in spaces $V_1 \otimes V_2$ which represent the two bulk particles. Here $\tilde{K}_{B2}(\gamma,\gamma_B)=(\mathds{1}\otimes\mathds{P})(\tilde{K}(\gamma,\gamma_B)\otimes \mathds{1})(\mathds{1}\otimes\mathds{P})$ and $\tilde{K}_{B1}(\gamma,\gamma_B)= \tilde{K}(\gamma,\gamma_B)\otimes\mathds{1}$ denote the reflection of the second and first particle with the boundary, with $\mathds{P}$ the graded permutation and $\tilde{K}(\gamma,\gamma_B)$ the complete vector boundary reflection matrix for the left wall. Thanks to the factorisation \eqref{Vector-K-matrix-factorisation}, the above BYBE can again be rewritten as a set eight equations
\begin{eqnarray}\label{left-wall-BYBE-rep-index-notation}
&&\tilde{K}_{B2}^{cb}(-\gamma_2,\gamma_B) R^{ab}_{+-}\Big(\gamma_1 - (-\gamma_2)\Big)\tilde{K}_{B1}^{ca}(-\gamma_1,\gamma_B)[R_{--}^{ba}]^{op}\Big(-\gamma_2-(-\gamma_1)\Big) \nonumber\\
&&\qquad \qquad = R^{ab}_{++}\Big(\gamma_1-\gamma_2\Big)\tilde{K}_{B1}^{ca}(-\gamma_1,\gamma_B)[R_{+-}^{ba}]^{op}\Big(\gamma_2 - (-\gamma_1)\Big)\tilde{K}_{B2}^{cb}(-\gamma_2,\gamma_B)
\end{eqnarray}
with $a,b,c \in \{ L,\tilde{L} \}$. We proceed, as for the right wall, by solving the boundary intertwining equations, which for a left wall take the form
\begin{equation}\label{left-wall-vector-intertwining-equation}
\bigl(\pi^b_B\otimes\pi_{p}^{a+}\bigr)\Delta(\mathfrak{b})\, \tilde{K}^{ba}\bigl(-\gamma(p)\bigr)=\tilde{K}^{ba}\bigl(-\gamma(p)\bigr)\, \Delta(\mathfrak{b})\bigl(\pi^b_B\otimes\pi_{-p}^{a-}) \qquad \forall \, \mathfrak{b}\in \tilde{\mathcal{B}}^\mathfrak{U}_{V} \qquad \text{and} \qquad a,b\in \{L,\tilde{L}\}\,\, ,
\end{equation}
for the generators of the boundary subalgebra 
\begin{eqnarray}\label{left_wall_vec_bound_alg}
\tilde{\mathcal{B}}^{\mathfrak{U}}_{V}= \langle \mathfrak{H}_{L}, \, \mathfrak{H}_{R}, \, \mathfrak{U}^{-\tfrac{1}{2}}\mathfrak{Q}_{L}, \, \mathfrak{U}^{-\tfrac{1}{2}}\mathfrak{Q}_{R}, \, \mathfrak{U}^{\tfrac{1}{2}}\mathfrak{G}_{L}, \, \mathfrak{U}^{\tfrac{1}{2}}\mathfrak{G}_{R}, \, \mathfrak{U}^{-1}\mathfrak{P}, \, \mathfrak{U}\mathfrak{K}\rangle.
\end{eqnarray}
The two boundary representations are again chosen to be the same as the bulk ones, for a fixed value of the momentum parameter: $\pi^b_B(\mathfrak{b})=\pi^{b-}_p(\mathfrak{b})|_{p=-\pi,h\rightarrow h/2}$ for all $\mathfrak{b} \in \tilde{\mathcal{B}}^\mathfrak{U}_{V}$ and $b \in \{ L,\tilde{L} \}$. These lead to the following solutions 
\begin{eqnarray}\label{special-sol-left-wall-vector-K-matrix}
    \tilde{K}^{LL}(\gamma,\gamma_B) &=& \tilde{k}^{LL}(\gamma,\gamma_B)\, \Big[ E_{11} \otimes E_{11} + \left( 1 + 2 \sech{\gamma} \right) \left(i E_{11} \otimes E_{22} +  E_{22} \otimes E_{11}\right) + i E_{22} \otimes E_{22} \nonumber \\ 
    && -2 \sqrt{2}\sinh{\frac{\gamma}{2}}\sech{\gamma} \left(i E_{12} \otimes E_{21} +  \, E_{21} \otimes E_{12} \right) \Big]  
\end{eqnarray}
\begin{eqnarray}
    \tilde{K}^{\Tilde{L}\Tilde{L}}(\gamma,\gamma_B) &=& \tilde{k}^{\Tilde{L}\Tilde{L}}(\gamma,\gamma_B)\, \Big[ E_{11} \otimes E_{11} + \left( 1 + 2 \sech{\gamma} \right) \left(-i E_{11} \otimes E_{22} +  E_{22} \otimes E_{11} \right) - i E_{22} \otimes E_{22} \nonumber \\ 
    && + 2 \sqrt{2}\sinh{\frac{\gamma}{2}}\sech{\gamma} \left( i  E_{12} \otimes E_{21} - \, E_{21} \otimes E_{12} \right) \Big]  
\end{eqnarray}
\begin{eqnarray}
    \tilde{K}^{L\Tilde{L}}(\gamma,\gamma_B) &=& \tilde{k}^{L\Tilde{L}}(\gamma,\gamma_B)\, \Big[ E_{11} \otimes E_{11} +  \left( \frac{1}{1 + 2 \sech{\gamma}} \right) \left(-i E_{11} \otimes E_{22} +\, E_{22} \otimes E_{11} \right) - i\, E_{22} \otimes E_{22} \nonumber \\ 
    && + \left( \frac{2\sqrt{2} \sinh{\frac{\gamma}{2}} }{\cosh{\gamma} - 2 } \right) \left( \, E_{12} \otimes E_{12} +  iE_{21} \otimes E_{21} \right) \Big]  
\end{eqnarray}
\begin{eqnarray}
    \tilde{K}^{\Tilde{L}L}(\gamma,\gamma_B) &=& \tilde{k}^{\Tilde{L}L}(\gamma,\gamma_B)\, \Big[ E_{11} \otimes E_{11} +  \left( \frac{1}{1+ 2 \sech{\gamma}} \right) \left(i E_{11} \otimes E_{22} +\, E_{22} \otimes E_{11} \right) + i\, E_{22} \otimes E_{22} \nonumber \\ 
    && + \left( \frac{2\sqrt{2} \sinh{\frac{\gamma}{2}} }{\cosh{\gamma} - 2 } \right) \left( \, E_{12} \otimes E_{12} + i E_{21} \otimes E_{21} \right) \Big]  
\end{eqnarray}\\
with $\gamma_B = \gamma|_{p \to -\pi}$. We have verified that all of the equations summarised in \eqref{left-wall-BYBE-rep-index-notation} are satisfied by this set of matrices.\\

\noindent
\textbf{Unitarity.} \quad The results for both physical and braiding unitarity are identical to the results for the right wall. In particular, the physical unitarity condition (\ref{unitarity-vector-boundary}) imposes the constraint (\ref{right-wall-vector-boundary-dressing-factors-conditions}) to the dressing factors of $\tilde{K}$. To satisfy braiding unitarity (\ref{Braiding_Unitarity_Vector}), a matrix $\tilde{K}'$ needs to be introduced as a solution to the equations obtained by replacing $p\rightarrow -p$ in (\ref{left-wall-vector-intertwining-equation}) and using the appropriate bulk representations. The constraint on the dressing factors of $\tilde{K}$ and $\tilde{K}'$ is identical to (\ref{right-wall-vec-bdry-braidU-dressing-factors-conditions}).

\section{The massive case with the symmetric coproduct}\label{massive-case}

In this section we summarise the bulk scattering theory, in the most symmetric frame which we have been using so far. The representation $\Tilde{L}$ while physical for the massless case, is unphysical for the massive one \cite{Borsato:2014hja}. In this section, we will confine ourselves to the unphysical representations for the purpose of comparison with the massless case. In Appendix \ref{App:B}, we study the physical massive representations and as an example compute the $K$-matrices for the singlet supersymmetric boundary. For massive particles we do not need to distinguish between right and left movers, so we can simply provide four $R$-matrices as the complete bulk information. 

We define 
\begin{eqnarray}
\eta_p = \sqrt{i\left(x^-_p - x^+_p\right)} \, \, ,
\end{eqnarray}
where the Zhukovsky variables $x^\pm_p$ are constrained to satisfy
\begin{eqnarray}
    x^+_p +\frac{1}{x^+_p} - x^-_p - \frac{1}{x^-_p} = \frac{2i}{h} \, .
\end{eqnarray}
This constraint can be solved by setting
\begin{eqnarray}
    x^\pm_p = e^{\pm \frac{i p }{2}} \frac{1+\sqrt{1+4 h^2 \sin^2 \frac{p}{2}}}{2 h \sin \frac{p}{2}}\, .
\end{eqnarray}

The $L$ representation reads
\begin{eqnarray}\label{L_rep}
&&\pi^L_p(\mathfrak{G}_L) = \sqrt{\frac{h}{2}} \, \eta_p \begin{pmatrix}0&1\\0&0\end{pmatrix}, \qquad \qquad \qquad \quad\,\,  \pi^L_p(\mathfrak{Q}_L) = \sqrt{\frac{h}{2}} \, \eta_p \begin{pmatrix}0&0\\1&0\end{pmatrix},\nonumber\\
&&\pi^L_p(\mathfrak{G}_R) = -\sqrt{\frac{h}{2}} \, \frac{\eta_p \, e^{-i\frac{p}{2}}}{x_p^-} \begin{pmatrix}0&0\\1&0\end{pmatrix}, \qquad\qquad \pi^L_p(\mathfrak{Q}_R) = -\sqrt{\frac{h}{2}} \, \frac{\eta_p \, e^{i\frac{p}{2}}}{x_p^+} \begin{pmatrix}0&1\\0&0\end{pmatrix}.
\end{eqnarray}
The $\tilde{L}$ representation reads
\begin{eqnarray}\label{Ltil_rep}
&&\pi^{\Tilde{L}}_p(\mathfrak{G}_L) = \sqrt{\frac{h}{2}} \, \eta_p \begin{pmatrix}0&0\\1&0\end{pmatrix}, \qquad \qquad \qquad \quad \,\, \pi^{\Tilde{L}}_p(\mathfrak{Q}_L) = \sqrt{\frac{h}{2}} \, \eta_p \begin{pmatrix}0&1\\0&0\end{pmatrix},\nonumber\\
&&\pi^{\Tilde{L}}_p(\mathfrak{G}_R) = -\sqrt{\frac{h}{2}} \, \frac{\eta_p \, e^{-i\frac{p}{2}}}{x_p^-} \begin{pmatrix}0&1\\0&0\end{pmatrix}, \qquad \qquad \pi^{\Tilde{L}}_p(\mathfrak{Q}_R) = -\sqrt{\frac{h}{2}} \, \frac{\eta_p \, e^{i\frac{p}{2}}}{x_p^+} \begin{pmatrix}0&0\\1&0\end{pmatrix}.
\end{eqnarray}

The coproduct is always the same as (\ref{simo}) but now with the different possible choices of (massive) representations. We have four possible combinations (ignoring the dressing factors which are denoted here by $\Phi_m$ for ``massive"):
\begin{itemize}

\item $L-L$

\begin{eqnarray}
R^{LL} = \Phi_m^{LL} \, \begin{pmatrix}
\frac{e^{\frac{i}{4}(p-q)}(x^+_q-x^-_p)}{(x^-_q-x^+_p)}&0&0&0\\
0&\frac{e^{-\frac{i}{4}(p+q)}(x^+_q-x^+_p)}{(x^-_q-x^+_p)}&\frac{i\eta_p \eta_q}{(x^-_q-x^+_p)}&0\\
0&\frac{i\eta_p \eta_q}{(x^-_q-x^+_p)}&\frac{e^{\frac{i}{4}(p+q)}(x^-_q-x^-_p)}{(x^-_q-x^+_p)}&0\\
0&0&0&e^{\frac{i}{4}(q-p)}\end{pmatrix}.
\end{eqnarray}    

\item $\tilde{L}-\tilde{L}$

\begin{eqnarray}
R^{\tilde{L}\tilde{L}} = \Phi_m^{\tilde{L}\tilde{L}} \,\begin{pmatrix}
e^{\frac{i}{4}(q-p)}&0&0&0\\
0&\frac{e^{\frac{i}{4}(p+q)}(x^-_q-x^-_p)}{(x^-_q-x^+_p)}&-\frac{i\eta_p \eta_q}{(x^-_q-x^+_p)}&0\\
0&-\frac{i\eta_p \eta_q}{(x^-_q-x^+_p)}&\frac{e^{-\frac{i}{4}(p+q)}(x^+_q-x^+_p)}{(x^-_q-x^+_p)}&0\\
0&0&0&\frac{e^{\frac{i}{4}(p-q)}(x^+_q-x^-_p)}{(x^-_q-x^+_p)}\end{pmatrix}.
\end{eqnarray}

\item $L-\tilde{L}$

\begin{eqnarray}
R^{L\tilde{L}} = \Phi^{L\tilde{L}}_m \, \begin{pmatrix}
\frac{e^{-\frac{i}{4}(p+q)}(x^+_q-x^+_p)}{(x^-_q-x^+_p)}&0&0&\frac{i\eta_p \eta_q}{(x^-_q-x^+_p)}\\
0&\frac{e^{\frac{i}{4}(p-q)}(x^+_q-x^-_p)}{(x^-_q-x^+_p)}&0&0\\
0&0&e^{\frac{i}{4}(q-p)}&0\\
\frac{i\eta_p \eta_q}{(x^-_q-x^+_p)}&0&0&\frac{e^{\frac{i}{4}(p+q)}(x^-_q-x^-_p)}{(x^-_q-x^+_p)}\end{pmatrix}.   
\end{eqnarray}

\item $\tilde{L}-L$

\begin{eqnarray}
R^{\tilde{L}L} = \Phi^{\tilde{L}L}_m \, \begin{pmatrix}
\frac{e^{\frac{i}{4}(p+q)}(x^-_q-x^-_p)}{(x^-_q-x^+_p)}&0&0&-\frac{i\eta_p \eta_q}{(x^-_q-x^+_p)}\\
0&e^{\frac{i}{4}(q-p)}&0&0\\
0&0&\frac{e^{\frac{i}{4}(p-q)}(x^+_q-x^-_p)}{(x^-_q-x^+_p)}&0\\
-\frac{i\eta_p \eta_q}{(x^-_q-x^+_p)}&0&0&\frac{e^{-\frac{i}{4}(p+q)}(x^+_q-x^+_p)}{(x^-_q-x^+_p)}
\end{pmatrix}.   
\end{eqnarray}

\end{itemize}

We have checked that in the massless right-right moving limit, namely when $x^\pm_p \to e^{\pm i \frac{p}{2}}$ and $x^\pm_q \to e^{\pm i \frac{q}{2}}$, and when adequately normalised, the four $R$ matrices above reduce exactly to the four $R^{ab}_{++}$ ($a,b \in \{L,\Tilde{L}\}$) $R$-matrices which we have displayed in the previous sections, upon using the variables $\gamma_p = \log \tan \frac{p}{4}$ and $\gamma_q = \log \tan \frac{q}{4}$, respectively, with $p,q \in (0,\pi)$.

\subsection{Singlet boundary}
In this section we calculate the boundary reflection matrices for the massive particles using the symmetric co-product (\ref{simo}). We use the representation $L$ (\ref{L_rep}) and the fictitious representation $\Tilde{L}$ (\ref{Ltil_rep}). In the massive case, we do not have the added complication of decoupled left and right moving modes. However, we do not have the luxury of a difference form either. The boundary Yang-Baxter equation is given as 
\begin{eqnarray}
K_2(q) R^{op}(q,-p)K_1(p)R(p,q) =  R^{op}(-q,-p)K_1(p)R(p,-q)K_2(q).
\label{BYBE_massive}
\end{eqnarray}
where $K_1(p) = K(p) \otimes \mathds{1} $ and $K_2(p) = \mathds{1} \otimes K(p)$ with $K(p) = K^L(p) \oplus K^{\Tilde{L}}(p)$ being the complete singlet reflection matrix.

As in the case of the massless particles, we have the bulk algebra $\mathcal{A}^{\mathfrak{U}}$ given by (\ref{new-algebra-basis-right-wall}) and two boundary subalgebras $\mathcal{B}^\mathfrak{U}_L$ and $\mathcal{B}^\mathfrak{U}_R$ given by (\ref{rescaled-susy-boundary-algebras}) satisfying the coideal condition (\ref{right-coideal-subalgebra}) and the symmetric pair condition (\ref{symmetric-space}). The boundary intertwining equation for the massive case is
\begin{equation}\label{right-wall-massive-BIE}
\pi^L_{-p}(\mathfrak{b})\, K^L(p)=K^L(p)\, \pi^{L}_p(\mathfrak{b}) \qquad \text{and} \qquad   \pi^{\tilde{L}}_{-p}(\mathfrak{b})\, K^{\tilde{L}}(p)=K^{\tilde{L}}(p)\, \pi^{\tilde{L}}_{p}(\mathfrak{b}) \qquad \forall \, \mathfrak{b} \in \mathcal{B}
\end{equation}
with the representations $\pi^L$ and $\pi^{\Tilde{L}}$ in (\ref{L_rep}) and (\ref{Ltil_rep}) respectively. We consider the boundary subalgebra $\mathcal{B}$ either $\mathcal{B}^\mathfrak{U}_L$ or $\mathcal{B}^\mathfrak{U}_R$. Solving the boundary intertwining equations for the rescaled generators, we find the following consistent solutions,
\begin{eqnarray}\label{KL-matrix-massive}
K^L_{\mathcal{B}^\mathfrak{U}_L}(p) = \begin{pmatrix}1&0\\0&e^{-\frac{i p}{2}}\end{pmatrix} \qquad \text{and} \qquad K^L_{\mathcal{B}^\mathfrak{U}_R}(p) = \begin{pmatrix}1&0\\0&-e^{\frac{i p}{2}}\end{pmatrix}
\end{eqnarray}
for the particle in the $L$ representation, and
\begin{eqnarray}\label{KLtil-matrix-massive}
K^{\Tilde{L}}_{\mathcal{B}^\mathfrak{U}_L}(p) = \begin{pmatrix}1&0\\0&e^{\frac{i p}{2}}\end{pmatrix} \qquad \text{and} \qquad K^{\Tilde{L}}_{\mathcal{B}^\mathfrak{U}_R}(p) = \begin{pmatrix}1&0\\0&-e^{-\frac{i p}{2}}\end{pmatrix}
\end{eqnarray}
for the particle in the $\Tilde{L}$ representation. We have verified that these satisfy the BYBE (\ref{BYBE_massive}). Note that these are the same solutions that we obtained for the supersymmetric singlet boundaries in the massless case. Both of these solutions satisfy the braiding and physical unitarity described in \eqref{Unitarity-relations-K-matrices} without the subtleties that plague the massless braiding unitarity case.

\subsection{Vector boundary}
As was the case with massless vector boundaries, the boundary subalgebra for the right wall is given by $\mathcal{B}^{\mathfrak{U}}_{V} = \langle \mathfrak{H}_{L}, \, \mathfrak{H}_{R}, \, \mathfrak{U}^{\tfrac{1}{2}}\mathfrak{Q}_{L}, \, \mathfrak{U}^{\tfrac{1}{2}}\mathfrak{Q}_{R}, \, \mathfrak{U}^{-\tfrac{1}{2}}\mathfrak{G}_{L}, \, \mathfrak{U}^{-\tfrac{1}{2}}\mathfrak{G}_{R}, \, \mathfrak{U}\mathfrak{P}, \, \mathfrak{U}^{-1}\mathfrak{K}\rangle$. Once again, there is enough supersymmetry in the boundary subalgebra that the boundary intertwining equations constrain all the coefficients of the reflection matrices (apart from the dressing factor). The intertwining equations for the massive case are given as 
\begin{eqnarray}\label{BIE_vec_mass}
(\pi^{a}_{-p}\otimes \pi^b_B)\Delta(\mathfrak{b}) \, K^{ab}(p) = K^{ab}(p)  \, (\pi^{a}_{p}\otimes \pi^b_B)\Delta(\mathfrak{b})
\quad \quad \forall \, \mathfrak{b} \in \mathcal{B}^{\mathfrak{U}}_{V} \quad \text{and} \quad a,b \in \{L,\Tilde{L} \}\, .
\end{eqnarray}
$K^{ab}$ again denote the partial reflection matrices with $a$ being the representation of the particle and $b$ being the representation of the boundary. As mentioned earlier, we are using the fictitious representation $\Tilde{L}$ instead of the physical massive representations given in Appendix \ref{App:B}, so as to be able to compare with the massless case.

In analogy with the massless case, we solve the above intertwining equation for a specific choice of the boundary representation: $\pi_B^a(\mathfrak{b}) = \pi^{a}_{p}(\mathfrak{b})|_{p=\pi,h\to h/2}$ for all $\mathfrak{b} \in \mathcal{B}_{V}^{\mathfrak{U}}$ and $a \in \{ L,\Tilde{L} \}$ as used in \cite{Prinsloo:2015apa}.

The partial reflection matrices obtained are given as follows,
\begin{eqnarray}
    K^{LL}(p) &=& k^{LL}(p)\left[ E_{11} \otimes E_{11} + \frac{x_p^+ - e^{-ip}x_B}{x_p^- - x_B}  E_{11} \otimes E_{22} + \frac{e^{-ip/2}x_p^- + e^{ip/2}x_B}{x_p^- - x_B}  E_{22} \otimes E_{11} \right. \nonumber \\ && \left. +  e^{\frac{-ip}{2}}\frac{x_p^+ + x_B}{x_p^- - x_B} E_{22} \otimes E_{22} 
     + \sqrt{2}e^{-i\frac{p-\pi}{4}}\frac{\eta_p \eta_B \cos{\frac{p}{2}}}{x_p^- - x_B} \left( i \,E_{12} \otimes E_{21} - E_{21} \otimes E_{12} \right) \right]  \label{Kll-solved_mass}
\end{eqnarray}
\begin{eqnarray}
    K^{\Tilde{L}\Tilde{L}}(p) &=& k^{\Tilde{L}\Tilde{L}}(p) \left[ E_{11} \otimes E_{11} + \left( \frac{x^-_p + e^{ip} x_B }{x^+_p + x_B} \right) E_{11} \otimes E_{22} + \left( \frac{e^{i\frac{p}{2}} x^+_p - e^{-i\frac{p}{2}} x_B }{x^+_p + x_B} \right) E_{22} \otimes E_{11} \right. \nonumber \\ 
    && + \left. \left( e^{i\frac{p}{2}}\frac{x^-_p -  x_B }{x^+_p + x_B} \right) E_{22} \otimes E_{22}   
    - \left( \frac{\sqrt{2} \cos{\frac{p}{2}} e^{i\frac{p-\pi}{4}} \eta_p \eta_B }{x^+_p + x_B} \right) \left( i \, E_{12} \otimes E_{21} +  E_{21} \otimes E_{12} \right) \right] \nonumber \\ \label{Kltlt-solved_mass}
\end{eqnarray}
\begin{eqnarray}
    K^{L\Tilde{L}}(p) &=& k^{L\Tilde{L}}(p) \left[E_{11} \otimes E_{11} +  \frac{e^{-i\frac{p}{2}}x^+_p - e^{i\frac{p}{2}}x_B}{e^{i\frac{p}{2}}x^+_p - e^{-i\frac{p}{2}}x_B}  E_{11} \otimes E_{22} + \frac{x^+_p + x_B}{ e^{i\frac{p}{2}}x^+_p - e^{-i\frac{p}{2}}x_B} E_{22} \otimes E_{11} \right. \nonumber \\ 
    && \left. + \frac{x^-_p + e^{ip}x_B}{e^{i\frac{p}{2}}x^+_p - e^{-i\frac{p}{2}}x_B} E_{22} \otimes E_{22}  + \left( \frac{\sqrt{2} e^{i\frac{p+\pi}{4}} \eta_p \eta_B \cos{\frac{p}{2}}}{e^{i\frac{p}{2}}x^+_p - e^{-i\frac{p}{2}}x_B} \right) \left( i\, E_{12} \otimes E_{12} -  E_{21} \otimes E_{21} \right) \right] \nonumber \\ \label{Kllt-solved_mass}
\end{eqnarray}
\begin{eqnarray}
    K^{\Tilde{L}L}(p) &=& k^{\Tilde{L}L}(p) \left[ E_{11} \otimes E_{11} +  \left( \frac{e^{i\frac{p}{2}} x^-_p + e^{-i\frac{p}{2}} x_B}{e^{-i\frac{p}{2}} x^-_p + e^{i\frac{p}{2}} x_B} \right)  E_{11} \otimes E_{22} + \left( \frac{x^-_p - x_B}{e^{-i\frac{p}{2}} x^-_p + e^{i\frac{p}{2}} x_B} \right) E_{22} \otimes E_{11} \right. \nonumber \\
    && \left. + \left( \frac{ x^+_p - e^{-ip} x_B}{e^{-i\frac{p}{2}} x^-_p + e^{i\frac{p}{2}} x_B} \right) E_{22} \otimes E_{22} - \left( \frac{\sqrt{2} e^{-i\frac{p+\pi}{4}} \eta_p \eta_B \cos{\frac{p}{2}}}{e^{-i\frac{p}{2}} x^-_p + e^{i\frac{p}{2}} x_B} \right) \left( i\, E_{12} \otimes E_{12} +  E_{21} \otimes E_{21} \right) \right] \nonumber \\  \label{Kltl-solved}
\end{eqnarray}
Here $x_B=x^+_p|_{p=\pi,h\to h/2}=\frac{i}{2}\eta^2_B$. The reflection matrices found above satisfy the braiding unitarity condition $K(p)K(-p)=\mathbbmss{1}$ subjected to the following constraints on the dressing factors,
\begin{equation}
\begin{aligned}
k^{ab}(p) k^{ab}(-p)  & = 1 \qquad &&\text{for} \qquad ab=\{ LL,\tilde{L}\tilde{L} \}
\\
k^{ab}(p) k^{ab}(-p)  & = 1 + \frac{4ix_B \eta^2_p\cos^2{\frac{p}{2}}}{x^+_p x^-_p - x_B^2 - ix_B \eta^2_p} 
\qquad &&\text{for} \qquad ab=\{ L\tilde{L},\tilde{L}L \}
\end{aligned}
\end{equation}
They also satisfy the physical unitarity condition $K(p)K^\dagger(p)=\mathbbmss{1}$ subjected to the following constraints on the dressing factors,
\begin{equation}
\begin{aligned}
k^{ab}(p) [k^{ab}(p)]^*  & = 1 \qquad &&\text{for} \qquad ab=\{ LL,\tilde{L}\tilde{L} \}
\\
k^{ab}(p) [k^{ab}(p)]^*  & = 1 + \frac{4ix_B \eta^2_p\cos^2{\frac{p}{2}}}{x^+_p x^-_p - x_B^2 - ix_B \eta^2_p} 
\qquad &&\text{for} \qquad ab=\{ L\tilde{L},\tilde{L}L \}
\end{aligned}
\end{equation}


\section{Generalised vector boundary representations} \label{Gen Vec Bound}
In previous sections, specifically \ref{subsection-right-wall-vector-boundary} and \ref{subsec:Left-Wall-Vector-Boundary}, we studied the case of right and left boundaries carrying vector representations of the coideal subalgebras \eqref{vec_bound_alg} and \eqref{left_wall_vec_bound_alg}. In analogy with \cite{Prinsloo:2015apa} we have exploited the physical representations of the bulk also for the boundaries, fixing the parameters to some constant values and solving the boundary intertwining equations. This allowed us to fully determine the $K$-matrices, checking a posteriori the satisfaction of the BYBE. As anticipated at the beginning of section \ref{subsection-right-wall-vector-boundary}, the choice of starting from the boundary intertwining equations rather than the BYBE - as done for the singlet case - is due to the complexity of the latter. However, while being much simpler than the BYBE, the intertwining equations have the feature of being strongly dependent on the choice of representation. Hence, in spite of their convenience, solving them naturally raises an important question: does the vector boundary need to be in the same representation as the bulk particles? 

In principle, following the logic of singlet boundaries, given a $K$-matrix which solves the BYBE and a choice of bulk representations, one may use any boundary representation - including ones differing from those of the bulk - to study the intertwining equations \eqref{BIE_vec} and \eqref{left-wall-vector-intertwining-equation}. Asking for a solution of the latter equations - after knowing, from previous sections, that certainly there exists one - would however constrain the boundary representations to be somewhat compatible with both the $K$-matrix and the bulk representations. Upon choosing a bulk representation and a $K$-matrix which solves the BYBE, in the next two subsections we shall try to use the above logic to answer the question \textit{what is an appropriate boundary representation for the coideal subalgebras \eqref{vec_bound_alg} and \eqref{left_wall_vec_bound_alg}?} To this aim, we shall take inspiration from the results of previous sections, carefully analysing the BYBE and finding generalised $K$-matrices which solve them.

\subsection{Right wall}\label{sec:generslised-rep-right-wall}
Having determined the specific solution \eqref{special-sol-right-wall-vector-K-matrix} for the $K$-matrix, which given a choice of bulk and boundary representations satisfies both BYBE and the intertwining equations for the subalgebra \eqref{vec_bound_alg}, we are in a somewhat good position to perform a step backward and try analysing the problem in a more systematic way, following the logic highlighted above, in analogy with the singlet case.

Ideally, after specifying the bulk $R$-matrices and the representations of the bulk excitations, consistent boundary representations could be determined by first finding the most general $K$-matrix satisfying the BYBE and successively determining what constraints this imposes, via the intertwining equations, on some arbitrary unconstrained boundary representation. Due to the complexity of the BYBE, it is extremely hard to find the most general $K$-matrix solving it. However the above specific $K$-matrix can be taken as a source of inspiration and used to construct a subset of solutions to the BYBE which might then be exploited to constrain an arbitrarily chosen boundary representation by means of the intertwining condition. Carefully inspecting the structure of the specific $K$-matrix \eqref{special-sol-right-wall-vector-K-matrix}, leads us to consider the following generalised ansatz:
\begin{align}\label{right-wall-generalised-K-matrix-ansatz}
K^{LL}(\gamma)=k^{LL}(\gamma &)  \biggl[ E_{11} \otimes E_{11} + a_{1}E_{22}\otimes E_{22}+
\\
& + \kappa(\gamma)(E_{11}\otimes E_{22} + a_{2}E_{22}\otimes E_{11}) + \eta(\gamma)(E_{12}\otimes E_{21} + a_{3} E_{21}\otimes E_{12}) \biggr]
\notag
\end{align}
\begin{align}
K^{\tilde{L}\tilde{L}}(\gamma)=k^{\tilde{L}\tilde{L}}(\gamma &)  \biggl[ E_{11} \otimes E_{11} + b_{1}E_{22}\otimes E_{22}+
\\
& + b_4\kappa(\gamma)(E_{11}\otimes E_{22} + b_{2}E_{22}\otimes E_{11}) + b_5\eta(\gamma)(E_{12}\otimes E_{21} + b_{3} E_{21}\otimes E_{12}) \biggr]
\notag
\end{align}
\begin{align}
K^{L\tilde{L}}(\gamma)=k^{L\tilde{L}}(\gamma &)  \biggl[ E_{11} \otimes E_{11} + c_{1}E_{22}\otimes E_{22}+
\\
& + \frac{c_4}{\kappa(\gamma)}(E_{11}\otimes E_{22} + c_{2}E_{22}\otimes E_{11}) + \frac{c_5\eta(\gamma)}{\kappa(\gamma)}(E_{12}\otimes E_{12} + c_{3} E_{21}\otimes E_{21}) \biggr]
\notag
\end{align}
\begin{align}
K^{\tilde{L}L}(\gamma)=k^{\tilde{L}L}(\gamma &)  \biggl[ E_{11} \otimes E_{11} + d_{1}E_{22}\otimes E_{22}+
\\
& + \frac{d_4}{\kappa(\gamma)}(E_{11}\otimes E_{22} + d_{2}E_{22}\otimes E_{11}) + \frac{d_5\eta(\gamma)}{\kappa(\gamma)}(E_{12}\otimes E_{12} + d_{3} E_{21}\otimes E_{21}) \biggr]
\notag
\end{align}
which depends on 2 functions, $\kappa(\gamma)$ and $\eta(\gamma)$, and 18 constant coefficients $a_{i}$ with $i=1,...,3$ and $b_{i},c_{i},d_{i}$ with $i=1,...,5$. Already at this level, with limited freedom in terms of functions of the rapidity $\gamma$, the BYBE \eqref{right-wall-BYBE-rep-indices} turns out to exhibit quite an intricate structure, and even small generalisations of the above ansatz lead to significantly more complicated equations to solve. Starting from the above expressions it is however still possible to carefully inspect the explicit equations - in this case each BYB equation is an $8\times 8$ matrix equation - and find two families of parametric solutions, distinguished by a sign factor $s_{1}= \pm1$, and characterised by two unconstrained integration constants $k_{1},k_{2}$ and the parameter $b_5$ introduced above, which remains completely free. The solutions can be written as follows:
\begin{equation}\label{right_wall_BYBE_solutions}
\begin{aligned}
\kappa(\gamma)=-1 +\frac{2(s_{1}+\cosh{\gamma})}{s_{1}+2e^{2k_{1}}+\cosh{\gamma}} \qquad \qquad & \qquad \qquad \eta(\gamma)=\frac{k_2(e^{\frac{\gamma}{2}}+s_{1}e^{-\frac{\gamma}{2}})}{s_{1}+2e^{2k_{1}}+\cosh{\gamma}}
\\
a_{1}=a_{2}=- s_{1}i \qquad\,\,\,\,\, b_{1}=b_{2}= s_{1}i \qquad \qquad\,\,\, & \qquad\qquad\,\, c_{1}=c_{2}=- s_{1}i \qquad \,\,\,\,\, d_{1}=d_{2}= s_{1}i
\\
b_4=c_4=d_4=1 \qquad a_3=-is_1\frac{4e^{2k_1}}{k_2^2} \qquad\quad\,\,\,  & \qquad\qquad\,\, b_3=-\frac{a_3}{b_5^2} \qquad d_3=\frac{1}{a_3} \qquad c_3=-\frac{b_5^2}{a_3}
\\
c_5=i s_1 \frac{a_3}{b_5} \qquad\quad \qquad\,\,\, & \qquad \qquad \, d_5=i s_1 a_3
\end{aligned}
\end{equation}
Notice that the specific $K$-matrices \eqref{special-sol-right-wall-vector-K-matrix}, found above by solving the intertwining equations, can be recovered by setting
\begin{equation}
s_{1}=-1 \qquad k_{1}=-\tfrac{1}{2}\log{2} \qquad k_{2}=\sqrt{2} \qquad b_5=1 \,\, .
\end{equation}
Having found two families of solutions to the BYBE, we are in a position to try exploiting the available parameters to learn about boundary representations which are consistent with the physically relevant bulk representations. As highlighted above, the strategy is to consider an unconstrained boundary representation and to use the intertwining equations for the generators \eqref{vec_bound_alg} to determine which boundary structures they are compatible with. Keeping this objective in mind, we consider the following general structure for two boundary representations, labelled as $L$ and $\tilde{L}$ for consistency with the bulk, but otherwise based on arbitrary functions to be determined:
\begin{equation}\label{boundary-rep-ansatz}
\begin{aligned}
\pi_{B}^{L}(\mathfrak{U}^{\tfrac{1}{2}}Q_{L}) = f_{QLL}(h) \begin{pmatrix}
0 & q_{12}^{LL} \\
q_{21}^{LL} & 0
\end{pmatrix} 
\quad & \quad \,\,\,\,
\pi_{B}^{L}(\mathfrak{U}^{\tfrac{1}{2}}Q_{R}) = f_{QRL}(h) \begin{pmatrix}
0 & q_{12}^{RL} \\
q_{21}^{RL} & 0
\end{pmatrix}
\\
\pi_{B}^{L}(\mathfrak{U}^{-\tfrac{1}{2}}G_{L}) = f_{GLL}(h) \begin{pmatrix}
0 & g_{12}^{LL} \\
g_{21}^{LL} & 0
\end{pmatrix} 
\quad & \quad
\pi_{B}^{L}(\mathfrak{U}^{-\tfrac{1}{2}}G_{R}) = f_{GRL}(h) \begin{pmatrix}
0 & g_{12}^{RL} \\
g_{21}^{RL} & 0
\end{pmatrix}
\end{aligned}
\end{equation}
\begin{equation}
\begin{aligned}
\pi_{B}^{\tilde{L}}(\mathfrak{U}^{\tfrac{1}{2}}Q_{L}) = f_{QL\tilde{L}}(h) \begin{pmatrix}
0 & q_{12}^{L\tilde{L}} \\
q_{21}^{L\tilde{L}} & 0
\end{pmatrix} 
\quad & \quad \,\,\,\,
\pi_{B}^{\tilde{L}}(\mathfrak{U}^{\tfrac{1}{2}}Q_{R}) = f_{QR\tilde{L}}(h) \begin{pmatrix}
0 & q_{12}^{R\tilde{L}} \\
q_{21}^{R\tilde{L}} & 0
\end{pmatrix}
\\
\pi_{B}^{\tilde{L}}(\mathfrak{U}^{-\tfrac{1}{2}}G_{L}) = f_{GL\tilde{L}}(h) \begin{pmatrix}
0 & g_{12}^{L\tilde{L}} \\
g_{21}^{L\tilde{L}} & 0
\end{pmatrix} 
\quad & \quad
\pi_{B}^{\tilde{L}}(\mathfrak{U}^{-\tfrac{1}{2}}G_{R}) = f_{GR\tilde{L}}(h) \begin{pmatrix}
0 & g_{12}^{R\tilde{L}} \\
g_{21}^{R\tilde{L}} & 0
\end{pmatrix}
\end{aligned}
\end{equation}
The representations of the central elements are then a result of the commutation relations. It is here important to remark that intending to constrain the above functions by solving the intertwining equations \eqref{BIE_vec} for the generators in the boundary algebra \eqref{vec_bound_alg}, one would naturally expect, at the end of the procedure, a dependence of the boundary representation on both the parameters of the $K$-matrices and the parameters which characterise the bulk representations - namely the momentum $p$ (or rapidity $\gamma$) and the coupling $h$. However, while it seems reasonable, if not natural, to characterise the boundary representations in terms of the $K$-matrix parameters and the $h$-coupling, finding a non-trivial dependence on the bulk momentum certainly looks more unphysical - since the boundaries should carry no information about motion. For this reason, a trivial bulk momentum dependence will be imposed on the boundary representations, and this will turn out to be crucial for our purposes.

One can now start studying the intertwining equations \eqref{BIE_vec} for the generators \eqref{vec_bound_alg}. Even without specifying the functions and parameters \eqref{right_wall_BYBE_solutions}, which characterise the ansatz for the $K$-matrix, it is possible to partially constrain the boundary representations by simply exploiting the matrix structure of the bulk representations. Solving (part of) the intertwining equations indeed requires that
\begin{equation}\label{intertwining-constraints-1}
\begin{aligned}
q^{LL}_{12}=q^{RL}_{21}=q^{L\tilde{L}}_{21}=q^{R\tilde{L}}_{12}&=0
\\
g^{LL}_{21}=g^{RL}_{12}=g^{L\tilde{L}}_{12}=g^{R\tilde{L}}_{21}&=0
\end{aligned}
\end{equation}
which in turn imply
\begin{equation}\label{intertwining-constraints-2}
q^{RL}_{12} = -\frac{i}{4n_1 q^{LL}_{21}} \quad 
q^{L\tilde{L}}_{12} = -\frac{i}{4 n_3 q^{R\tilde{L}}_{21}}
\quad 
g^{RL}_{21}=\frac{i}{4 n_2 g^{LL}_{12}}
\quad 
g^{L\tilde{L}}_{21}=\frac{i}{4 n_4 g^{L\tilde{L}}_{12}}
\end{equation}
and
\begin{equation}\label{intertwining-constraints-3}
f_{QRL}(h)=\frac{n_1 h}{f_{QLL}(h)}
\quad
f_{GRL}(h)=\frac{n_2 h}{f_{GLL}(h)} 
\quad 
f_{QL\tilde{L}}(h)=\frac{n_3 h}{f_{QR\tilde{L}}(h)}
\quad
f_{GL\tilde{L}}(h)=\frac{n_4 h}{f_{GR\tilde{L}}(h)}
\end{equation}
with $n_1,...,n_4$ undetermined proportionality constants. Given the above constraints, the remaining equations only have a chance of being solved provided that 
\begin{equation}\label{intertwining-constraints-4}
f_{QLL}(h)=m_{1} \sqrt{h} 
\qquad
f_{GLL}(h)=m_{2} \sqrt{h}
\qquad 
f_{QR\tilde{L}}(h)=m_{3} \sqrt{h}
\qquad
f_{GR\tilde{L}}(h)=m_{4}\sqrt{h}
\end{equation}
with $m_{1},...,m_{4}$ other constants to be determined. The resulting boundary representations then read
\begin{equation}\label{boundary-rep-intermediate-stage}
\begin{aligned}
\pi_{B}^{L}(\mathfrak{U}^{\tfrac{1}{2}}Q_{L}) = m_{1}\sqrt{h} q_{21}^{LL} \begin{pmatrix}
0 & 0 \\
1 & 0
\end{pmatrix} 
\quad & \quad \,\,\,\,\,\,\,
\pi_{B}^{L}(\mathfrak{U}^{\tfrac{1}{2}}Q_{R}) = -\frac{\sqrt{h}}{4m_{1}} \frac{i}{q^{LL}_{21}}\begin{pmatrix}
0 & 1 \\
0 & 0
\end{pmatrix}
\\
\pi_{B}^{L}(\mathfrak{U}^{-\tfrac{1}{2}}G_{L}) = m_{2}\sqrt{h} g_{12}^{LL} \begin{pmatrix}
0 & 1 \\
0 & 0
\end{pmatrix} 
\quad & \quad\,\,\,
\pi_{B}^{L}(\mathfrak{U}^{-\tfrac{1}{2}}G_{R}) = \frac{\sqrt{h}}{4m_{2}} 
\frac{i}{g^{LL}_{12}} 
\begin{pmatrix}
0 & 0 \\
1 & 0
\end{pmatrix}
\end{aligned}
\end{equation}
\begin{equation}
\begin{aligned}
\pi_{B}^{\tilde{L}}(\mathfrak{U}^{\tfrac{1}{2}}Q_{L}) = -\frac{\sqrt{h}}{4m_{3}}\frac{i}{q^{R\tilde{L}}_{21}}
\begin{pmatrix}
0 & 1 \\
0 & 0
\end{pmatrix} 
\quad  & \quad \,\,\,
\pi_{B}^{\tilde{L}}(\mathfrak{U}^{\tfrac{1}{2}}Q_{R}) = m_{3}\sqrt{h}q_{21}^{R\tilde{L}}\begin{pmatrix}
0 & 0 \\
1 & 0
\end{pmatrix}
\\
\pi_{B}^{\tilde{L}}(\mathfrak{U}^{-\tfrac{1}{2}}G_{L}) = \frac{\sqrt{h}}{4m_{4}}\frac{i}{g^{R\tilde{L}}_{12}}\begin{pmatrix}
0 & 0 \\
1 & 0
\end{pmatrix} 
\quad \,\,\,\,\, & \quad
\pi_{B}^{\tilde{L}}(\mathfrak{U}^{-\tfrac{1}{2}}G_{R}) = m_{4}\sqrt{h}g_{12}^{R\tilde{L}} \begin{pmatrix}
0 & 1 \\
0 & 0
\end{pmatrix}
\end{aligned}
\end{equation}
At this point one has found the boundary representations whose matrix structure is compatible with those of the bulk representations and of the ansatz for the $K$-matrix. However, this does not entirely solve the intertwining equations, which still exhibit non-vanishing entries depending on the unspecified functions and the parameters appearing in the $K$-matrix ansatz, as well as on those characterising the boundary. To proceed further it is necessary to make explicit use of the solutions \eqref{right_wall_BYBE_solutions}, so that the intertwining equations only depend on the truly free parameters $s_{1},k_{1},k_{2},b_{5}$, the bulk rapidity $\gamma$ and coupling $h$ and the above undetermined functions. The hope, at this stage, is that one might be able to fix the free functions and the free parameters of the $K$-matrix so as to solve all the intertwining equations. While the latter exhibit non-trivial dependence on the above quantities, which make it non straightforward to find a solution, the physical intuition - that the boundary should not carry information about motion in the bulk - will turn out to be extremely powerful in revealing a special choice of parameters. 

Having specified the solution \eqref{right_wall_BYBE_solutions} to the BYBE, we proceed with our program by sampling a random subset of non-vanishing entries in the intertwining equations, so as to study them in detail. After solving four of the sampled equations in terms of the remaining functions $q^{LL}_{21},q^{R\tilde{L}}_{21},g^{LL}_{12},g^{R\tilde{L}}_{12}$ one finds, as expected, complicated expressions depending on the constant parameters of the $K$-matrix and the bulk rapidity $\gamma$. We shall here display one such function, to better convey the idea
\begin{equation}
q_{21}^{R\tilde{L}}=
\frac{b_{5}k_{2}e^{-2k_{1}+\tfrac{\gamma}{2}-i\arctan{e^{\gamma}}}[(s_{1}+\cosh{\gamma})(1-e^{4i\arctan{e^{\gamma}}})-2e^{2k_{1}}(1+e^{4i\arctan{e^{\gamma}}})]}{8\sqrt{2}m_{3}s_{1}(s_{1}+e^{\gamma})\sqrt{\sin{(2\arctan{e^{\gamma}})}}} \,\, .
\end{equation}
In light of our physical intuition, if the solutions found for $q^{LL}_{21},q^{R\tilde{L}}_{21},g^{LL}_{12},g^{R\tilde{L}}_{12}$ make any sense, it should now be possible to find a convenient choice of constant parameters which removes their dependence on $\gamma$. To search for these, it is natural to look for values of the free parameters which solve $df/d\gamma = 0$ for $f$ any of the above four functions. For the example displayed above one should solve the condition
\begin{equation}
\frac{dq^{R\tilde{L}}_{21}}{d\gamma}\propto\bigl(-i+2[1+is_{1}]e^{\gamma}+i[1-is_{1}]e^{2\gamma}+s_{1}[1+2is_{1}]+4e^{2k_{1}+\gamma}[i(1-is_{1})+(1+is_{1})\cosh{\gamma}]\bigr) = 0 \, .
\end{equation}
By recalling that $s_{1}=\pm 1$ is just a sign distinguishing between the two families of $K$-matrices, it is then straightforward to find the following two cases
\begin{equation}\label{fixing-k1}
\begin{aligned}
s_{1}=+1:& \quad \frac{dq^{R\tilde{L}}_{21}}{d\gamma}\propto(1+2e^{2k_{1}})= 0  \qquad \Rightarrow \qquad k_{1}=i\pi z -\tfrac{1}{2}\log{2}+\tfrac{i}{2}\pi \qquad z\in \mathbb{Z}
\\
s_{1}=-1:& \quad \frac{dq^{R\tilde{L}}_{21}}{d\gamma}\propto (1-2e^{2k_{1}})= 0 \qquad \Rightarrow \qquad k_{1}=i\pi z -\tfrac{1}{2}\log{2} \qquad \qquad\,\,\,   z\in \mathbb{Z}
\end{aligned}
\end{equation}
Remarkably, similar factorisations with the exact same solutions for $s_{1}$ and $k_{1}$ take place in all the four functions above: these not only allow us to get rid of the $\gamma$-dependence, as desired, but also to completely solve all the remaining intertwining equations with no need to tune further parameters. At the end of the story, one finds two families of parametric representations, distinguished by the sign factor $s_{1}=\pm 1$ and characterised by the free constants $k_{2}$ and $b_{5}$. The boundary representation for the supercharges take the explicit form
\begin{align}\label{right-vector-boundary-final-reps}
\pi_{B}^{L}(\mathfrak{U}^{\tfrac{1}{2}}Q_{L}) = (1-is_{1})\frac{k_{2}}{4}\sqrt{h}  \begin{pmatrix}
0 & 0 \\
1 & 0
\end{pmatrix} 
\qquad\quad\,\,\,\,\, & \quad \,\,\,\,
\pi_{B}^{L}(\mathfrak{U}^{\tfrac{1}{2}}Q_{R}) = \frac{s_{1}}{2k_{2}}(1-is_{1})\sqrt{h}
\begin{pmatrix}
0 & 1 \\
0 & 0
\end{pmatrix}
\notag \\
\pi_{B}^{L}(\mathfrak{U}^{-\tfrac{1}{2}}G_{L}) = -\frac{s_{1}}{2k_{2}}(1+is_{1})\sqrt{h}
\begin{pmatrix}
0 & 1 \\
0 & 0
\end{pmatrix} 
\qquad\,\,\, & \quad
\pi_{B}^{L}(\mathfrak{U}^{-\tfrac{1}{2}}G_{R}) = -(1+is_{1})\frac{k_{2}}{4}\sqrt{h} 
\begin{pmatrix}
0 & 0 \\
1 & 0
\end{pmatrix}
\notag \\
\\
\pi_{B}^{\tilde{L}}(\mathfrak{U}^{\tfrac{1}{2}}Q_{L}) = -\frac{s_{1}}{2b_{5}k_{2}}(1-is_{1})\sqrt{h}
\begin{pmatrix}
0 & 1 \\
0 & 0
\end{pmatrix} 
\quad \,\,\,\, & \quad \,\,\,\,
\pi_{B}^{\tilde{L}}(\mathfrak{U}^{\tfrac{1}{2}}Q_{R}) = -\frac{b_{5}k_{2}}{4}(1-is_{1})\sqrt{h}
\begin{pmatrix}
0 & 0 \\
1 & 0
\end{pmatrix}
\notag \\
\pi_{B}^{\tilde{L}}(\mathfrak{U}^{-\tfrac{1}{2}}G_{L}) = \frac{b_{5}k_{2}}{4}(1+is_{1})\sqrt{h} \begin{pmatrix}
0 & 0 \\
1 & 0
\end{pmatrix} 
\qquad\quad  & \quad
\pi_{B}^{\tilde{L}}(\mathfrak{U}^{-\tfrac{1}{2}}G_{R}) = \frac{s_{1}}{2b_{5}k_{2}}(1+is_{1})\sqrt{h} 
\begin{pmatrix}
0 & 1 \\
0 & 0
\end{pmatrix}
\notag
\end{align}
and in turn lead to the following central elements, for both $a=L$ and $a=\tilde{L}$
\begin{equation}\label{generalised-right-wall-vector-boundary-central-elements-rep}
\begin{aligned}
\pi_{B}^{a}(\mathfrak{H}_{L}) = 
-s_{1}\frac{h}{4}
\begin{pmatrix}
1 & 0 \\
0 & 1
\end{pmatrix} 
\qquad\,\,\,\, & \qquad \,\,\,\,
\pi_{B}^{a}(\mathfrak{H}_{R}) = -s_{1}\frac{h}{4} \begin{pmatrix}
1 & 0 \\
0 & 1
\end{pmatrix}
\\
\pi_{B}^{a}(\mathfrak{U}\mathfrak{P}) = -\frac{i}{4}h 
\begin{pmatrix}
1 & 0 \\
0 & 1
\end{pmatrix} 
\quad \quad \quad & \quad\,\,\,\,
\pi_{B}^{a}(\mathfrak{U}^{-1}\mathfrak{K}) = \frac{i}{4}h 
\begin{pmatrix}
1 & 0 \\
0 & 1
\end{pmatrix}
\end{aligned}
\end{equation}
We conclude the analysis by noting that
\begin{itemize}
\item Different families of $K$-matrices solving the BYBE determine different boundary representations in terms of the sign factor $s_{1}=\pm1$. Additionally, each $K$-matrix in the families is parametrised by two parameters, $k_2$ and $b_5$, the choice of which also affect the boundary representations.
\item Setting $s_{1}=-1,\, k_{2}=\sqrt{2}, \, b_{5}=1$ one correctly recovers the representations used to derive the $K$-matrices \eqref{special-sol-right-wall-vector-K-matrix}, namely $\pi_B^a(\mathfrak{b}) = \pi^{a+}_{p}(\mathfrak{b})|_{p=\pi,h\to h/2}$ for all generators $\mathfrak{b}$ in the boundary subalgebra \eqref{vec_bound_alg} and $a \in \{ L,\Tilde{L} \}$, using \eqref{Lrep+} for $L$ and \eqref{Ltilrep+} for $\Tilde{L}$.
\item Identifying, in analogy with equation \eqref{energy} for the bulk, the energy of the boundary with the sum of eigenvalues of the generators $\mathfrak{H}_L$ and $\mathfrak{H}_R$ one finds $E_B = -s_1 h$. This seems to suggest that the physically relevant family of $K$-matrices should be the one characterised by $s_1=-1$, and is indeed in agreement with the solution found using the same representations for bulk and boundary.
\item Demanding independence of the boundary representation on the bulk momentum was crucial in finding the values of $k_1$ that solve the intertwining equations. While in principle there might be other choices of free parameters solving such equations, none of these would lead to momentum-independent boundary representations and even assuming to know all the possible choices of parameters solving the intertwining equations, the momentum-independent one would single out as a very special case, signaling an intrinsic need for the boundary to enjoy such a property.
\\
\end{itemize}

\noindent
\textbf{Unitarity and other relations.} \quad
It is important to look again at the unitarity conditions \eqref{Unitarity-relations-K-matrices}, that physically relevant $K$-matrices should satisfy, using the rewriting  \eqref{unitarity-vector-boundary}, reported here for clarity
\begin{equation}\label{unitarity-vector-boundary-1}
K^{ab}(\gamma)\bigl[K^{ab}(\gamma)\bigr]^{\dagger}=\mathbbmss{1} \qquad \forall \,\, a,b \in \{ L,\tilde{L} \} \, \, .
\end{equation}
Exploiting the solution \eqref{right_wall_BYBE_solutions} to the BYBE and recalling that the intertwining relations are satisfied provided one chooses $s_{1}=+1$ and $k_{1}=i\pi z-\tfrac{1}{2}\log{2}+\tfrac{i}{2}\pi$, or alternatively $s_{1}=-1$ and $k_{1}=i\pi z-\tfrac{1}{2}\log{2}$, with the parameters $k_{2},b_{5}$ remaining free, it is easy to verify that:
\begin{itemize}
\item For $s_{1}=+1$ and $k_{1}=i\pi z-\tfrac{1}{2}\log{2}+\tfrac{i}{2}\pi$ there exists no choice of the free parameters $k_{2},b_{5}$ which makes the relation \eqref{unitarity-vector-boundary-1} satisfied for any of the $K^{ab}(\gamma)$ matrices.
\item For $s_{1}=-1$ and $k_{1}=i\pi z-\tfrac{1}{2}\log{2}$ the relation \eqref{unitarity-vector-boundary-1} requires the free parameters to satisfy
\begin{equation}\label{right-vector-boundary-physical-unitarity-constraints-1}
|k_{2}|^2 = 2 \quad \text{for} \quad ab=\{LL,\tilde{L}L \} 
\qquad \text{and} \qquad 
|k_{2}|^2|b_{5}|^2 = 2 \quad \text{for} \quad ab=\{\tilde{L}\tilde{L},L\tilde{L} \}
\end{equation}
and at the same time enforces the following conditions on the unconstrained dressing factors
\begin{equation}\label{right-wall-vector-boundary-dressing-factors-conditions-1}
\begin{aligned}
k^{ab}(\gamma)\bigl[ k^{ab}(\gamma) \bigr]^{*} & = 1 \qquad &&\text{for} \qquad ab=\{ LL,\tilde{L}\tilde{L} \}
\\
k^{ab}(\gamma)\bigl[ k^{ab}(\gamma) \bigr]^{*} & = \frac{(\cosh{\gamma}-2)^2}{(\cosh{\gamma})^2} 
\qquad &&\text{for} \qquad ab=\{ L\tilde{L},\tilde{L}L \}
\end{aligned}
\end{equation}
\end{itemize}
Hence, while the BYBE and intertwining equations are consistent with two families of $K$-matrices, each depending on two parameters, physical unitarity completely rules out one of the two families, and at the same time fixes the free parameters almost entirely. Once again, the $K$-matrices \eqref{special-sol-right-wall-vector-K-matrix} found at the beginning of the section respect all the above conditions and thus satisfy physical unitarity provided the dressing factors satisfy \eqref{right-wall-vector-boundary-dressing-factors-conditions}, which is the same as \eqref{right-wall-vector-boundary-dressing-factors-conditions-1}. This result is also in agreement with the above statement regarding the positivity of the boundary energy for the family with $s_1 = -1$.   \\
\indent

As we discussed in the unitarity paragraph of section \ref{subsection-right-wall-vector-boundary}, due to the presence of different representations for left and right moving particles the braiding unitarity relation (\ref{Unitarity-relations-K-matrices}) is not satisfied by the above $K$-matrices. To overcome this problem one needs to introduce a new matrix, denoted $K'$, solving  the right wall vector boundary intertwining equation (\ref{BIE_vec}) for opposite particle momentum, namely
\begin{eqnarray}\label{BIE_vec_unitarity_repeated}
(\pi^{a+}_{+p}\otimes \pi^b_B)\Delta(\mathfrak{b}) \, K'^{ab}(-\gamma(p)) = K'^{ab}(-\gamma(p))  \, (\pi^{a-}_{-p}\otimes \pi^b_B)\Delta(\mathfrak{b})\,\, .
\end{eqnarray}
In the spirit of this section, the latter equation should be solved for $K'$ using the generalised boundary representations \eqref{right-vector-boundary-final-reps}. The resulting $K'$ matrices consequently depend on the parameters \eqref{right_wall_BYBE_solutions}, characterising the families of $K$-matrices solving BYBE, and correctly satisfy the braiding unitarity relations
\begin{equation}
K^{ab}(\gamma)K'^{ab}(-\gamma)=\mathbbmss{1} \qquad \forall \,\, a,b \in \{ L,\tilde{L} \} \, \, .
\end{equation}

In addition to the above unitarity conditions, there are some other interesting relations that the generalised $K$-matrices seem to satisfy. To begin, one can notice that 
\begin{equation}
K^{LL}(\gamma)K^{\tilde{L}\tilde{L}}(-\gamma)^{T}=\mathbbmss{1} \qquad \text{and} \qquad K^{L\tilde{L}}(\gamma)K^{\tilde{L}L}(-\gamma)^{T}=\mathbbmss{1} 
\end{equation}
hold for both families of $K$-matrices, provided one fixes the parameter $b_{5}$ as
\begin{equation}
b_{5} = -\frac{2}{k_{2}^2}
\end{equation}
and imposes the following relations on the dressing factors
\begin{equation}\label{right-wall-vector-braiding-conditions-on-dressing-factors}
k^{LL}(\gamma)k^{\tilde{L}\tilde{L}}(-\gamma)=1 \qquad \text{and} \qquad k^{L\tilde{L}}(\gamma)k^{\tilde{L}L}(-\gamma)=\frac{(\cosh{\gamma}+2s_{1})^2}{(\cosh{\gamma})^2} \,\, .
\end{equation}
Notice how the above condition on $b_{5}$ is consistent with the requirements \eqref{right-vector-boundary-physical-unitarity-constraints-1} of physical unitarity. Finally, both families of $K$-matrices satisfy the following mixed braiding-like unitarity relations
\begin{equation}
K^{LL}(\gamma)K^{\tilde{L}\tilde{L}}(-\gamma)=\mathbbmss{1} \qquad \text{and} \qquad K^{L\tilde{L}}(\gamma)K^{\tilde{L}L}(-\gamma)=\mathbbmss{1} \,\, , 
\end{equation}
provided that
\begin{equation}
b_{5} = -i
\end{equation}
and \eqref{right-wall-vector-braiding-conditions-on-dressing-factors} hold true for the dressing factors. The latter condition on $b_{5}$ is again consistent with \eqref{right-vector-boundary-physical-unitarity-constraints-1}.
\\

\noindent
\textbf{Short summary of the generalised right wall vector boundary.} \quad Boundaries carrying non-trivial representations of the symmetry algebra exhibit a quite complicated structure of the BYBE and for this reason it is very hard to find the most general $K$-matrix solving it, as done for the singlet case. While in sections \ref{subsection-right-wall-vector-boundary} and \ref{subsec:Left-Wall-Vector-Boundary} - after deciding to use for the boundary the same representations as for the bulk - we determined the $K$-matrices starting from the simpler boundary intertwining equations, in \ref{sec:generslised-rep-right-wall} we changed our perspective. Upon constructing the $K$-matrix ansatz \eqref{right-wall-generalised-K-matrix-ansatz}, which generalises the result \eqref{special-sol-right-wall-vector-K-matrix}, we solved the BYBE determining two families of parametric solutions, distinguished by a sign factor $s_1 = \pm1$ and characterised by the three free parameters $k_1, k_2, b_5$, as given in \eqref{right_wall_BYBE_solutions}. We then chose not to use for the boundary the same representation as for the bulk: considering an arbitrary boundary representation of the form \eqref{boundary-rep-ansatz} for the boundary coideal subalgebra \eqref{vec_bound_alg}, we constrained its structure by solving the intertwining equations \eqref{BIE_vec} and demanding independence on the bulk momentum - which require fixing $k_1$ as in \eqref{fixing-k1} - thus obtaining the generalised vector boundary representations \eqref{right-vector-boundary-final-reps}. While the intertwining and boundary Yang-Baxter equations turn out to be compatible with any $K$-matrix and boundary representation depending on the sign $s_1$ and the free parameters $k_2,b_5$, physical unitarity \eqref{unitarity-vector-boundary-1} rules out the $s_1 = +1$ family, further constraining $|k_2|^2=2$ and $|b_5|^2=1$. This restriction also leads to a positive boundary energy $E_B = h$, identified with the bulk energy as in \eqref{energy}.
In conclusion, the $K$-matrices \eqref{right-wall-generalised-K-matrix-ansatz} respect intertwining and BYB equations, as well as physical unitarity, provided the free parameters $s_{1},k_{1},k_{2},b_{5}$ in \eqref{right_wall_BYBE_solutions} are chosen as
\begin{equation}
s_{1}= -1 \qquad k_{1}=i\pi z-\tfrac{1}{2}\log{2} \qquad k_{2}=\sqrt{2} e^{i\alpha} \qquad b_{5}=e^{i\beta} \qquad \text{with} \qquad z\in \mathbb{Z}, \quad \alpha,\beta \in \mathbb{R}
\end{equation}
where $\alpha,\beta$ are free phases.

\subsection{Left wall}\label{subsec:generalised-vector-boundary-left-wall}
Let us finally try to generalise the vector boundary representations for the case of a left wall, repeating the analysis performed in the previous subsection. Also in this case we begin by considering an ansatz for the $K$-matrices, inspired by the results \eqref{special-sol-left-wall-vector-K-matrix}, found by adopting the bulk representations also for the boundary. The ansatz has formally the same structure as the one used for the right wall
\begin{align}\label{generalised-vector-boundary-K-matrix-ansatz}
\tilde{K}^{LL}(\gamma)=\tilde{k}^{LL}(\gamma&)  \biggl[ E_{11} \otimes E_{11} + \tilde{a}_{1}E_{22}\otimes E_{22}+
\\
& + \tilde{\kappa}(\gamma)(E_{11}\otimes E_{22} + \tilde{a}_{2}E_{22}\otimes E_{11}) + \tilde{\eta}(\gamma)(E_{12}\otimes E_{21} + \tilde{a}_{3} E_{21}\otimes E_{12}) \biggr]
\notag
\end{align}
\begin{align}
\tilde{K}^{\tilde{L}\tilde{L}}(\gamma)=\tilde{k}^{\tilde{L}\tilde{L}}(\gamma &)  \biggl[ E_{11} \otimes E_{11} + \tilde{b}_{1}E_{22}\otimes E_{22}+
\\
& + \tilde{b}_4\tilde{\kappa}(\gamma)(E_{11}\otimes E_{22} + \tilde{b}_{2}E_{22}\otimes E_{11}) + \tilde{b}_5\tilde{\eta}(\gamma)(E_{12}\otimes E_{21} + \tilde{b}_{3} E_{21}\otimes E_{12}) \biggr]
\notag
\end{align}
\begin{align}
\tilde{K}^{L\tilde{L}}(\gamma)=\tilde{k}^{L\tilde{L}}(\gamma &)  \biggl[ E_{11} \otimes E_{11} + \tilde{c}_{1}E_{22}\otimes E_{22}+
\\
& + \frac{\tilde{c}_4}{\tilde{\kappa}(\gamma)}(E_{11}\otimes E_{22} + \tilde{c}_{2}E_{22}\otimes E_{11}) + \frac{\tilde{c}_5\tilde{\eta}(\gamma)}{\tilde{\kappa}(\gamma)}(E_{12}\otimes E_{12} + \tilde{c}_{3} E_{21}\otimes E_{21}) \biggr]
\notag
\end{align}
\begin{align}
\tilde{K}^{\tilde{L}L}(\gamma)=\tilde{k}^{\tilde{L}L}(\gamma &)  \biggl[ E_{11} \otimes E_{11} + \tilde{d}_{1}E_{22}\otimes E_{22}+
\\
& + \frac{\tilde{d}_4}{\tilde{\kappa}(\gamma)}(E_{11}\otimes E_{22} + \tilde{d}_{2}E_{22}\otimes E_{11}) + \frac{\tilde{d}_5\tilde{\eta}(\gamma)}{\tilde{\kappa}(\gamma)}(E_{12}\otimes E_{12} + \tilde{d}_{3} E_{21}\otimes E_{21}) \biggr]
\notag
\end{align}
and hence depends on 2 functions, $\tilde{\kappa}(\gamma)$ and $\tilde{\eta}(\gamma)$, and 18 constant coefficients $\tilde{a}_{i}$ with $i=1,...,3$ and $\tilde{b}_{i},\tilde{c}_{i},\tilde{d}_{i}$ with $i=1,...,5$. This leads again to two families of parametric solutions to the BYBE \eqref{left-wall-BYBE-rep-index-notation}, distinguished by a sign factor $\tilde{s}_{1} = \pm1$, and characterised by two unconstrained integration constants $\tilde{k}_{1},\tilde{k}_{2}$ and the parameter $\tilde{b}_5$, which remains free. The solutions can be written as follows:
\begin{align}\label{left_wall_BYBE_solutions}
\tilde{\kappa}(\gamma)=-\tilde{s}_{1}i\biggl(-1 +\frac{2(\tilde{s}_{1}+\cosh{\gamma})}{\tilde{s}_{1}+2e^{2\tilde{k}_{1}}+\cosh{\gamma}}\biggr) \quad \quad & \quad \quad \tilde{\eta}(\gamma)=\frac{\tilde{k}_2(e^{\frac{\gamma}{2}}+\tilde{s}_{1}e^{-\frac{\gamma}{2}})}{\tilde{s}_{1}+2e^{2\tilde{k}_{1}}+\cosh{\gamma}} 
\notag \\
\tilde{a}_{1}=\tilde{b}_{2}=- \tilde{s}_{1}i \qquad \quad \,\,\,\,\,\,\,\,\,\,  \tilde{a}_{2}=\tilde{b}_{1}= \tilde{s}_{1}i \qquad\qquad  & \qquad\quad \tilde{c}_{2}=\tilde{d}_{1}=- \tilde{s}_{1}i \qquad \tilde{c}_{1}=\tilde{d}_{2}= \tilde{s}_{1}i
\\
\tilde{a}_3=\tilde{c}_3=-\tilde{s}_{1}\frac{4ie^{2\tilde{k_{1}}}}{\tilde{k}_2^2} \qquad \,\, \tilde{b}_3=\tilde{d}_3=-\frac{\tilde{a}_3}{\tilde{b}_5^2}
\qquad\quad\,   & \qquad\quad 
\tilde{b}_4=\tilde{d}_4=-1 \qquad \,\,\,\,\,
\tilde{c}_4=\tilde{c}_5=1 \qquad \tilde{d}_5=-\tilde{b}_5
\notag
\end{align}
We then proceed as in section \ref{sec:generslised-rep-right-wall}, by considering an unconstrained boundary representation and looking for the constraints that the left wall vector boundary intertwining equation \eqref{left-wall-vector-intertwining-equation} imposes on this. We recall that for a left wall one should use the convenient basis $\tilde{A}^{\mathfrak{U}}$ in \eqref{new-algebra-basis-left-wall}, while the general form of the boundary representation can again be taken as \eqref{boundary-rep-ansatz}, with all functions appropriately dressed by a tilde, which characterise the left wall quantities. The intertwining equation then imposes constraints similar to \eqref{intertwining-constraints-1}, \eqref{intertwining-constraints-2} and \eqref{intertwining-constraints-3}, leading to:
\begin{equation}
\begin{aligned}
\tilde{\pi}_{B}^{L}(\mathfrak{U}^{-\tfrac{1}{2}}Q_{L}) = \tilde{m}_{1}\sqrt{h} \tilde{q}_{21}^{LL} \begin{pmatrix}
0 & 0 \\
1 & 0
\end{pmatrix} 
\quad & \quad \,\,\,\,\,\,\,\,
\tilde{\pi}_{B}^{L}(\mathfrak{U}^{-\tfrac{1}{2}}Q_{R}) = \frac{\sqrt{h}}{4\tilde{m}_{1}} \frac{i}{\tilde{q}^{LL}_{21}}\begin{pmatrix}
0 & 1 \\
0 & 0
\end{pmatrix}
\\
\tilde{\pi}_{B}^{L}(\mathfrak{U}^{\tfrac{1}{2}}G_{L}) = \tilde{m}_{2}\sqrt{h} \tilde{g}_{12}^{LL}
\begin{pmatrix}
0 & 1 \\
0 & 0
\end{pmatrix} 
\quad & \quad \,\,\,\,\,\,\,\,\,\,\,\,
\tilde{\pi}_{B}^{L}(\mathfrak{U}^{\tfrac{1}{2}}G_{R}) = -\frac{\sqrt{h}}{4\tilde{m}_{2}} 
\frac{i}{\tilde{g}^{LL}_{12}} 
\begin{pmatrix}
0 & 0 \\
1 & 0
\end{pmatrix}
\end{aligned}
\end{equation}
\begin{equation}
\begin{aligned}
\tilde{\pi}_{B}^{\tilde{L}}(\mathfrak{U}^{-\tfrac{1}{2}}Q_{L}) = \frac{\sqrt{h}}{4\tilde{m}_{3}}\frac{i}{\tilde{q}^{R\tilde{L}}_{21}}
\begin{pmatrix}
0 & 1 \\
0 & 0
\end{pmatrix} 
\quad \quad \,\,\,\,\, & \quad 
\tilde{\pi}_{B}^{\tilde{L}}(\mathfrak{U}^{-\tfrac{1}{2}}Q_{R}) = \tilde{m}_{3}\sqrt{h}\tilde{q}_{21}^{R\tilde{L}}\begin{pmatrix}
0 & 0 \\
1 & 0
\end{pmatrix}
\\
\tilde{\pi}_{B}^{\tilde{L}}(\mathfrak{U}^{\tfrac{1}{2}}G_{L}) = -\frac{\sqrt{h}}{4\tilde{m}_{4}}\frac{i}{\tilde{g}^{R\tilde{L}}_{12}}\begin{pmatrix}
0 & 0 \\
1 & 0
\end{pmatrix} 
\quad \,\,\,\,\,\,  & \quad\,\,\,\,
\tilde{\pi}_{B}^{\tilde{L}}(\mathfrak{U}^{\tfrac{1}{2}}G_{R}) = \tilde{m}_{4}\sqrt{h}\tilde{g}_{12}^{R\tilde{L}} \begin{pmatrix}
0 & 1 \\
0 & 0
\end{pmatrix}
\end{aligned}
\end{equation}
The latter provides the form of left boundary representations whose matrix structure is compatible with those of the bulk representations and of the ansatz for the $K$-matrix. As already happened for the right wall, at this point it is necessary to characterise the ansatz for the $K$-matrices using the solutions \eqref{left_wall_BYBE_solutions}, so that the intertwining equations exhibit dependence on the free parameters $\tilde{s}_{1},\tilde{k}_{1},\tilde{k}_{2},\tilde{b}_{5}$ and one may search for special values which would solve them. We thus proceed by sampling a subset of non-vanishing entries in the intertwining equation (which is again an $8\times 8$ matrix equation) and solving them for the functions $\tilde{q}^{LL}_{21},\tilde{q}^{R\tilde{L}}_{21},\tilde{g}^{LL}_{12},\tilde{g}^{R\tilde{L}}_{12}$. This leads to expressions which depend on the parameters introduced above, as well as the coupling $h$ and bulk rapidity $\gamma$. The latter dependence should however disappear and it is indeed possible to find a convenient choice of parameters which completely gets rid of it. Enforcing $df/d\gamma = 0$ for $f$ any of the above four functions and specifying a family of $K$-matrices in terms of $\tilde{s}_{1}=\pm 1$, one finds that all of the four functions lose the $\gamma$ dependence provided that
\begin{equation}
\begin{aligned}
\tilde{s}_{1}=+1:& \quad \frac{df}{d\gamma}\propto(1+2e^{2\tilde{k}_{1}})= 0  \qquad  \Rightarrow \qquad \tilde{k}_{1}=i \pi \tilde{z} -\tfrac{1}{2}\log{2}+\tfrac{i}{2}\pi \qquad \tilde{z}\in \mathbb{Z}
\\
\tilde{s}_{1}=-1:& \quad \frac{df}{d\gamma}\propto (1-2e^{2\tilde{k}_{1}})= 0 \qquad \Rightarrow \qquad \tilde{k}_{1}=i \pi \tilde{z} - \tfrac{1}{2}\log{2} \qquad \qquad\,\,\,   \tilde{z}\in \mathbb{Z}
\end{aligned}
\end{equation}
As for the right wall, the above conditions not only allow us to get rid of the $\gamma$-dependence, as desired, but also to completely solve all the remaining intertwining equations with no need to tune further parameters. In conclusion, after performing calculations along the lines of section \ref{sec:generslised-rep-right-wall} one finds two families of parametric representations, distinguished by the sign factor $\tilde{s}_{1}=\pm 1$ and characterised by the free constants $\tilde{k}_{2}$ and $\tilde{b}_{5}$. The boundary representation for the supercharges takes the explicit form
\begin{align}
\tilde{\pi}_{B}^{L}(\mathfrak{U}^{-\tfrac{1}{2}}Q_{L}) = -(1+i\tilde{s}_{1})\frac{1}{2\tilde{k}_{2}}\sqrt{h}  \begin{pmatrix}
0 & 0 \\
1 & 0
\end{pmatrix} 
\qquad\,\, & \quad 
\tilde{\pi}_{B}^{L}(\mathfrak{U}^{-\tfrac{1}{2}}Q_{R}) = -\tilde{s}_{1}(1+i\tilde{s}_{1})\frac{\tilde{k}_{2}}{4}\sqrt{h}
\begin{pmatrix}
0 & 1 \\
0 & 0
\end{pmatrix}
\notag \\
\tilde{\pi}_{B}^{L}(\mathfrak{U}^{\tfrac{1}{2}}G_{L}) = 
\tilde{s}_{1}(1-i\tilde{s}_{1})\frac{\tilde{k}_{2}}{4}\sqrt{h} 
\begin{pmatrix}
0 & 1 \\
0 & 0
\end{pmatrix} 
\qquad \,\,\,\, & \quad \,\,\,\,
\tilde{\pi}_{B}^{L}(\mathfrak{U}^{\tfrac{1}{2}}G_{R}) = (1-i\tilde{s}_{1})\frac{1}{2\tilde{k}_{2}}\sqrt{h} 
\begin{pmatrix}
0 & 0 \\
1 & 0
\end{pmatrix}
\notag \\
\\
\tilde{\pi}_{B}^{\tilde{L}}(\mathfrak{U}^{-\tfrac{1}{2}}Q_{L}) = \tilde{s}_{1}(1+i\tilde{s}_{1})\frac{\tilde{b}_{5}\tilde{k}_{2}}{4}\sqrt{h}
\begin{pmatrix}
0 & 1 \\
0 & 0
\end{pmatrix} 
\qquad & \quad 
\tilde{\pi}_{B}^{\tilde{L}}(\mathfrak{U}^{-\tfrac{1}{2}}Q_{R}) = (1+i\tilde{s}_{1})\frac{1}{2\tilde{b}_{5}\tilde{k}_{2}}\sqrt{h}
\begin{pmatrix}
0 & 0 \\
1 & 0
\end{pmatrix}
\notag \\
\tilde{\pi}_{B}^{\tilde{L}}(\mathfrak{U}^{\tfrac{1}{2}}G_{L}) = -(1-i\tilde{s}_{1})\frac{1}{2\tilde{b}_{5}\tilde{k}_{2}}\sqrt{h}
\begin{pmatrix}
0 & 0 \\
1 & 0
\end{pmatrix} 
\quad \,\,\,\, & \quad \,\,\,\,
\tilde{\pi}_{B}^{\tilde{L}}(\mathfrak{U}^{\tfrac{1}{2}}G_{R}) = -\tilde{s}_{1}(1-i\tilde{s}_{1})\frac{\tilde{b}_{5}\tilde{k}_{2}}{4}\sqrt{h}
\begin{pmatrix}
0 & 1 \\
0 & 0
\end{pmatrix}
\notag
\end{align}
and in turn leads to the following central elements, for $a\in \{L,\tilde{L} \}$
\begin{equation}
\begin{aligned}
\tilde{\pi}_{B}^{a}(\mathfrak{H}_{L}) = 
-\tilde{s}_{1}\frac{h}{4}
\begin{pmatrix}
1 & 0 \\
0 & 1
\end{pmatrix} 
\quad\, & \quad \,\,
\tilde{\pi}_{B}^{a}(\mathfrak{H}_{R}) = -\tilde{s}_{1}\frac{h}{4} \begin{pmatrix}
1 & 0 \\
0 & 1
\end{pmatrix}
\\
\tilde{\pi}_{B}^{a}(\mathfrak{U}^{-1}\mathfrak{P}) = \frac{i}{4}h 
\begin{pmatrix}
1 & 0 \\
0 & 1
\end{pmatrix} 
\quad \,\,\,\,\,\,\,\, & \quad\,\,\,
\tilde{\pi}_{B}^{a}(\mathfrak{U}\mathfrak{K})
= -\frac{i}{4}h
\begin{pmatrix}
1 & 0 \\
0 & 1
\end{pmatrix}
\end{aligned}
\end{equation}
Comments similar to those below equation \eqref{generalised-right-wall-vector-boundary-central-elements-rep} hold at this stage. \\

\noindent
\textbf{Unitarity and other relations.} \quad
We can now look at unitarity \eqref{Unitarity-relations-K-matrices}, starting from the physical one
\begin{equation}\label{unitarity-left-wall-vector-boundary}
\tilde{K}^{ab}(\gamma)\bigl[\tilde{K}^{ab}(\gamma)\bigr]^{\dagger}=\mathbbmss{1} \qquad \forall \,\, a,b \in \{ L,\tilde{L} \}\,\, .
\end{equation}
Exploiting the solution \eqref{left_wall_BYBE_solutions} to the BYBE and recalling that the intertwining relations are satisfied provided one chooses $\tilde{s}_{1}=+1$ and $\tilde{k}_{1}=i\pi \tilde{z}-\tfrac{1}{2}\log{2}+\tfrac{i}{2}\pi$, or alternatively $\tilde{s}_{1}=-1$ and $\tilde{k}_{1}=i\pi \tilde{z}-\tfrac{1}{2}\log{2}$, with the parameters $\tilde{k}_{2},\tilde{b}_{5}$ remaining free, it is easy to verify that:
\begin{itemize}
\item For $\tilde{s}_{1}=+1$ and $\tilde{k}_{1}=i\pi \tilde{z}-\tfrac{1}{2}\log{2}+\tfrac{i}{2}\pi$ there exists no choice of the free parameters $\tilde{k}_{2},\tilde{b}_{5}$ which makes the relation \eqref{unitarity-left-wall-vector-boundary} satisfied for any of the $\tilde{K}^{ab}(\gamma)$ matrices.
\item For $\tilde{s}_{1}=-1$ and $\tilde{k}_{1}=i\pi \tilde{z}-\tfrac{1}{2}\log{2}$ the relation \eqref{unitarity-left-wall-vector-boundary} requires the free parameters to satisfy
\begin{equation}\label{left-wall-vector-physical-unitarity-conditions-on-parameters}
|\tilde{k}_{2}|^2 = 2 \quad \text{for} \quad ab=\{LL,L\tilde{L} \} 
\qquad \text{and} \qquad 
|\tilde{k}_{2}|^2|\tilde{b}_{5}|^2 = 2 \quad \text{for} \quad ab=\{\tilde{L}\tilde{L},\tilde{L}L \}
\end{equation}
and at the same time enforces the following conditions on the unconstrained dressing factors
\begin{equation}\label{left-wall-vector-boundary-dressing-factors-conditions}
\begin{aligned}
\tilde{k}^{ab}(\gamma)\bigl[ \tilde{k}^{ab}(\gamma) \bigr]^{*} & = 1 \qquad &&\text{for} \qquad ab=\{ LL,\tilde{L}\tilde{L} \}
\\
\tilde{k}^{ab}(\gamma)\bigl[ \tilde{k}^{ab}(\gamma) \bigr]^{*} & = \frac{(\cosh{\gamma}-2)^2}{(\cosh{\gamma})^2} 
\qquad &&\text{for} \qquad ab=\{ L\tilde{L},\tilde{L}L \}
\end{aligned}
\end{equation}
\end{itemize}
Hence, as for the right wall, while the BYBE and intertwining relations are consistent with both families of $K$-matrices, the physical unitarity requirement rules out one of them and almost completely fixes the remaining free parameters. This is again in agreement with positivity of the energy under the identification \eqref{energy}.  Concerning braiding unitarity, one encounters the same problems as for the right wall and these can be overcome by introducing a new matrix $\tilde{K}'$, determined by solving the left wall intertwining equation \eqref{left-wall-vector-intertwining-equation} with the opposite momentum
\begin{equation}
\bigl(\pi^b_B\otimes\pi_{-p}^{a-}\bigr)\Delta(\mathfrak{b})\tilde{K}'^{ba}\bigl(\gamma(p)\bigr)=\tilde{K}'^{ba}\bigl(\gamma(p)\bigr)\Delta(\mathfrak{b})\bigl(\pi^b_B\otimes\pi_{+p}^{a+}) \qquad \forall \, \mathfrak{b}\in \tilde{\mathcal{B}}^\mathfrak{U}_{V} \qquad \text{and} \qquad a,b\in \{L,\tilde{L}\}\,\, .
\end{equation}
This matrix then satisfies the braiding unitarity relations
\begin{equation}
\tilde{K}^{ab}(\gamma)\tilde{K}'^{ab}(-\gamma)=\mathbbmss{1} \qquad \forall \,\, a,b \in \{ L,\tilde{L} \} \, \, .
\end{equation}\\
\noindent
Also for the left wall one then observes that the $K$-matrices satisfy extra relations similar to those found for the right wall. Indeed it can be checked that
\begin{equation}
\tilde{K}^{LL}(\gamma)\tilde{K}^{\tilde{L}\tilde{L}}(-\gamma)^{T}=\mathbbmss{1} \qquad \text{and} \qquad \tilde{K}^{L\tilde{L}}(\gamma)\tilde{K}^{\tilde{L}L}(-\gamma)^{T}=\mathbbmss{1} 
\end{equation}
provided the parameter $\tilde{b}_{5}$ is fixed to be
\begin{equation}\label{constraint-b5-left-wall}
\tilde{b}_{5} = -\frac{2}{\tilde{k}_{2}^2}\,\, ,
\end{equation}
and the dressing factors satisfy
\begin{equation}\label{left-vector-boundary-conditions-on-dressing-factors-from-braiding-unitarity}
\tilde{k}^{LL}(\gamma)\tilde{k}^{\tilde{L}\tilde{L}}(-\gamma)=1 \qquad \text{and} \qquad \tilde{k}^{L\tilde{L}}(\gamma)\tilde{k}^{\tilde{L}L}(-\gamma)=\frac{(\cosh{\gamma}+2\tilde{s}_{1})^2}{(\cosh{\gamma})^2} \,\, .
\end{equation}
The condition \eqref{constraint-b5-left-wall} on $\tilde{b}_{5}$ is consistent with the requirements of physical unitarity \eqref{left-wall-vector-physical-unitarity-conditions-on-parameters}. One can then also check the satisfaction, for both families of $K$-matrices, of mixed braiding-like unitarity relations
\begin{equation}
\tilde{K}^{LL}(\gamma)\tilde{K}^{\tilde{L}\tilde{L}}(-\gamma)=\mathbbmss{1} \qquad \text{and} \qquad \tilde{K}^{L\tilde{L}}(\gamma)\tilde{K}^{\tilde{L}L}(-\gamma)=\mathbbmss{1} \,\, , 
\end{equation}
provided that
\begin{equation}\label{condition-b5tilde}
\tilde{b}_{5}=-i
\end{equation}
and the dressing factors satisfy \eqref{left-vector-boundary-conditions-on-dressing-factors-from-braiding-unitarity}. The latter condition on $\tilde{b}_{5}$ is again consistent with \eqref{left-wall-vector-physical-unitarity-conditions-on-parameters}.

We conclude by noting that, given the ansatze for the right and left wall $K$-matrices and the choice of parameters \eqref{right_wall_BYBE_solutions}, \eqref{left_wall_BYBE_solutions} which solve the BYBE for respectively the right and left walls, setting
\begin{equation}\label{vector-K-matrices-parameter-relations}
\tilde{s}_{1}=s_{1} \qquad\qquad  \tilde{k}_{1}=k_{1} \qquad\qquad  \tilde{k}_{2}=s_{1}i\, \frac{4e^{2k_{1}}}{k_{2}} \qquad \qquad \tilde{b}_{5}=-\frac{1}{b_{5}}
\end{equation}
one obtains the following relations between left and right wall $K$-matrices
\begin{equation}\label{vector-K-matrices-relations}
\begin{aligned}
\tilde{K}^{aa}(\gamma)&=\bigl[K^{aa}(\gamma)\bigr]^{op} \qquad \text{iff} \qquad \tilde{k}^{aa}(\gamma)=k^{aa}(\gamma)
\qquad \text{for} \qquad a=\{ 
L,\tilde{L} \}
\\
\tilde{K}^{ab}(\gamma)&=\bigl[K^{ba}(\gamma)\bigr]^{op} \qquad \text{iff} \qquad \tilde{k}^{ab}(\gamma)=k^{ba}(\gamma)
\qquad \text{for} \qquad ab=\{ 
L\tilde{L}, \tilde{L}L \}
\end{aligned}
\end{equation}
The above conditions are also consistent with physical unitarity and energy positivity under the identification \eqref{energy}, as for both the left and right walls the unitary family of $K$-matrices has $\tilde{s}_{1}=s_{1}=-1$ and using \eqref{vector-K-matrices-parameter-relations} it is not hard to verify that
\begin{equation}
|\tilde{k}_{2}|^2 = 2 \quad \Leftrightarrow \quad |k_{2}|^2 = 2
\qquad \text{and} \qquad 
|\tilde{k}_{2}|^2|\tilde{b}_{5}|^2 = 2 \quad \Leftrightarrow \quad |k_{2}|^2|b_{5}|^2 = 2 \,\, .
\end{equation}
In light of \eqref{vector-K-matrices-parameter-relations}, the condition \eqref{constraint-b5-left-wall} then imposes, on both families of $K$-matrices
\begin{equation}
b_{5}=-\frac{8e^{4k_{1}}}{k_{2}^2}=
-\frac{2}{k_{2}^2} \qquad \text{for} \qquad s_{1}=\pm 1
\end{equation}
which is again, in both cases, compatible with the unitarity conditions $|k_{2}|^2=2$ and $|b_{5}|^2=1$. Finally, the restriction \eqref{condition-b5tilde} on $\tilde{b}_{5}$ is also consistent with the relation \eqref{vector-K-matrices-parameter-relations} between $\tilde{b}_{5}$ and $b_{5}$, since for the right wall we obtained $b_{5}=-i$ as well.
\\

\noindent
\textbf{Short summary of the generalised left wall vector boundary.} \quad For a short summary of the approach used to construct generalised left wall vector boundary representations, we refer to the right wall summary above subsection \ref{subsec:generalised-vector-boundary-left-wall}. We highlight here that left wall $K$-matrices satisfying intertwining and BYB equations, as well as physical unitarity, can be obtained starting from the ansatz \eqref{generalised-vector-boundary-K-matrix-ansatz} and requiring the unfixed parameters $\tilde{s}_{1},\tilde{k}_{1},\tilde{k}_{2},\tilde{b}_{5}$ in the solution \eqref{left_wall_BYBE_solutions} to take the form
\begin{equation}
\tilde{s}_{1}= -1 \qquad \tilde{k}_{1}=i\pi \tilde{z}-\tfrac{1}{2}\log{2} \qquad \tilde{k}_{2}=\sqrt{2} e^{i\tilde{\alpha}} \qquad \tilde{b}_{5}=e^{i\tilde{\beta}} \qquad \text{with} \qquad \tilde{z}\in \mathbb{Z}, \quad \tilde{\alpha},\tilde{\beta} \in \mathbb{R}
\end{equation}
where $\tilde{\alpha},\tilde{\beta}$ are free phases. We remark again that such restrictions also lead to a positive boundary energy $\tilde{E}_B = h$, identified with the bulk energy as in \eqref{energy}.


\section{Conclusions}

In this paper we have extended the study of reflection matrices and boundary scattering to the massless sector of $AdS_3 \times S^3 \times T^4$ string theory. This is a first study of this kind in $AdS/CFT$ integrability, since all the earlier treatments necessarily focused on massive excitations - see for instance the vast literature following from \cite{Hof}, and the literature cited in \cite{Prinsloo:2015apa}. We analyse the new setup offered by massless modes and their supersymmetric representations, and explore it extensively finding a huge wealth of boundary reflection $K$-matrices. These correspond to different boundary co-ideal subalgebras, which we specify in detail.

Throughout the paper we worked with a symmetric co-product in which the bulk $R$-matrices can be expressed in terms of the difference in the rapidity variables of the two scattering magnons. The co-ideal subalgebra condition, \eqref{right-coideal-subalgebra} and \eqref{left-coideal-subalgebra} for right and left boundaries respectively, coupled with the symmetric co-product, forces us to rescale the generators of the Hopf algebra to construct the boundary subalgebras. We computed the reflection matrices for both the singlet and vector boundaries which satisfy the Boundary Yang-Baxter Equations. Furthermore, we specified the boundary coideal subalgebra intertwined by each $K$-matrix. For the case of the singlet boundaries, we found a hidden symmetry while solving the BYBE for non-supersymmetric chiral boundaries which, while not coideal in our symmetric coproduct, we expect should form an enhanced boundary coideal subalgebra in some co-product as was the case for the massive sector discussed in \cite{Prinsloo:2015apa}. We verified that the $K$-matrices satisfy physical unitarity as well as braiding unitarity modulo a subtlety arising from having a different bulk representation for right and left moving particles in the massless limit.

For the vector boundary, for comparison with \cite{Prinsloo:2015apa}, we initially fixed the boundary representation to be the bulk representation with a particular value for the momentum $p = \pm \pi$ (for right and left boundary respectively) and a rescaling $h \to h/2$. This choice of boundary representation is physically motivated as the group velocity of the wavepacket for this choice of representation vanishes, and hence the boundary wavepacket is stationary. However, we can consider a more generic choice for the boundary representation which may or may not have a physical interpretation. Hence in section \ref{Gen Vec Bound}, we generalize our boundary representations which gives slightly more generic $K$-matrices that satisfy the BYBE. For this generic class of $K$-matrices, we see some interesting relations obeyed by the different partial $K$-matrices. 

This work is the first step in a series of studies which we intend to carry out, including a derivation of the boundary Bethe ansatz and boundary thermodynamic Bethe ansatz. The next step would be to compute the dressing factors for the $K$-matrices using crossing symmetry. The massless sector has extremely rich physics, particularly in conjunction with boundary massless flows and boundary conformal field theory. We plan to investigate these connections in future work.


\section*{Acknowledgments}
We very much thank Vidas Regelskis for illuminating discussions. AT thanks Bogdan Stefa\'nski for discussions about the  physics of the vector boundary. AT thanks the Isaac Newton Institute in Cambridge and the organisers of the programme {\it Machine Learning Toolkits and Integrability Techniques in Gravity}, EPSRC grant nr. EP/R014604/1, for their kind support and hospitality. The work of AT has been
supported by EPSRC-SFI under the grant EP/S020888/1 {\it Solving Spins and Strings} in the initial phase of this project. VG thanks STFC for the Doctoral Training Programme funding (ST/W507854-2021 Maths DTP). VM is supported by the STFC under the grant ST/X508809/1. The work of DB has been supported by Thailand NSRF via PMU-B, grant number B13F670063. DB thanks the School of Mathematics and Physics at the University of Surrey for the generous support and hospitality during the early stages of this work, as well as the Physics Department at the University of Milano-Bicocca for the warm hospitality.

\section*{Data and Licence Management}
No additional research data beyond the data presented and cited in this work are needed to validate the research findings in this work. For the purpose of open access, the authors have applied a Creative Commons Attribution (CC BY) licence to any Author Accepted Manuscript version arising.


\begin{appendix}

\section{Different frame factors}\label{App:A}

In this appendix, for the ease of comparison with  \cite{Prinsloo:2015apa}, we report the calculation, for the singlet boundary, in a different convention for the frame factors (equivalently, a different convention for the coproduct). The $L$ representation of the supercharges is slightly modified as well, it is taken to be the massless limit of the $L$ representation considered in \cite{Prinsloo:2015apa}.
\begin{eqnarray}
&&\pi_{p_i}^{L+}(\mathfrak{Q}_L) = - \pi_{p_i}^{L+}(\mathfrak{G}_R) e^{i\frac{p_i}{2}}= \sqrt{\frac{h}{2}\sin \frac{p_i}{2}}\begin{pmatrix}0&0\\1&0\end{pmatrix}, \nonumber\\
&&\pi_{p_i}^{L+}(\mathfrak{G}_L) = - \pi_{p_i}^{L+}(\mathfrak{Q}_R) e^{-i\frac{p_i}{2}}=\sqrt{\frac{h}{2}\sin \frac{p_i}{2}}\begin{pmatrix}0&1\\0&0\end{pmatrix}, \qquad p_i\in (0,\pi),
\end{eqnarray}
\begin{eqnarray}
&&\pi_{p_i}^{L-}(\mathfrak{Q}_L) = \pi_{p_i}^{L-}(\mathfrak{G}_R) e^{i\frac{p_i}{2}}= \sqrt{\frac{h}{2}\sin \frac{-p_i}{2}}\begin{pmatrix}0&0\\1&0\end{pmatrix},\nonumber\\
&&\pi_{p_i}^{L-}(\mathfrak{G}_L) =  \pi_{p_i}^{L-}(\mathfrak{Q}_R) e^{-i\frac{p_i}{2}}= \sqrt{\frac{h}{2}\sin \frac{-p_i}{2}}\begin{pmatrix}0&1\\0&0\end{pmatrix}, \qquad p_i\in (-\pi,0).
\end{eqnarray}
We then change the coproduct assignment as follows:
\begin{eqnarray}
&&\Delta(\mathfrak{Q}_L)=\mathfrak{Q}_L\otimes \mathbbmss{1} + e^{i\frac{\mathfrak{p}}{2}}\otimes \mathfrak{Q}_L, \qquad \Delta(\mathfrak{G}_L)=\mathfrak{G}_L\otimes \mathbbmss{1} + e^{-i\frac{\mathfrak{p}}{2}}\otimes \mathfrak{G}_L,\nonumber\\
&&\Delta(\mathfrak{Q}_R)=\mathfrak{Q}_R\otimes \mathbbmss{1} + e^{i\frac{\mathfrak{p}}{2}}\otimes \mathfrak{Q}_R, \qquad \Delta(\mathfrak{G}_R)=\mathfrak{G}_R\otimes \mathbbmss{1} + e^{-i\frac{\mathfrak{p}}{2}}\otimes \mathfrak{G}_R.\label{nuo}
\end{eqnarray}
This should ultimately not change the physics, meaning the energy spectrum of the theory obtained by solving the Bethe ansatz equations. It has however the consequence that the $S$-matrix in the massless sector ceases to be of manifest difference form, precisely because of the different frame factors which spoil it. The use of the pseudo-relativistic variable $\gamma_i$ becomes therefore less relevant and it is actually simpler to use the individual momenta. Because of the lack of manifest difference form it is therefore in general more desirable to adopt the conventions of the main text. We will only describe the level-$0$ coproducts and not the higher Yangian levels. 

The $R$-matrices are modified as follows (ignoring the dressing phases):
\begin{eqnarray}
&&R_{++} \sim \begin{pmatrix}1&0&0&0\\0&-\mbox{csc}\frac{p_1+p_2}{4}\sin \frac{p_1-p_2}{4}&\frac{e^{i\frac{p_1-p_2}{4}} \sqrt{\sin \frac{p_1}{2}\sin\frac{p_2}{2}}}{\sin \frac{p_1+p_2}{4}}&0\\0&\frac{e^{-i\frac{p_1-p_2}{4}} \sqrt{\sin \frac{p_1}{2}\sin\frac{p_2}{2}}}{\sin \frac{p_1+p_2}{4}}&\mbox{csc}\frac{p_1+p_2}{4}\sin \frac{p_1-p_2}{4}&0\\0&0&0&-1\end{pmatrix}, \qquad p_1,p_2 \in (0,\pi),\nonumber\\ 
&&R_{--} \sim \begin{pmatrix}1&0&0&0\\0&-\mbox{csc}\frac{p_1+p_2}{4}\sin \frac{p_1-p_2}{4}&\frac{-e^{i\frac{p_1-p_2}{4}} \sqrt{-\sin \frac{p_1}{2}}\sqrt{-\sin\frac{p_2}{2}}}{\sin \frac{p_1+p_2}{4}}&0\\0&\frac{-e^{-i\frac{p_1-p_2}{4}} \sqrt{-\sin \frac{p_1}{2}}\sqrt{-\sin\frac{p_2}{2}}}{\sin \frac{p_1+p_2}{4}}&\mbox{csc}\frac{p_1+p_2}{4}\sin \frac{p_1-p_2}{4}&0\\0&0&0&-1\end{pmatrix}, \qquad p_1,p_2 \in (-\pi,0),\nonumber\\ 
&&R_{+-} \sim \begin{pmatrix}1&0&0&0\\0&\mbox{sec}\frac{p_1+p_2}{4}\cos \frac{p_1-p_2}{4}&\frac{- ie^{i \frac{p_1-p_2}{4}} \sqrt{-\sin \frac{p_1}{2}\sin\frac{p_2}{2}}}{\cos\frac{p_1+p_2}{4}}&0\\0&\frac{- ie^{-i \frac{p_1-p_2}{4}} \sqrt{-\sin \frac{p_1}{2}\sin\frac{p_2}{2}}}{\cos\frac{p_1+p_2}{4}}&\mbox{sec}\frac{p_1+p_2}{4}\cos \frac{p_1-p_2}{4}&0\\0&0&0&1\end{pmatrix}, \qquad p_1 \in (0,\pi), \quad p_2 \in (-\pi,0),\nonumber\\
&&R_{-+} \sim \begin{pmatrix}1&0&0&0\\0&\mbox{sec}\frac{p_1+p_2}{4}\cos \frac{p_1-p_2}{4}&\frac{ie^{i \frac{p_1-p_2}{4}} \sqrt{-\sin \frac{p_1}{2}\sin\frac{p_2}{2}}}{\cos\frac{p_1+p_2}{4}}&0\\0&\frac{ie^{-i \frac{p_1-p_2}{4}} \sqrt{-\sin \frac{p_1}{2}\sin\frac{p_2}{2}}}{\cos\frac{p_1+p_2}{4}}&\mbox{sec}\frac{p_1+p_2}{4}\cos \frac{p_1-p_2}{4}&0\\0&0&0&1\end{pmatrix}, \qquad p_1 \in (-\pi,0), \quad p_2 \in (0,\pi).\nonumber\\
\label{for}\end{eqnarray} 
The frame factors all become equal to $1$ in the BMN limit, therefore one can ignore the different conventions in that case and recover the relativistic limit exactly as we did in the main text. When working with the momentum variables it is easy to see that the BMN limit $p_i \to 0^\pm$ is indeed not trivial for $R_{++}$ and $R_{--}$ because of the ratio of sines, while it trivialises $R_{+-}$ and $R_{-+}$ which instead have cosines in various places.

\subsection{Right wall singlet boundary}
We have solved the boundary Yang-Baxter equation for the right wall in the case of a singlet boundary in these conventions, and again we find a family of solutions parameterised by one complex constant. The equation for complete $K$ and $R$-matrices ($K(p) = \oplus K^a (p)$ and $R(p_1,p_2) = \oplus R^{ab}(p_1,p_2)$ with $a,b \in \{L,R\}$) reads
\begin{eqnarray}
K_2(p_2)  R^{op}_{+-}(p_2,-p_1)K_1(p_1)R_{++}(p_1,p_2) = R^{op}_{--}(-p_2,-p_1)K_1(p_1)R_{+-}(p_1,-p_2)K_2(p_2), 
\end{eqnarray}
where $p_1,p_2 \in (0,\pi)$. The solution is given by
\begin{eqnarray}
K^L(p_i) =  \begin{pmatrix}1&0\\0&\sigma(p_i)\end{pmatrix}, \qquad \sigma(p_i) = \frac{e^{ip_i}+2c_0 \, e^{i\frac{p_i}{2}}}{-1+2c_0 \, e^{i\frac{p_i}{2}}},\qquad i=1,2,
\end{eqnarray}
with $c_0$ any complex number. We can see that for $c_0 \to \pm \infty$ the reflection matrix becomes the identity. Also, both for $c_0 \to \pm \infty$ and $c_0 = 0$ it is unitary, with the condition on the reflection dressing factors being $k^L(-p_i)k^L(p_i)=1$.

The next step is to impose the boundary intertwining relations to fix the value of the constant $c_0$. For a generator $\mathfrak{b}$ in the boundary algebra $\mathcal{B}$ these are of the form
\begin{eqnarray}
    \pi^{L-}_{-p_i}(\mathfrak{b})\, K^L(p_i)=K^L(p_i)\, \pi^{L+}_{p_i}(\mathfrak{b}),
\end{eqnarray}
with $ \pi^{L\pm}_{p_i}(\mathfrak{b})$ denoting the representation of $\mathfrak{b} \in \mathcal{B}$. 

For the singlet chiral boundary algebras presented in \cite{Prinsloo:2015apa}, we have
\begin{itemize}
   \item 
   Non-supersymmetric chiral, $\mathcal{B}_{NC}=\langle \mathfrak{H}_L,\mathfrak{H}_R\rangle$:\\
   The relations impose no constraints on the $K$-matrix, because $\pi_{p_i}^{\pm}(\mathfrak{H}_{L,R})=\pi_{-p_i}^{\mp}(\mathfrak{H}_{L,R})\propto I$. This seems to be in agreement with the massive case, where the solution also includes a free parameter. Denoting by $\tilde{c}_0$ the constant that appears in $\mathbb{K}_{\mathcal{B}_{NC}}^L(p_i)$  of \cite{Prinsloo:2015apa}, our solution agrees with its massless limit if $\tilde{c}_0=-2 c_0$.
   \item
   Left half-supersymmetric, $\mathcal{B}_L=\langle\mathfrak{Q}_L,\mathfrak{G}_L,\mathfrak{H}_L,\mathfrak{H}_R \rangle$:\\
   Satisfying the symmetry relations requires $c_0\rightarrow\infty$, implying $K^L_{\mathcal{B}_L}(p_i)=I$
   \item
   Right half-supersymmetric, $\mathcal{B}_R=\langle\mathfrak{Q}_R,\mathfrak{G}_R,\mathfrak{H}_L,\mathfrak{H}_R \rangle$:\\
   Satisfying the symmetry relations requires $c_0=0$, implying 
   \begin{eqnarray}
   K^L_{\mathcal{B}_{R}}(p)=
    \begin{pmatrix}
     1 && 0 \\
     0 && -e^{ip_i}
    \end{pmatrix}\label{inth}
    \end{eqnarray}
\end{itemize}
In the last two cases, the result agrees with the massless limit of the corresponding  $\mathbb{K}_{\mathcal{B}}^L(p_i)$ in \cite{Prinsloo:2015apa}. These solutions can also be obtained by imposing the intertwining relations on a generic diagonal $K$-matrix $K(p_i)=E_{11}+\sigma (p_i)E_{22}$ and solving for $\sigma(p_i)$ (without using the BYBE). 


We have also computed all the $R$-matrices for the situation where either one or both particles are in the massless limit of the $R$ representation of \cite{Prinsloo:2015apa}, which reads
\begin{eqnarray}
&&\pi^{R+}_{p_i}(\mathfrak{Q}_L) = - \pi^{R+}_{p_i}(\mathfrak{G}_R) e^{i\frac{p_i}{2}}
= \sqrt{\sin \frac{p_i}{2}}\begin{pmatrix}0&1\\0&0\end{pmatrix}, \nonumber\\ 
&&\pi^{R+}_{p_i}(\mathfrak{G}_L) = - \pi^{R+}_{p_i}(\mathfrak{Q}_R) e^{-i\frac{p_i}{2}}= 
\sqrt{\sin \frac{p_i}{2}}\begin{pmatrix}0&0\\1&0\end{pmatrix}, \qquad p_i\in (0,\pi),
\end{eqnarray}
\begin{eqnarray}
&&\pi^{R-}_{p_i}(\mathfrak{Q}_L) = \pi^{R-}_{p_i}(\mathfrak{G}_R) e^{i\frac{p_i}{2}}= 
\sqrt{\sin \frac{-p_i}{2}}\begin{pmatrix}0&1\\0&0\end{pmatrix}, \nonumber\\ 
&&\pi^{R-}_{p_i}(\mathfrak{G}_L) =  \pi^{R-}_{p_i}(\mathfrak{Q}_R) e^{-i\frac{p_i}{2}}= 
\sqrt{\sin \frac{-p_i}{2}}\begin{pmatrix}0&0\\1&0\end{pmatrix}, \qquad p_i\in (-\pi,0).
\end{eqnarray}
The coproduct is the one in (\ref{nuo}).
We do not report the $R$-matrices explicitly for brevity - with this choice of coproduct all the $16$ bulk $R$-matrices\footnote{The total of $16$ comes from $L$,$R$ and $+$,$-$ for either particle $1$ or $2$.}, of which $4$ are shown in (\ref{for}) are different by some frame factors, and we do not have simple relations as in  (\ref{compa}). We have then computed the singlet-boundary reflection matrix for $R$ particles, checking the boundary Yang-Baxter equation when two bulk particles are of type $R$. We obtain 
\begin{eqnarray}
K^{R}(p_i) =  \begin{pmatrix}1&0\\0&\sigma^{R}(p_i)\end{pmatrix}, \qquad \sigma^{R}(p_i) = 1-\frac{2\cos\frac{p_i}{2}}{e^{i\frac{p_i}{2}}-2c_0^{R}},\qquad i=1,2,    
\end{eqnarray}
with $c_0^{R}$ any constant. Once again, both for $c^{R}_0 \to \pm \infty$ and $c^{R}_0 = 0$ it is unitary, with the condition on the reflection dressing factors being $k^{R}(-p_i)k^{R}(p_i)=1$. 

We find that we can solve the mixed boundary Yang-Baxter equations when one bulk particle is $L$ and one is $R$ by specifying the values of the constants, and in particular
\begin{eqnarray}
c_0 = - c_0^{R}    
\end{eqnarray}
is a choice that works, under the analogue of the conditions (\ref{think1}) and (\ref{think2}) on the respective dressing factors.

\subsection{Left wall singlet boundary}

We have also repeated the calculation for a left wall in the case of a singlet boundary in the same conventions, and found that the $L$ particle reflects of the left wall with
\begin{eqnarray}
\tilde{K}^L(p_i) =  \begin{pmatrix}1&0\\0&\tilde{\sigma}(p_i)\end{pmatrix}, \qquad \tilde{\sigma}(p_i) = \frac{e^{-i\frac{p_i}{2}}-\tilde{c}_{0}}{e^{i\frac{p_i}{2}}+\tilde{c}_{0}},\qquad i=1,2,    
\end{eqnarray}
with $\tilde{c}_{0}$ any constant.
One $R$ particles instead reflects off the left wall with 
\begin{eqnarray}
\tilde{K}^{R}(p_i) =  \begin{pmatrix}1&0\\0&\tilde{\sigma}^{R}(p_i)\end{pmatrix}, \qquad \tilde{\sigma}^{R}(p_i) = -\frac{e^{i\frac{p_i}{2}}+\tilde{c}_{0}^{R} \, e^{i p_i}}{e^{i\frac{p_i}{2}}-\tilde{c}_{0}^{R}},\qquad i=1,2,    
\end{eqnarray}
with $\tilde{c}_{0}^{R}$ any constant. Unitarity works for both $\tilde{c}_{0}$ equal to $0$ or $\pm \infty$, and the same for $\tilde{c}_{0}^{R}$. This implies the standard conditions
\begin{eqnarray}
    \tilde{k}^L(-p_i)\tilde{k}^L(p_i)=1=\tilde{k}^{R}(-p_i)\tilde{k}^{R}(p_i)
\end{eqnarray}
on the reflection dressing.

To solve the mixed boundary Yang-Baxter equations with at least one $L$ and one $\tilde{L}$ particle we need to set
\begin{eqnarray}
\tilde{c}_{0}=\frac{1}{\tilde{c}_{0}^{R}}.    \end{eqnarray}
We have also in this case the same conditions on the dressing factors as those coming from the right wall analysis.

\section{Physical representations for the massive case: $L$ and $R$}\label{App:B}

We now compute the bulk $R$-matrices for the physical massive representations $L$ and $R$, with the symmetric coproduct.

The massive $L$ representation reads as before
\begin{eqnarray}\label{massive-phys-rep-L}
&&\pi^L_p(\mathfrak{G}_L) = \sqrt{\frac{h}{2}} \, \eta_p \begin{pmatrix}0&1\\0&0\end{pmatrix}, \qquad \pi^L_p(\mathfrak{Q}_L) = \sqrt{\frac{h}{2}} \, \eta_p \begin{pmatrix}0&0\\1&0\end{pmatrix},\nonumber\\
&&\pi^L_p(\mathfrak{G}_R) = -\sqrt{\frac{h}{2}} \, \frac{\eta_p \, e^{-i\frac{p}{2}}}{x_p^-} \begin{pmatrix}0&0\\1&0\end{pmatrix}, \qquad \pi^L_p(\mathfrak{Q}_R) = -\sqrt{\frac{h}{2}} \, \frac{\eta_p \, e^{i\frac{p}{2}}}{x_p^+} \begin{pmatrix}0&1\\0&0\end{pmatrix}.
\end{eqnarray}
The massive $R$ representation reads
\begin{eqnarray}\label{massive-phys-rep-R}
&&\pi^R_p(\mathfrak{G}_L) = -\sqrt{\frac{h}{2}} \, \frac{\eta_p}{\sqrt{x^+_p x^-_p}} \begin{pmatrix}0&0\\1&0\end{pmatrix}, \qquad \pi^R_p(\mathfrak{Q}_L) = -\sqrt{\frac{h}{2}} \, \frac{\eta_p}{\sqrt{x^+_p x^-_p}} \begin{pmatrix}0&1\\0&0\end{pmatrix},\nonumber\\
&&\pi^R_p(\mathfrak{G}_R) = \sqrt{\frac{h}{2}} \, \eta_p \, \begin{pmatrix}0&1\\0&0\end{pmatrix}, \qquad \pi^R_p(\mathfrak{Q}_R) = \sqrt{\frac{h}{2}} \, \eta_p \, \begin{pmatrix}0&0\\1&0\end{pmatrix}.
\end{eqnarray}
Both the $L$ and $R$ representations coincide with the $\rho_L$, resp. $R$, rep of \cite{Borsato:2014hja} modulo a different choice of phases multiplying the supercharges - recalling that $\big(\frac{x^+}{x^-}\big)^\lambda = e^{i \lambda p}$.

We have already displayed in the main text the $LL$ bulk $R$-matrix, but we repeat it here for completeness:

\begin{itemize}

\item $L-L$

\begin{eqnarray}
R^{LL} = \Phi_m^{LL} \, \begin{pmatrix}
\frac{e^{\frac{i}{4}(p-q)}(x^+_q-x^-_p)}{(x^-_q-x^+_p)}&0&0&0\\
0&\frac{e^{-\frac{i}{4}(p+q)}(x^+_q-x^+_p)}{(x^-_q-x^+_p)}&\frac{i\eta_p \eta_q}{(x^-_q-x^+_p)}&0\\
0&\frac{i\eta_p \eta_q}{(x^-_q-x^+_p)}&\frac{e^{\frac{i}{4}(p+q)}(x^-_q-x^-_p)}{(x^-_q-x^+_p)}&0\\
0&0&0&e^{\frac{i}{4}(q-p)}\end{pmatrix}.
\end{eqnarray}    

\item $R-R$

\begin{eqnarray}
R^{RR} = \Phi_m^{RR} \,\begin{pmatrix}
e^{\frac{3i}{4}(p-q)}\frac{x^-_p-x^+_q}{x^+_p-x^-_q}&0&0&0\\
0&\frac{e^{\frac{i}{4}(p-3q)}(x^+_q-x^+_p)}{(x^-_q-x^+_p)}&\frac{ie^{\frac{i}{2}(p-q)}\eta_p \eta_q}{(x^-_q-x^+_p)}&0\\
0&\frac{ie^{\frac{i}{2}(p-q)}\eta_p \eta_q}{(x^-_q-x^+_p)}&\frac{e^{\frac{i}{4}(3p-q)}(x^-_q-x^-_p)}{(x^-_q-x^+_p)}&0\\
0&0&0&e^{\frac{i}{4}(p-q)}\end{pmatrix}.
\end{eqnarray}

\item $L-R$

\begin{eqnarray}
R^{LR} = \Phi^{LR}_m \, \begin{pmatrix}
1&0&0&\frac{e^{\frac{i}{4}(p-q)}i \eta_p \eta_q}{(x^-_q x^+_p-1)}\\
0&\frac{e^{\frac{ip}{2}}(x^-_p x^-_q-1)}{(x^-_q x^+_p-1)}&0&0\\
0&0&\frac{e^{-\frac{i q}{2}}(x^+_p x^+_q-1)}{(x^-_q x^+_p-1)}&0\\
\frac{e^{-\frac{i}{4}(-p+q)}i \eta_p \eta_q}{(x^-_q x^+_p-1)}&0&0&\frac{e^{-\frac{i}{2}(-p+q)}(x^-_px^+_q-1)}{(x^-_q x^+_p-1)}\end{pmatrix}.   
\end{eqnarray}

\item $R-L$

\begin{eqnarray}
R^{RL} = \Phi^{RL}_m \, \begin{pmatrix}
1&0&0&\frac{e^{\frac{i}{4}(p-q)}i \eta_p \eta_q}{(x^-_q x^+_p-1)}\\
0&\frac{e^{\frac{ip}{2}}(x^-_p x^-_q-1)}{(x^-_q x^+_p-1)}&0&0\\
0&0&\frac{e^{\frac{-i q}{2}}(x^+_p x^+_q-1)}{(x^-_q x^+_p-1)}&0\\
\frac{e^{\frac{i}{4}(p-q)}i \eta_p \eta_q}{(x^-_q x^+_p-1)}&0&0&\frac{e^{-\frac{i}{2}(-p+q)}(x^-_px^+_q-1)}{(x^-_q x^+_p-1)}\end{pmatrix}.   
\end{eqnarray}

\end{itemize}

In order to find the reflection matrices, we again take the boundary subalgebras $\mathcal{B}^\mathfrak{U}_L$ and $\mathcal{B}^\mathfrak{U}_R$ as given in (\ref{rescaled-susy-boundary-algebras}), and solve the boundary intertwining equations for the right wall:  
\begin{equation}\label{right-wall-phys-massive-BIE}
\pi^L_{-p}(\mathfrak{b}) \, K^L(p) = K^L(p) \, \pi^{L}_p(\mathfrak{b}) \qquad \text{and} \qquad   \pi^R_{-p}(\mathfrak{b}) \, K^R(p) =K^R(p) \, \pi^R_{p}(\mathfrak{b}) \qquad \forall \, \mathfrak{b} \in \mathcal{B}.
\end{equation}
Here we use the physical representations $\pi^L$ and $\pi^R$ in (\ref{massive-phys-rep-L})
and (\ref{massive-phys-rep-R}) respectively, and we consider the boundary subalgebra $\mathcal{B}$ either $\mathcal{B}^\mathfrak{U}_L$ or $\mathcal{B}^\mathfrak{U}_R$. Solving the boundary intertwining equations yields us the following reflection matrices (upto an overall dressing factor):
\begin{eqnarray}\label{KL-phys-matrix-massive}
K^L_{\mathcal{B}^\mathfrak{U}_L}(p) = \begin{pmatrix}1&0\\0&e^{-\frac{i p}{2}}\end{pmatrix} \qquad \text{and} \qquad K^L_{\mathcal{B}^\mathfrak{U}_R}(p) = \begin{pmatrix}1&0\\0&-e^{\frac{i p}{2}}\end{pmatrix}.
\end{eqnarray}
for the particle in $L$ representation and
\begin{eqnarray}\label{KR-phys-matrix-massive}
K^R_{\mathcal{B}^\mathfrak{U}_L}(p) = \begin{pmatrix}1&0\\0&e^{\frac{i p}{2}}\end{pmatrix} \qquad \text{and} \qquad K^R_{\mathcal{B}^\mathfrak{U}_R}(p) = \begin{pmatrix}1&0\\0&e^{-\frac{i p}{2}}\end{pmatrix}.
\end{eqnarray}
for the particle in the $R$ representation. We have checked that this reflection matrix satisfies the massive Boundary Yang-Baxter equation
\begin{eqnarray}
&&K_2(p_2) R^{op}\big(p_2,-p_1\big)K_1(p_1)R(p_1,p_2) =  R^{op}\big(-p_2,-p_1\big)K_1(p_1)R\big(p_1,-p_2\big)K_2(p_2),
\end{eqnarray}
with the massive bulk $R$-matrices of this section.

\section{Twist}\label{App:C}

We can introduce a twist which connects the symmetric coproduct which we have been using so far, to the one used in appendix \ref{App:A} and in \cite{Prinsloo:2015apa}. First, let us introduce an additional bosonic generator $\mathfrak{B}$, with commutation relations
\begin{eqnarray}
&&[\mathfrak{B},\mathfrak{Q}_L] = \mathfrak{Q}_L, \qquad [\mathfrak{B},\mathfrak{G}_L] = -\mathfrak{G}_L, \nonumber\\
&&[\mathfrak{B},\mathfrak{Q}_R] = -\mathfrak{Q}_R, \qquad [\mathfrak{B},\mathfrak{G}_R] = \mathfrak{G}_R.
\end{eqnarray}

We now define the twist as
\begin{eqnarray}
&&T_L = e^{\tau_L}, \qquad \tau_L = \frac{i}{4}\Big[\mathfrak{p} \otimes \mathfrak{B} + \mathfrak{B} \otimes \mathfrak{p}\Big],\nonumber\\
&&T_R = e^{\tau_R}, \qquad \tau_R = -\frac{i}{4}\Big[\mathfrak{p} \otimes \mathfrak{B} + \mathfrak{B} \otimes \mathfrak{p}\Big].
\end{eqnarray}
We can easily calculate (only based on the commutation relations)
\begin{eqnarray}
&&[\tau_L,\mathfrak{Q}_L \otimes \mathbbmss{1}] = \frac{i}{4} \mathfrak{Q}_L \otimes \mathfrak{p}, \qquad  [\tau_L,\mathfrak{G}_L \otimes \mathbbmss{1}] =  -\frac{i}{4} \mathfrak{G}_L \otimes \mathfrak{p},\nonumber\\
&&[\tau_R,\mathfrak{Q}_R \otimes \mathbbmss{1}] =  \frac{i}{4} \mathfrak{Q}_R \otimes \mathfrak{p}, \qquad  [\tau_R,\mathfrak{G}_R \otimes \mathbbmss{1}] = -\frac{i}{4} \mathfrak{G}_R \otimes \mathfrak{p},\nonumber\\
&&[\tau_L,\mathbbmss{1}\otimes \mathfrak{Q}_L] = \frac{i}{4} \mathfrak{p} \otimes \mathfrak{Q}_L, \qquad [\tau_L,\mathbbmss{1}\otimes \mathfrak{G}_L] = -\frac{i}{4} \mathfrak{p} \otimes \mathfrak{G}_L,\nonumber\\
 &&[\tau_R,\mathbbmss{1}\otimes \mathfrak{Q}_R] = \frac{i}{4} \mathfrak{p} \otimes \mathfrak{Q}_R, \qquad [\tau_R,\mathbbmss{1}\otimes \mathfrak{G}_R] = -\frac{i}{4} \mathfrak{p} \otimes \mathfrak{G}_R.
\end{eqnarray}
If $c$ is any central element, and $\mathfrak{b},\mathfrak{x}$ are abstract generators satisfying 
\begin{eqnarray}
[\mathfrak{b},\mathfrak{x}]=c \, \mathfrak{x},
\end{eqnarray}
then one deduces
\begin{eqnarray}
e^\mathfrak{b} \mathfrak{x} = \sum_{n=0}^\infty \frac{1}{n!} \mathfrak{b}^n \, \mathfrak{x} = \sum_{n=0}^\infty \frac{1}{n!} \mathfrak{x} \, (\mathfrak{b} + c)^n = \mathfrak{x} \, e^\mathfrak{b} \, e^c.
\end{eqnarray}
This means that
\begin{eqnarray}
&&e^{\tau_L} [\mathfrak{Q}_L \otimes \mathbbmss{1}]= [\mathfrak{Q}_L \otimes \mathbbmss{1}]e^{\tau_L} e^{\frac{i}{4} \mathbbmss{1} \otimes \mathfrak{p}}, \qquad e^{\tau_L} [\mathfrak{G}_L \otimes \mathbbmss{1}]= [\mathfrak{G}_L \otimes \mathbbmss{1}]e^{\tau_L} e^{-\frac{i}{4} \mathbbmss{1} \otimes \mathfrak{p}}, \nonumber\\
&&e^{\tau_R} [\mathfrak{Q}_R \otimes \mathbbmss{1}]= [\mathfrak{Q}_R \otimes \mathbbmss{1}]e^{\tau_R} e^{\frac{i}{4} \mathbbmss{1} \otimes \mathfrak{p}}, \qquad e^{\tau_R} [\mathfrak{G}_R\otimes \mathbbmss{1}] = [\mathfrak{G}_R \otimes \mathbbmss{1}]e^{\tau_R} e^{-\frac{i}{4} \mathbbmss{1} \otimes \mathfrak{p}}\nonumber\\
&&e^{\tau_L} [\mathbbmss{1} \otimes \mathfrak{Q}_L]= [\mathbbmss{1}\otimes \mathfrak{Q}_L]e^{\tau_L} e^{\frac{i}{4} \mathfrak{p} \otimes \mathbbmss{1}}, \qquad e^{\tau_L} [\mathbbmss{1} \otimes \mathfrak{G}_L]= [\mathbbmss{1}\otimes \mathfrak{G}_L]e^{\tau_L} e^{-\frac{i}{4} \mathfrak{p} \otimes \mathbbmss{1}},\nonumber\\
&&e^{\tau_R} [\mathbbmss{1} \otimes \mathfrak{Q}_R]= [\mathbbmss{1}\otimes \mathfrak{Q}_R]e^{\tau_L} e^{\frac{i}{4} \mathfrak{p} \otimes \mathbbmss{1}}, \qquad e^{\tau_R} [\mathbbmss{1} \otimes \mathfrak{G}_R]= [\mathbbmss{1}\otimes \mathfrak{G}_L]e^{\tau_R} e^{-\frac{i}{4} \mathfrak{p} \otimes \mathbbmss{1}}.
\label{use}
\end{eqnarray}
At this point we write the symmetric coproduct as
\begin{eqnarray}
&&\Delta(\mathfrak{Q}_L) = \Big[\mathfrak{Q}_L \otimes \mathbbmss{1}\Big]\Big[\mathbbmss{1} \otimes e^{-i \frac{\mathfrak{p}}{4}}\Big] + \Big[e^{i \frac{\mathfrak{p}}{4}}\otimes \mathbbmss{1}\Big]\Big[\mathbbmss{1}\otimes \mathfrak{Q}_L\Big], \nonumber\\
&&\Delta(\mathfrak{G}_L) = \Big[\mathfrak{G}_L \otimes \mathbbmss{1}\Big]\Big[\mathbbmss{1} \otimes e^{i \frac{\mathfrak{p}}{4}}\Big] + \Big[e^{-i \frac{\mathfrak{p}}{4}}\otimes \mathbbmss{1}\Big]\Big[\mathbbmss{1}\otimes \mathfrak{G}_L\Big], \nonumber\\
&&\Delta(\mathfrak{Q}_R) = \Big[\mathfrak{Q}_R \otimes \mathbbmss{1}\Big]\Big[\mathbbmss{1} \otimes e^{-i \frac{\mathfrak{p}}{4}}\Big] + \Big[e^{i \frac{\mathfrak{p}}{4}}\otimes \mathbbmss{1}\Big]\Big[\mathbbmss{1}\otimes \mathfrak{Q}_R\Big], \nonumber\\
&&\Delta(\mathfrak{G}_R) = \Big[\mathfrak{G}_R \otimes \mathbbmss{1}\Big]\Big[\mathbbmss{1} \otimes e^{i \frac{\mathfrak{p}}{4}}\Big] + \Big[e^{-i \frac{\mathfrak{p}}{4}}\otimes \mathbbmss{1}\Big]\Big[\mathbbmss{1}\otimes \mathfrak{G}_R\Big],
\end{eqnarray}
and use the fact that any expression only involving the generator $\mathfrak{p}$ will commute through everything. We therefore write, using (\ref{use}),
\begin{eqnarray}
&&e^{\tau_L} \Delta(\mathfrak{Q}_L)e^{-\tau_L} = \Delta(\mathfrak{Q}_L) = \Big[\mathfrak{Q}_L \otimes \mathbbmss{1}\Big]\Big[\mathbbmss{1} \otimes \mathbbmss{1}\Big] + \Big[e^{i \frac{\mathfrak{p}}{2}}\otimes \mathbbmss{1}\Big]\Big[\mathbbmss{1}\otimes \mathfrak{Q}_L\Big], \nonumber\\
&&e^{\tau_L} \Delta(\mathfrak{G}_L)e^{-\tau_L} = \Delta(\mathfrak{G}_L) = \Big[\mathfrak{G}_L \otimes \mathbbmss{1}\Big]\Big[\mathbbmss{1} \otimes \mathbbmss{1}\Big] + \Big[e^{-i \frac{\mathfrak{p}}{2}}\otimes \mathbbmss{1}\Big]\Big[\mathbbmss{1}\otimes \mathfrak{G}_L\Big], \nonumber\\
&&e^{\tau_R} \Delta(\mathfrak{Q}_R)e^{-\tau_R} = \Delta(\mathfrak{Q}_R) = \Big[\mathfrak{Q}_R \otimes \mathbbmss{1}\Big]\Big[\mathbbmss{1} \otimes \mathbbmss{1}\Big] + \Big[e^{i \frac{\mathfrak{p}}{2}}\otimes \mathbbmss{1}\Big]\Big[\mathbbmss{1}\otimes \mathfrak{Q}_R\Big], \nonumber\\
&&e^{\tau_R} \Delta(\mathfrak{G}_R)e^{-\tau_R} = \Delta(\mathfrak{G}_R) = \Big[\mathfrak{G}_R \otimes \mathbbmss{1}\Big]\Big[\mathbbmss{1} \otimes \mathbbmss{1}\Big] + \Big[e^{-i \frac{\mathfrak{p}}{2}}\otimes \mathbbmss{1}\Big]\Big[\mathbbmss{1}\otimes \mathfrak{G}_R\Big].\label{see}
\end{eqnarray}
We see that (\ref{see}) is the coproduct used in appendix \ref{App:A} and in \cite{Prinsloo:2015apa}. Furthermore, since we never rely on the dispersion relation for any of the arguments contained in this subsection, the twist which we have found works for massive as well as massless representations.

\end{appendix}


\end{document}